\documentclass[webpdf,contemporary,large]{oup-authoring-template}%

\graphicspath{{Fig/}}
\usepackage{natbib}
\usepackage{subfigure}
\usepackage{bm}
\usepackage{subcaption}

\theoremstyle{thmstyleone}%
\newtheorem{theorem}{Theorem}
\theoremstyle{thmstyletwo}%
\theoremstyle{thmstylethree}%

\expandafter\def\expandafter\normalsize\expandafter{%
    \normalsize%
    \setlength\abovedisplayskip{0pt}%
    \setlength\belowdisplayskip{8pt}%
    \setlength\abovedisplayshortskip{-6pt}%
    \setlength\belowdisplayshortskip{2pt}%
}

\begin{document}

\journaltitle{Bioinformatics}
\DOI{DOI HERE}
\copyrightyear{2026}
\pubyear{2026}
\access{Advance Access Publication Date: Day Month Year}
\appnotes{Original Paper}

\firstpage{1}

\subtitle{Data and text mining}

\title[B-MASTER]{B-MASTER: Scalable Bayesian Multivariate Regression for Master Predictor Discovery in Colorectal Cancer Microbiome-Metabolite Profiles}

\author[1,$\ast$]{Priyam Das\ORCID{0000-0003-2384-0486}}
\author[2]{Tanujit Dey\ORCID{0000-0001-5559-211X}}
\author[3]{Christine B.\ Peterson\ORCID{0000-0003-3316-0468}}
\author[4]{Sounak Chakraborty\ORCID{0000-0003-2022-1735}}

\address[1]{\orgdiv{Department of Biostatistics}, \orgname{Virginia Commonwealth University}}
\address[2]{\orgdiv{Center for Surgery and Public Health, Brigham and Women’s Hospital}, \orgname{Harvard Medical School}}
\address[3]{\orgdiv{Department of Statistics}, \orgname{Rice University}}
\address[4]{\orgdiv{Department of Statistics}, \orgname{University of Missouri}}

\corresp[$\ast$]{Corresponding author. \href{dasp4@vcu.edu}{dasp4@vcu.edu}}

\received{Date}{0}{2026}
\revised{Date}{0}{2026}
\accepted{Date}{0}{2026}


\abstract{\textbf{Motivation:}
The gut microbiome shapes cancer therapy response through its influence on host metabolism. While prior studies examine pairwise associations between individual genera and metabolites, there is limited methodology for identifying microbial genera that systematically regulate the overall metabolome. Scalable statistical tools are needed to uncover such system-level ``master predictors'' in high-dimensional microbiome-metabolome data.\\
\textbf{Results:}
We introduce B-MASTER, a scalable Bayesian multivariate regression framework combining $\ell_1$ sparsity and $\ell_2$ group shrinkage to identify essential cross-metabolite regulators. A Gibbs sampler enables near-linear computational scaling, supporting models with millions of parameters. The method is supported by theoretical guarantees, including posterior contraction and selection consistency. Analysis of colorectal cancer microbiome-metabolome data reveals key microbial genera that govern global and cancer-associated metabolite patterns, highlighting system-level regulatory structure.\\
\textbf{Availability:} The B-MASTER code, including demonstration scripts, is available at \href{https://github.com/priyamdas2/B-MASTER}{https://github.com/priyamdas2/B-MASTER}. An archived snapshot of the code corresponding to this manuscript is available on Zenodo with DOI: \href{https://doi.org/10.5281/zenodo.20484958}{10.5281/zenodo.20484958}.
\\ 
\textbf{Contact:} dasp4@vcu.edu\\
\textbf{Supplementary information:} Supplementary data are available at \textit{Bioinformatics} online.
}
\maketitle
\vspace{-3.8cm}
\section{Introduction} 
Advances in sequencing technologies and bioinformatic methods have revealed the critical role of the microbiome in human health and disease \citep{Xu2015b, li_microbiome}. In colorectal cancer (CRC), gut microbial communities not only contribute to tumor initiation and progression but also influence therapeutic response \citep{Garrett2015}. High-throughput sequencing now enables detailed analyses of microbial richness and abundance \citep{Caporaso2011}, and when combined with metabolomic profiling, offers a window into the microbial metabolites that shape immune regulation, inflammation, and cancer development \citep{Belkaid2014, Zhang2021}.  
\begin{figure}[h]
    \centering
    \includegraphics[width=0.43\textwidth]{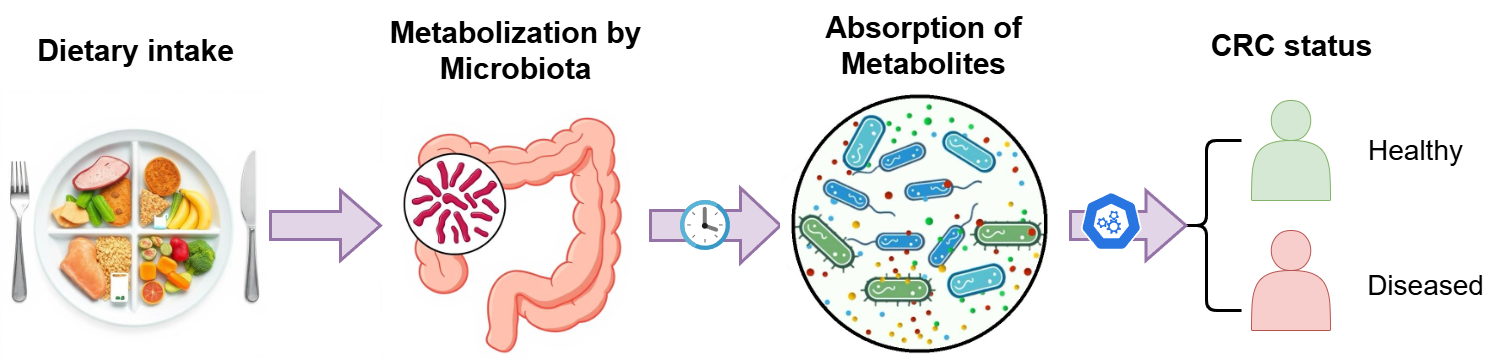}\vspace{-0.4cm}
    \caption{Conceptual illustration of how gut microbial activity influences metabolite production, which in turn may affect colorectal cancer progression.}\vspace{-0.7cm}
    \label{fig:concept}
\end{figure}
The microbiome exerts its effects largely through metabolism. Metabolites such as short-chain fatty acids, amino acids, and fermentation products have been linked to both protective and pro-carcinogenic processes \citep{Zhang2023, Lu2024, Song2022}. Yet most existing studies focus on isolated pathways or pairwise associations, leaving open the broader question of which microbial genera systematically regulate the overall metabolite profile. This question is particularly important for metabolites that are differentially abundant in CRC, where identifying the microbial drivers may provide mechanistic insight and therapeutic opportunities \citep{Yachida2019}. A conceptual illustration of how microbial activity shapes metabolites and their downstream influence on colorectal cancer is provided in Figure~\ref{fig:concept}.

Our study aims to identify ``master predictors'': microbial genera exerting coordinated influence across metabolite pathways. Statistically, this requires selecting regressors affecting multiple outcomes simultaneously while controlling false positives in high-dimensional settings. Variable selection has progressed from Ridge regression \citep{Hoerl1970}, LASSO \citep{Tibshirani1996}, and Elastic-Net \citep{Zou2005} to structured extensions such as group and adaptive Lasso \citep{Yuan2006}. Bayesian formulations, including the Bayesian Lasso and Spike-and-Slab approaches \citep{Park2008,Rockova2018}, further enhanced shrinkage-based inference, with scalable adaptations emerging for modern data \citep{Qian2020}. In multivariate regression, methods such as Bayesian group Lasso \citep{Liquet2017}, Spike-and-Slab Lasso \citep{Deshpande2019}, and reduced-rank regression \citep{Chakraborty2020} improve sparse estimation across correlated outcomes. However, many remain computationally burdensome in ultra-high dimensions or do not explicitly target elimination of non-essential regressors. Frequentist approaches such as remMap \citep{Peng2010} attempt master predictor identification but exhibit scalability limitations and instability under correlated predictors. These challenges motivate a scalable Bayesian framework tailored to high-dimensional microbiome-metabolome integration.

We address these challenges by proposing \textit{Bayesian Multivariate regression Analysis for Selecting Targeted Essential Regressors} (B-MASTER), which reformulates the combined $l_1$ and $l_2$ penalization strategy within a fully Bayesian framework. Through a hierarchical prior specification, B-MASTER yields a Gibbs sampling algorithm whose computation time remains nearly invariant with sample size, permitting full posterior inference even in ultra-high-dimensional models. Compared to existing approaches, B-MASTER offers several advantages: (i) tuning parameters are automatically adjusted through hyperpriors, avoiding repeated cross-validation; (ii) posterior samples provide direct measures of uncertainty for regression coefficients, eliminating the need for computationally intensive bootstrapping; (iii) computation time is predictable and robust across datasets of similar dimension; and (iv) the algorithm scales efficiently to ultra-high-dimensional models, with demonstrated performance on problems involving up to four million parameters, while maintaining predictable linear computation time. Beyond computational efficiency, B-MASTER is supported by rigorous high-dimensional theory, establishing posterior contraction, selection consistency for master predictors, and robustness to correlated errors and mild model misspecification.

The remainder of the paper is organized as follows. Section~\ref{sec:method} introduces the B-MASTER model, the associated Gibbs sampling scheme, and highlights its theoretical properties. Section~\ref{sim_study} evaluates performance in simulation studies. Section~\ref{sec:case_study} applies B-MASTER to identify master predictors regulating the metabolomics of CRC subjects. Section~\ref{sec:discussion} concludes with a discussion of implications and future directions.
\vspace{-0.5cm}
\section{B-MASTER} \label{sec:method}
\vspace{-0.1cm}
\subsection{Model} 
\begin{figure}[!t]
\centering
\includegraphics[width=.7\linewidth]{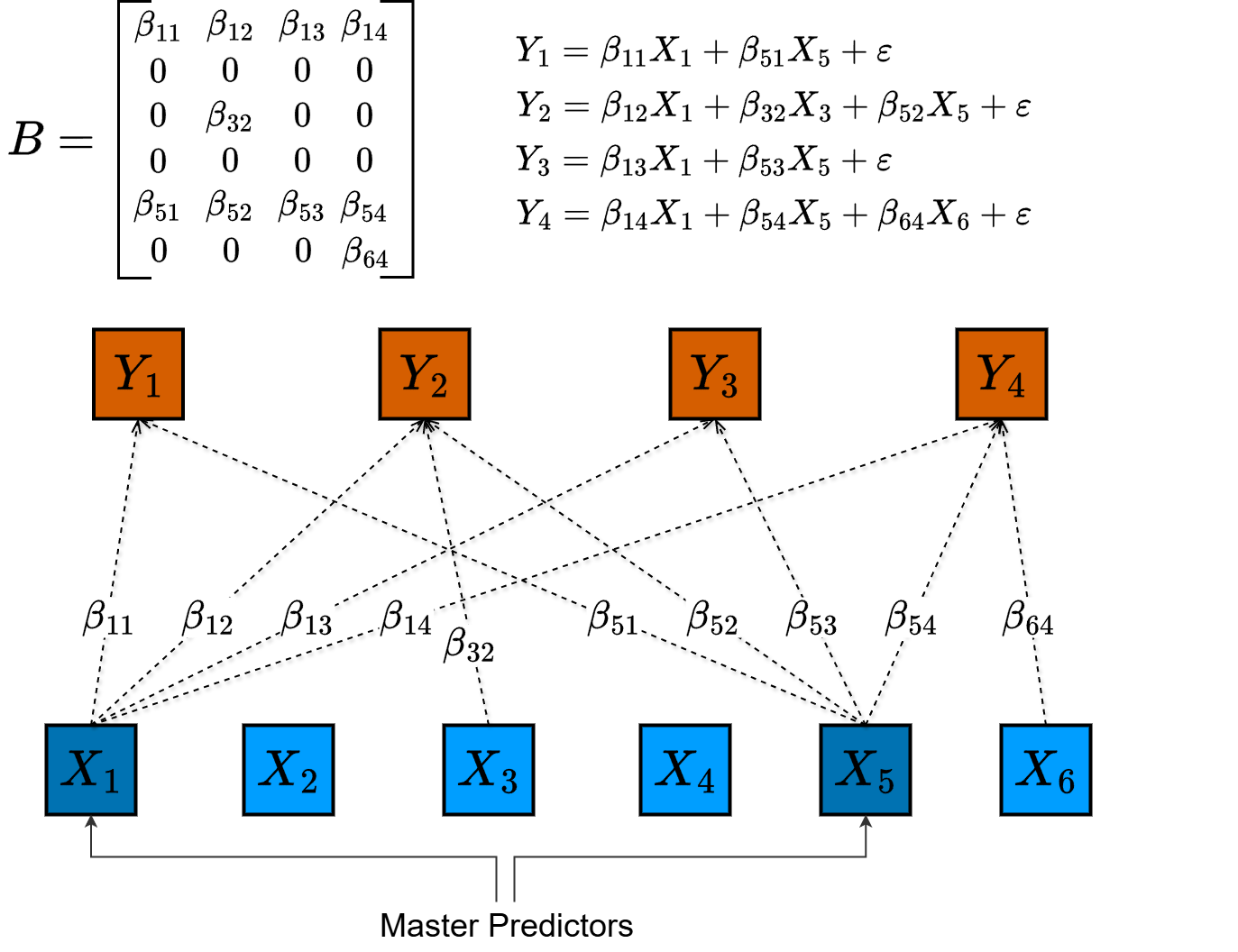}\vspace{-0.2cm}
\caption{Conceptual diagram illustrating `master predictors': predictors such as $X_1$ and $X_5$
exert influence on multiple response variables, highlighting their role as master predictors.}
\label{Bmaster_concept}\vspace{-0.5cm}
\end{figure}

\vspace{-0.2cm}
Suppose we have $N$ observations with data $(\bm{x}^i, \bm{y}^i)$ for $i = 1, \ldots, N$, where $\bm{x}^i = (x^i_1, \ldots, x^i_P)$ is a $1 \times P$ vector of predictors, and $\bm{y}^i = (y^i_1, \ldots, y^i_Q)$ is a $1 \times Q$ vector of responses. The multivariate regression model is given by
\begin{align}
\bm{y}^i = \bm{x}^i \bm{B} + \bm{\epsilon}^i, \quad i = 1, \ldots, N,
\label{model1}
\end{align}
where $\bm{\epsilon}^i \stackrel{\text{i.i.d.}}{\sim} N_Q(\mathbf{0}, \mathrm{diag}(\sigma_1^2, \ldots, \sigma_Q^2))$. Here, $\bm{B} = \big(\beta_{pq}\big)$ denotes the $P \times Q$ matrix of regression coefficients. In our case study, we aim to identify the key components of the microbiome that exert the most significant influence on the overall microbiome-metabolite interaction. This hypothesis is grounded in recent studies, such as \citet{Liu2022}, which explore the relationship between microbiome composition and metabolite production, demonstrating that specific microbiome species substantially affect the types and concentrations of metabolites within a host organism. Consequently, it is essential to impose sparsity on the $\bm{B}$-matrix in a structured manner. It is also important to ensure that the proposed model not only achieves overall sparsity but also facilitates the identification of critical predictors, enabling a biologically interpretable understanding of the microbiome-metabolite relationship. A conceptual diagram of `master predictors', which are subsequently identified using B-MASTER, is presented in Figure \ref{Bmaster_concept}.
\vspace{-0.3cm}
\subsection{Prior Formulation and Gibbs Sampling} 
Extensive research on variable selection has shown that, in high-dimensional settings, predictor-response relationships are often driven by a small subset of influential predictors \citep{Qian2020}. Penalized regression methods, such as LASSO \citep{Tibshirani1996}, are designed to capture these sparse structures, even in the presence of multicollinearity. The $l_1$-penalty applied to the coefficients of $\bm{B}$ induces sparsity by shrinking many coefficients to zero. Additionally, to identify predictors with dominant effects across multiple response components, an $l_2$-norm can be applied to the coefficient vectors of each predictor \citep{Peng2010}. This hybrid penalization balances model parsimony and shrinkage for less significant predictors. Following \citet{Kyung2010}, we adopt the conditional prior
\begin{align}
\pi(\bm{B}|\sigma_1^2,\ldots,\sigma_Q^2)\propto
\exp\bigg[-\lambda_1\sum_{p=1}^P\sum_{q=1}^{Q}\frac{|\beta_{pq}|}{2\sigma_q^2}
-
\lambda_2\sum_{p=1}^P\bigg(\sum_{q=1}^{Q} \frac{\beta_{pq}^2}{2\sigma^2_q}\bigg)^{1/2}\bigg].
\label{kyung_prior}
\end{align}

The parameter $\lambda_1$ controls overall coefficient shrinkage, whereas $\lambda_2$ penalizes the $\ell_2$-norm of coefficients within each predictor across responses. Consequently, the $l_1$ component encourages sparsity among individual coefficients $\beta_{pq}$, while the group-wise $l_2$ component acts on the entire coefficient vector associated with predictor $p$. Predictors exhibiting weak effects across all responses are therefore shrunk toward zero, whereas predictors with broad influence across multiple responses are more likely to be retained. This structure is particularly appealing for identifying \emph{master predictors}, namely predictors that influence many response variables simultaneously. Although the prior in \eqref{kyung_prior} combines $l_1$- and $l_2$-type shrinkage, it differs fundamentally from the Elastic-Net penalty \citep{Zou2005}. A detailed discussion of these differences is provided in Supplement Section~S1.

To further characterize the prior structure, let $\bm{\beta}^{(p)} = (\beta_{p1}, \ldots, \beta_{pQ})$ denote the coefficient vector corresponding to predictor $p$. Then \eqref{kyung_prior} factorizes as
\begin{align}
\label{factorized_prior}
\pi(\bm{B}|\sigma_1^2,\ldots,\sigma_Q^2)  =& \prod_{p=1}^P \pi(\bm{\beta}^{(p)}|\sigma_1^2,\ldots,\sigma_Q^2),\;\; \text{where,} \nonumber \\
\pi(\bm{\beta}^{(p)}|\sigma_1^2,\ldots,\sigma_Q^2)  \propto &
	\exp\bigg[-\lambda_1\sum_{q=1}^{Q}\frac{|\beta_{pq}|}{2\sigma_q^2}
	- \lambda_2\bigg\{\sum_{q=1}^{Q} \frac{\beta_{pq}^2}{2\sigma^2_q}\bigg\}^{1/2}\bigg].
\end{align}

Although the prior in \eqref{factorized_prior} directly encodes the desired sparsity structure, posterior computation under this formulation is not conjugate. Following \citet{Xu2015}, we therefore employ an equivalent Gaussian scale-mixture representation obtained through the introduction of latent variance components. This reparameterization preserves the original prior while enabling efficient Gibbs sampling through closed-form conditional updates. Specifically, we introduce latent parameters $\bm{T}^2 = (\bm{\tau}_1^2, \ldots,  \bm{\tau}_P^2)^T = \big(\tau_{pq}^2\big)_{P \times Q}$ and $\bm{G}^2 = \big(\gamma_{p}^2\big)_{P \times 1}$. Here, $\tau_{pq}^2$ is a coefficient-specific local variance parameter associated with $\beta_{pq}$, whereas $\gamma_p^2$ is a predictor-level variance parameter shared across all responses for predictor $p$. Small values of $\tau_{pq}^2$ encourage shrinkage of individual coefficients, while small values of $\gamma_p^2$ shrink the entire coefficient vector $\bm{\beta}^{(p)}$, thereby reinforcing the master-predictor structure encoded by the prior. Under this hierarchical representation,
\begin{align*}
    \bm{\beta}^{(p)}|\bm{\tau}_p^2 ,\gamma_p^2,\sigma_1^2,\ldots,\sigma_Q^2 \sim 
	N_Q\bigg(\bm{0}, diag\left\{\sigma^2_q\left(\frac{1}{\tau_{pq}^2} + \frac{1}{\gamma^2_p}\right)^{-1}\right\}\bigg),
\end{align*}

The mixing distribution of $(\bm{\tau}_p^2,\gamma_p^2)$, the proof that integrating out the latent variables recovers the prior in \eqref{factorized_prior}, and the corresponding propriety results are provided in Supplementary Sections~S1 and S2 (Lemma S2.1). We place a non-informative prior on $\sigma_q^2$, namely $\pi(\sigma_q^2)\propto 1/\sigma_q^2$ for $q=1,\ldots,Q$, and Gamma priors on the squared penalty parameters $\lambda_1^2$ and $\lambda_2^2$. The latent variables are introduced solely as an equivalent hierarchical representation of the prior and are not parameters of primary scientific interest; their role is to facilitate efficient posterior computation within the Gibbs sampling framework.

Combining the likelihood from model \eqref{model1} with the aforementioned prior specification yields the full posterior distribution, from which a Gibbs sampling scheme can be derived. While the structure of the posterior and the detailed conditional distributions are deferred to Supplementary Section~S1, we provide below a skeletal outline of the proposed Gibbs sampler:
\vspace{-0.2cm}
\begin{itemize}
\item Sample $\bm{\beta}_q = (\beta_{1q},\ldots,\beta_{Pq})^T$ for $q=1,\ldots,Q$, from a multivariate normal distribution. 
\item Sample $\tau_{pq}^2$ for $p = 1,\ldots,P;q = 1,\ldots,Q$, such that $\tau_{pq}^2 = \frac{1}{\delta_{pq}^{(1)}}$, where $\delta_{pq}^{(1)}$ is sampled from inverse-Gaussian distribution.
\item Sample $\gamma_{p}^2$ for $p = 1,\ldots,P$, such that $\gamma_{p}^2 = \frac{1}{\delta_{p}^{(2)}}$, where $\delta_{p}^{(2)}$ is sampled from inverse-Gaussian distribution.
\item Sample $\sigma_q^2$ for $q=1,\ldots,Q$, from inverse-Gamma distribution.
\item Sample $\lambda_1^2$ and $\lambda_2^2$ from their Gamma full conditional distributions.
\end{itemize}

In B-MASTER, variable selection is performed as a post-processing step based on posterior summaries of the regression coefficients. Specifically, genus--metabolite edges are selected using credible-interval-based posterior evidence, rather than by thresholding posterior means directly; details of the selection rule and associated Bayesian posterior $p$-values are provided in Supplement Section~S3.1.

\vspace{-0.3cm}
\subsection{Computational Scalability}
\begin{figure}[H]
\centering
    \includegraphics[width=0.45\textwidth]{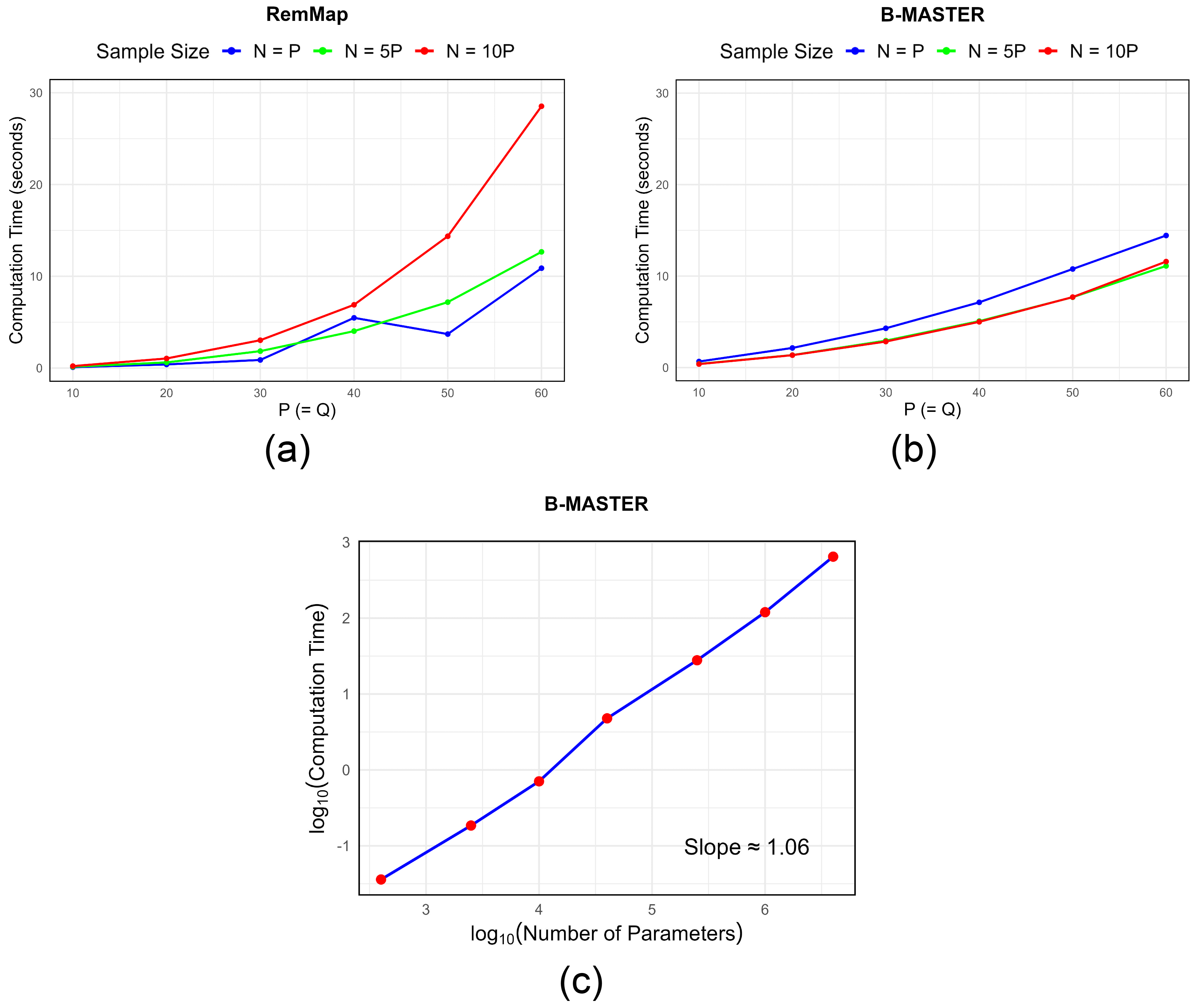}\vspace{-0.3cm}
\caption{(a) Computation times for remMap with $P = Q = 10, 20, 30, 40, 50, 60$ and $N = P, 5P, 10P$. 
(b) Computation times for B-MASTER under the same settings. While remMap runtime increases rapidly with $N$, B-MASTER shows near invariance to sample size for fixed $(P, Q)$. 
(c) Log--log plot of the number of parameters versus computation time for B-MASTER. The slope $\approx 1.06$ indicates near-linear scaling.}
\label{sample_size_effect}\vspace{-0.4cm}
\end{figure}

\vspace{-0.2cm}
A major motivation for B-MASTER is the development of a scalable Bayesian framework for identifying master predictors in high-dimensional microbiome--metabolome studies. While predictive performance and variable-selection accuracy are evaluated in the simulation studies of Section~\ref{sim_study}, an equally important consideration is computational feasibility as the dimensions of the problem increase. To provide context for the scalability results presented later in Section~\ref{sim_study_2}, we briefly contrast the computational structure of B-MASTER with that of remMap \citep{Peng2010}, a widely used frequentist approach for master predictor identification.

remMap optimizes $\bm{B}$ via the ``active-shooting'' algorithm \citep{Peng2009} with computational complexity $O(NPQ)$. As sample size increases, this leads to substantial computational burden. In contrast, the proposed Gibbs sampler performs most updates through parameter-wise conditional distributions, with only limited dataset-dependent operations that are efficiently handled by MATLAB's multi-threading \citep{MathWorks2024}. Consequently, for fixed $P$ and $Q$, computation time remains largely invariant as $N$ grows. Figures~\ref{sample_size_effect}(a) and \ref{sample_size_effect}(b) (see Section~\ref{sim_study_2}) illustrate that remMap runtime scales strongly and irregularly with $N$, whereas B-MASTER exhibits stable and predictable scaling.

\vspace{-0.5cm}

Moreover, implementation of remMap typically involves an additional tuning-parameter selection step, often based on cross-validation, while uncertainty assessment may require bootstrap-based procedures, both of which can increase overall computational cost. In contrast, the Bayesian framework underlying B-MASTER provides coefficient estimation and uncertainty quantification within a single model fit. Figure~\ref{sample_size_effect}(a) further illustrates that remMap runtime can vary noticeably across datasets even when $(P,Q)$ are held fixed. For example, when $P=Q=40$, the runtime for $N=40$ exceeded that for $N=200$, with similar behavior observed for $P=Q=50$. By comparison, Figure~\ref{sample_size_effect}(b) shows that B-MASTER exhibits more stable scaling behavior, with only modest runtime variation attributable to implementation-level factors such as increased multi-threading efficiency in higher-dimensional settings \citep{MathWorks2024}. The near-linear scaling observed in Figure~\ref{sample_size_effect}(c) is an empirical property of the proposed implementation. In practice, most updates in the proposed Gibbs sampler involve closed-form conditional distributions together with matrix operations that can be efficiently handled through MATLAB's optimized multi-threaded linear algebra routines \citep{MathWorks2024}. As a result, the observed wall-clock runtime may benefit substantially from parallel execution of low-level matrix computations, with the degree of improvement depending on the extent to which individual operations can be effectively multi-threaded. These implementation characteristics contribute to the stable computational performance observed across the high-dimensional settings considered in this study.

\vspace{-0.7cm}
\section{Simulation Study}\vspace{-0.2cm}
\label{sim_study}
In this section, we consider two complementary simulation studies. The first study is a data-driven simulation designed to closely mimic the structural characteristics of the motivating microbiome--metabolome dataset, including the observed predictor dependence and sparsity patterns. The second study is a fully synthetic high-dimensional simulation in which sparse coefficient structures are generated independently under both independent and correlated predictor settings. Together, these studies allow us to assess B-MASTER under realistic data conditions as well as under controlled settings where the true generating mechanism is known.
\vspace{-0.2cm}
\subsection{Data-driven Simulation Based on the CRC Dataset}\vspace{-0.2cm}
\label{sim_study_1}
We base the first simulation study on the Curated Gut Microbiome-Metabolome Data Resource \citep{Muller2022}, matching the case study dimensions after preprocessing ($P=287$, $Q=249$, $N=220$). Using B-MASTER, we estimate a coefficient matrix $B'$ and treat it as the ground truth. Synthetic responses are generated as $Y^{i}_q = X^{i}B'_q + \epsilon_{iq}$ with $\epsilon_{iq}\sim N(0,1)$, using the observed $X$ to preserve its empirical covariance structure (see Supplement Figure~S9). Ten replicate datasets are generated.

B-MASTER is implemented in MATLAB and run for 1,000 iterations with 100 burn-in (see Supplement~S1.5 for parameter settings). Competing methods include remMap \citep{Peng2010} and univariate/multivariate Spike-and-Slab LASSO (SSLASSO, \citealp[]{Rockova2018}; mSSL, \citealp[]{Deshpande2019}) via their R packages. Since the default cross-validation procedures did not terminate within 48 hours for remMap and mSSL under the considered high-dimensional settings, computationally feasible package-supported alternatives were adopted. For remMap, tuning parameters were selected by minimizing the Bayesian Information Criterion (BIC) over an $11\times11$ grid with $\lambda_1 \in [1,1000]$ evaluated on a logarithmic scale and $\lambda_2 \in [0,1000]$ evaluated on a uniform scale, consistent with the parameter-grid construction illustrated in the package demonstration code. For SSLASSO and mSSL, we used the package's default approximate fitting and tuning strategy, as adopted in the corresponding package demonstrations. Additional details on implementation, computational environment, tuning procedures, evaluation metrics (e.g., MCC and AUC20), and posterior edge selection are provided in Supplement Section~S3.
\begin{table}[!t]
\centering
\resizebox{0.96\columnwidth}{!}{%
\begin{tabular}{ccccccc}
\hline
Methods & TPR & FPR & MCC & AUC & AUC20 & \begin{tabular}[c]{@{}c@{}}Sparsity\\ (True = 0.76)\end{tabular} \\ \hline
B-MASTER & 0.84 (0.0012) & 0.01 (0.0003) & 0.87 (0.0001) & 0.98 (0.0004) & 0.94 (0.0014) & 0.79 (0.0003) \\
SSLASSO & 0.23 (0.0008) & 0.01 (0.0004) & 0.41 (0.0013) & 0.89 (0.0004) & 0.54 (0.0007) & 0.94 (0.0002) \\
mSSL-dcpe & 0.14 (0.0018) & 0.00 (0.0001) & 0.31 (0.0024) & 0.76 (0.0006) & 0.44 (0.0013) & 0.96 (0.0005) \\
remMap-bic & 0.00 (0.0001) & 0.00 (0.0000) & 0.02 (0.0008) & 0.58 (0.0015) & 0.23 (0.0009) & 1.00 (0.0000) \\ \hline
\end{tabular}}
\vspace{0.15cm}
\caption{Comparative study of B-MASTER, SSLASSO, mSSL, and remMap in the first simulation study based on true positive rate (TPR), false positive rate (FPR), Matthews Correlation Coefficient (MCC), AUC, and AUC20. Results are averages over 10 replicates; standard errors in parentheses.}
\label{simulation_table}
\end{table}\vspace{-0.3cm}
\begin{figure}[h!]
\centering
\includegraphics[width=.98\linewidth]{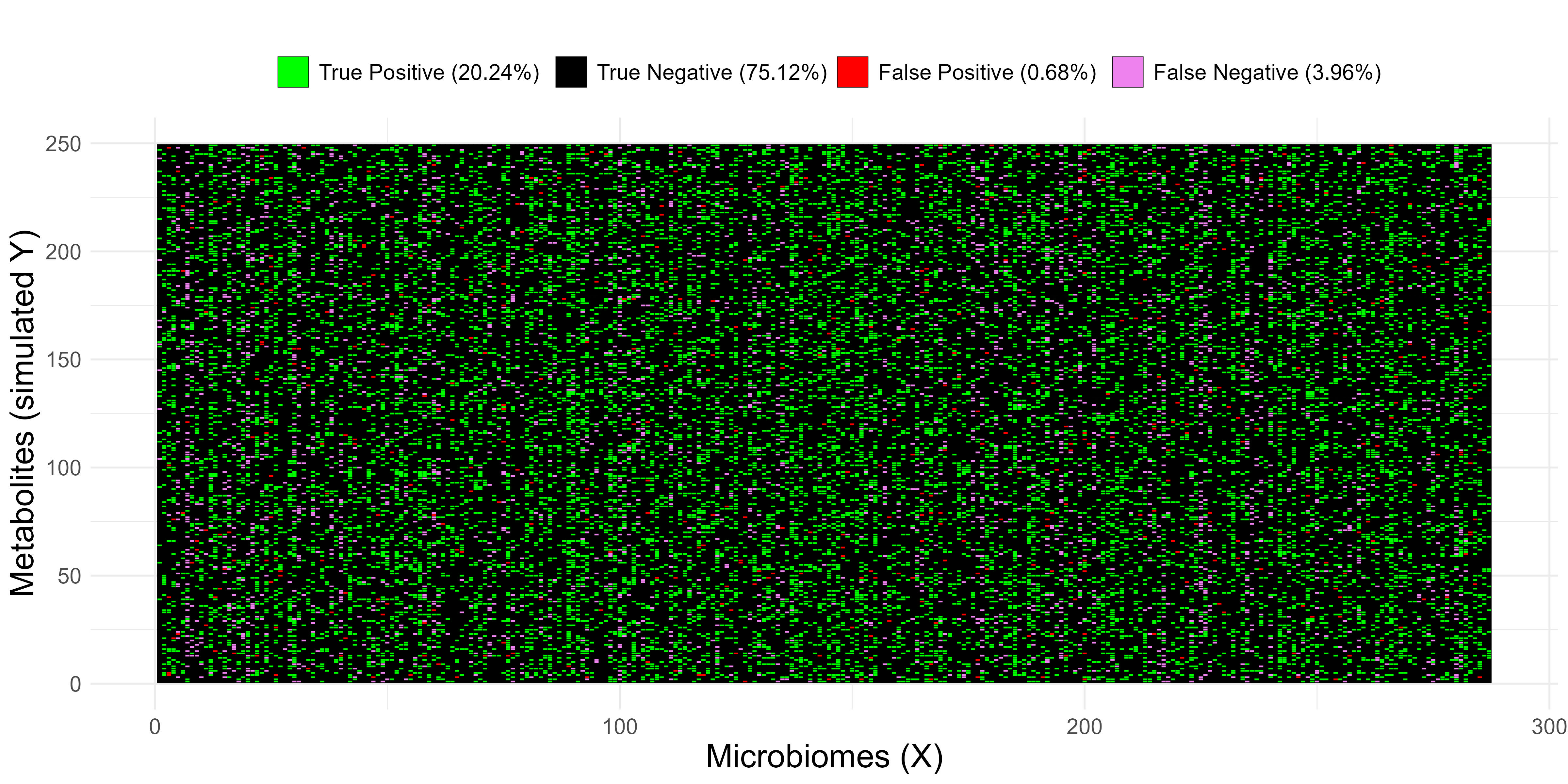} \\
\includegraphics[width=.45\linewidth]{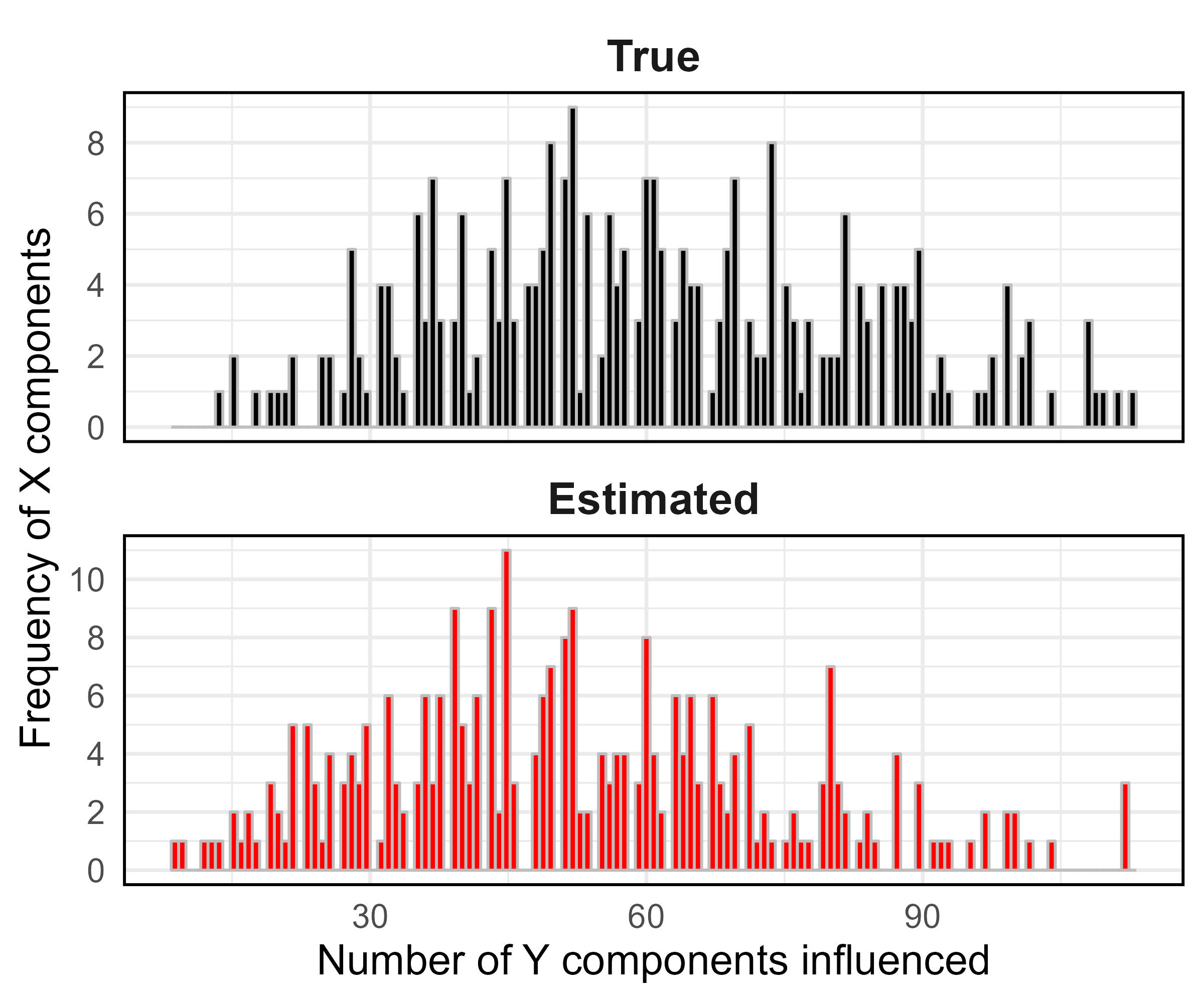} 
\hspace{0.04\linewidth}
\includegraphics[width=.45\linewidth]{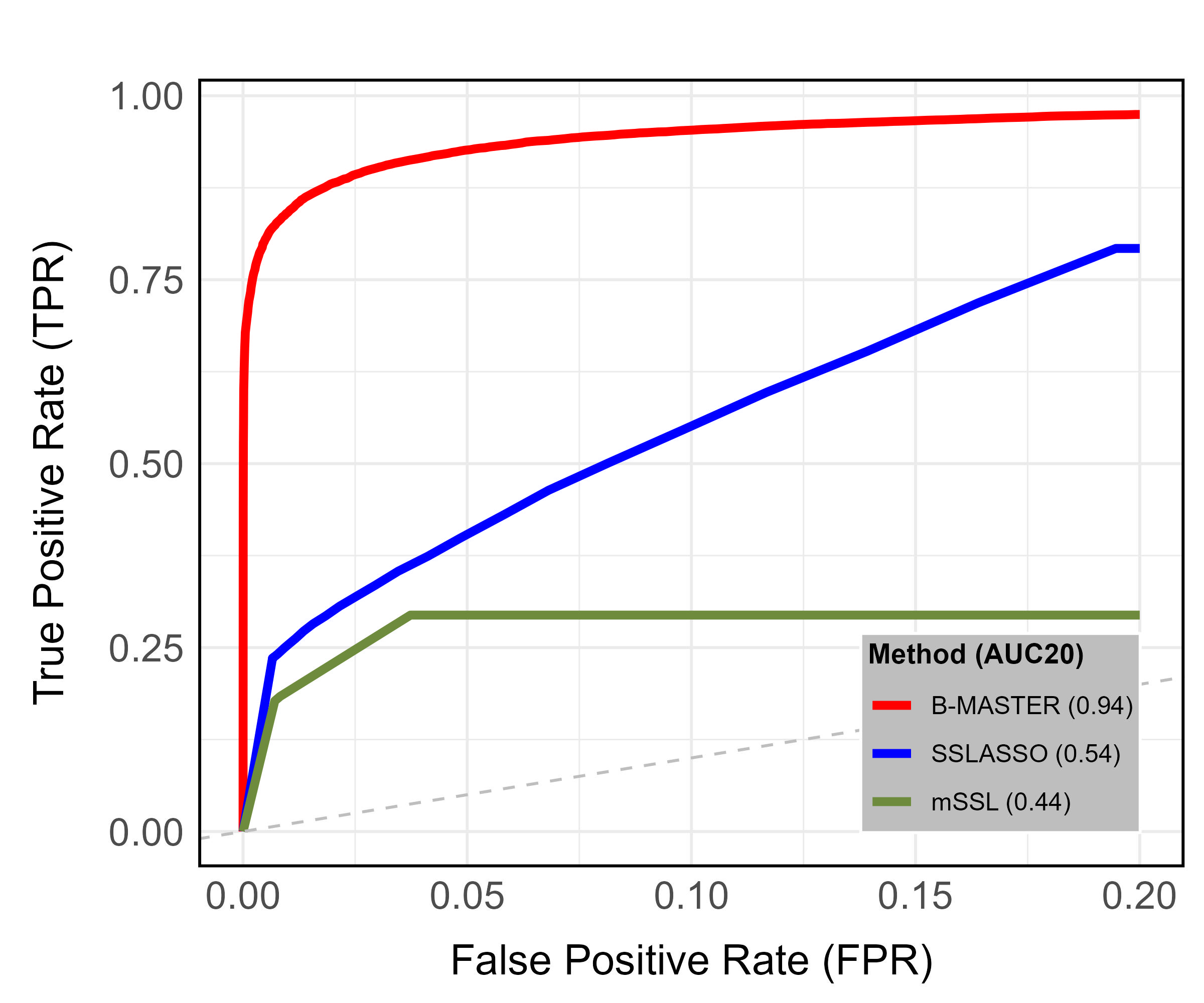} 
\vspace{-0.4cm}
\caption{\textit{Top:} Signal detection by B-MASTER, with TP, TN, FP, FN categories. 
\textit{Bottom-left:} Distribution of number of $Y$ influenced per $X$, true vs.~B-MASTER estimates. 
\textit{Bottom-right:} AUC curves up to 20\% FPR for B-MASTER, SSLASSO, and mSSL.}\vspace{-0.3cm}
\label{BMASTER_sim}
\end{figure}

Convergence diagnostics and sensitivity analyses are reported in Supplement Section~S4, including Geweke statistics and MCSE/SD\% ratios (Supplement Table~S4) with supporting trace plots (Supplement Figure~S3), robustness to prior hyperparameter choices (Supplement Table~S5), and sensitivity to the total number of MCMC iterations (Supplement Figures~S4--S6 and Supplement Table~S6). In the iteration-length analysis, B-MASTER produced nearly identical TPR, FPR, MCC, AUC, AUC20, and sparsity estimates across 500, 1000, and 2000 total iterations, each with 100 burn-in iterations, supporting the stability of the reported posterior selection results. Overall, B-MASTER achieves accurate sparsity recovery with strong signal detection, remains slightly conservative in the number of outcomes influenced per predictor, and consistently outperforms SSLASSO and mSSL, particularly in the low-FPR regime.
\vspace{-0.3cm}
\subsection{Higher-dimensional Evaluation}\vspace{-0.15cm}
\label{sim_study_2}
To assess B-MASTER in higher-dimensional regimes, we conducted an additional simulation study under settings with $P=Q=N$ ranging from 20 to 2000, with both independent ($\rho=0$) and correlated ($\rho=0.5$) predictor structures. The data-generating mechanism was designed to mimic the master-predictor setting, where 20\% of predictors influenced all responses, an additional 20\% influenced a randomly selected subset of responses, and the remaining 60\% had no effect. Full simulation details are provided in Supplement Section~S3.2.
\vspace{-0.3cm}
\begin{table}[!h]
\centering
\resizebox{0.8\columnwidth}{!}{%
\begin{tabular}{cccccccc}
\hline
$\rho$ & $P,Q,N$ & TPR & FPR & MCC & AUC & AUC20 & \begin{tabular}[c]{@{}c@{}}Sparsity\\ (True sparsity)\end{tabular} \\ \hline
\multirow{7}{*}{$\rho = 0$} & 20 & 0.974 & 0.000 & 0.982 & 0.999 & 0.999 & 0.715 (0.708) \\
 & 50 & 0.991 & 0.000 & 0.993 & 1.000 & 1.000 & 0.708 (0.705) \\
 & 100 & 0.978 & 0.012 & 0.964 & 0.997 & 0.986 & 0.699 (0.700) \\
 & 200 & 0.972 & 0.026 & 0.937 & 0.995 & 0.980 & 0.692 (0.702) \\
 & 500 & 0.996 & 0.029 & 0.951 & 0.998 & 0.990 & 0.681 (0.699) \\
 & 1000 & 0.998 & 0.006 & 0.989 & 0.999 & 0.999 & 0.697 (0.700) \\
 & 2000 & 0.998 & 0.000 & 0.998 & 1.000 & 1.000 & 0.702 (0.701) \\ \hline
\multirow{7}{*}{$\rho = 0.5$} & 20 & 0.966 & 0.000 & 0.976 & 1.000 & 1.000 & 0.718 (0.708) \\
 & 50 & 0.999 & 0.002 & 0.996 & 1.000 & 1.000 & 0.704 (0.705) \\
 & 100 & 0.982 & 0.025 & 0.947 & 0.994 & 0.970 & 0.689 (0.700) \\
 & 200 & 0.963 & 0.033 & 0.921 & 0.992 & 0.969 & 0.690 (0.702) \\
 & 500 & 0.990 & 0.021 & 0.959 & 0.998 & 0.991 & 0.688 (0.699) \\
 & 1000 & 0.984 & 0.001 & 0.988 & 1.000 & 1.000 & 0.705 (0.700) \\
 & 2000 & 0.947 & 0.000 & 0.963 & 1.000 & 0.999 & 0.717 (0.701) \\ \hline
\end{tabular}}
\vspace{0.2cm}
\caption{Performance evaluation of B-MASTER in the higher-dimensional simulation study based on true positive rate (TPR), false positive rate (FPR), Matthews correlation coefficient (MCC), AUC, AUC20, and sparsity recovery. Results are reported for $P=Q=N\in\{20,50,100,200,500,1000,2000\}$ under scenarios $\rho=0$ and $\rho=0.5$. Estimated sparsity values are reported with the corresponding true sparsity shown in parentheses.}
\label{tab:high_dim_main}\vspace{-0.5cm}
\end{table}

The results are summarized in Table~\ref{tab:high_dim_main}. Across all considered dimensions and correlation structures, B-MASTER achieved consistently high TPR values, near-zero FPR values, and excellent overall discrimination with AUC values. The corresponding MCC values remained high, indicating accurate recovery of the underlying sparse coefficient structure. Estimated sparsity closely matched the true sparsity across all settings, with only a slight tendency toward conservative selection. Although performance was marginally stronger under independent predictors ($\rho=0$), the degradation under correlated predictors ($\rho=0.5$) was modest, even at the largest dimensions. Notably, B-MASTER maintained strong variable-selection performance when the coefficient matrix contained up to four million parameters ($P=Q=2000$), demonstrating its ability to accurately identify master predictors in large-scale multivariate settings. Complete results for all scenarios are reported in Supplement Table~S3.

\vspace{-0.5cm}
\section{Theoretical properties}\label{sec:theory}
The mathematical framework underlying B-MASTER is presented in detail in Supplement Section~S2, where all assumptions, lemmas, and proofs are elaborated. Here we provide a concise summary of the key conditions and main results most relevant to the implemented B-MASTER model. The theoretical results are established for the same continuous shrinkage prior on $\bm{B}$ introduced in Section~\ref{sec:method}, namely the combined element-wise $\ell_1$ and predictor-level $\ell_2$ shrinkage prior. The latent variables $T^2$ and $G^2$ introduced in Section~\ref{sec:method} provide an equivalent Gaussian scale-mixture representation of this prior for Gibbs sampling; after marginalizing over these latent variables, the induced prior on $\bm{B}$ is the prior used in the theoretical analysis. Thus, separate assumptions on $T^2$ and $G^2$ are not required in (A1)--(A5) noted below, because they are auxiliary variables used to represent the same marginal prior rather than additional scientific components of the model. 

The mathematical framework underlying B-MASTER is presented in detail in Supplement Section~S2, where all assumptions, lemmas, and proofs are elaborated. Here we provide a concise summary of the key conditions and main results most relevant to B-MASTER. For convenience, only in this subsection we adopt the notation $Y = XB_0 + E$ with $X \in \mathbb{R}^{N \times P}$, $Y \in \mathbb{R}^{N \times Q}$, coefficient matrix $B_0$, and sparsity level $s=\#\{(p,q):\beta_{0,pq}\neq 0\}$. Let $\delta_N = \sqrt{s\log(PQ)/N}$. The theoretical results rely on standard high-dimensional regularity conditions: (A1) restricted eigenvalue design; (A2) sparsity scaling $s\log(PQ) = o(N)$; (A3) bounded diagonal error variances; (A4) bounded eigenvalues with sparse precision under correlated errors; and (A5) hierarchical shrinkage priors, including (i) combined $\ell_1$-$\ell_2$ continuous shrinkage with Gamma hyperpriors, (ii) Jeffreys or heavy-tailed variance priors, and (iii) sparse precision priors for correlated-error extensions (see Supplement~S2 for precise formulations). The B-MASTER implementation is justified under assumptions (A1)--(A3) and (A5)(i)--(ii), and therefore the implemented method is directly covered by Theorems~\ref{thm:diag}, \ref{thm:misspec}, and \ref{thm:rowsel}. Assumptions (A4) and (A5)(iii), by contrast, are not part of the current implementation; they are introduced only for the correlated-error theoretical extension and are used in Theorem~\ref{thm:generalSigma}.
\vspace{-0.4cm}
\begin{theorem}[Posterior contraction and consistency (diagonal errors)]\label{thm:diag}
Under (A1)--(A3) and (A5)(i)--(ii), for any fixed $M>0$, as $N\to\infty$,
\[
\Pi\!\left(\, \|B-B_0\|_F \;>\; M\,\delta_N \,\middle|\, X,Y \right)\;\to\; 0
\quad\text{in $P_{B_0,\Sigma_0}$-probability}.
\]
\end{theorem}
\vspace{-0.8cm}
\begin{theorem}[Posterior contraction under misspecification]\label{thm:misspec}
Under (A1), (A2), and (A5)(i),(ii), the posterior contracts at rate $\sqrt{s\log(PQ)/N}$ even under model misspecification. Specifically, when the true errors are sub-Gaussian with covariance $\Sigma_0$ (not necessarily diagonal), the posterior concentrates around the KL-optimal projection $(B^\star,\Sigma^\star)$ of the working diagonal Gaussian model. In the linear-mean setting, $B^\star = B_0$, implying consistent recovery of the true coefficient matrix despite covariance misspecification.
\end{theorem}
\vspace{-0.8cm}
\begin{theorem}[Posterior contraction under correlated errors]\label{thm:generalSigma}
Under assumptions (A1), (A2), (A4) and (A5)(i),(iii), the posterior contracts jointly for $(B,\Omega)$ under correlated errors. Specifically, $B$ contracts at rate $\delta_N = \sqrt{s\log(PQ)/N}$ and the precision matrix $\Omega$ at rate $\varepsilon_\Omega \asymp \sqrt{s_\Omega \log Q / N}$ in Frobenius norm. Thus, even with correlated residuals, estimation of $B$ achieves the same rate as in the diagonal-error case (up to constants), while consistently recovering the sparse precision structure.
\end{theorem}
\vspace{-0.8cm}
\begin{theorem}[Sure screening and selection consistency]\label{thm:rowsel}
Under (A1)--(A3) and (A5)(i)--(ii), let $\tau_N:=C\sqrt{\log(PQ)/N}$ and $\alpha_N:=e^{-c N \tau_N^2}$ for fixed $C,c>0$. Then:
\vspace{-0.2cm}
\begin{itemize}
\item[(i)] (Sure screening) $S_{\mathrm{row}}\subseteq \widehat{S}$ with probability $\to 1$.  
\item[(ii)] (Exact selection under beta-min) If $\beta_{\min}:=\min_{p\in S_{\mathrm{row}}}\|\beta_{0,p\cdot}\|_2 \;\ge\; K \sqrt{\log(PQ)/N}$ for sufficiently large $K$, then $\widehat{S}=S_{\mathrm{row}}$ with probability $\to 1$.  
\end{itemize}
\end{theorem}
\vspace{-1.0cm}
\section{Master Predictors in CRC Microbiome-Metabolome Analysis}\label{sec:case_study}
\begin{figure}[h]
    \centering
    \includegraphics[width=0.95\linewidth]{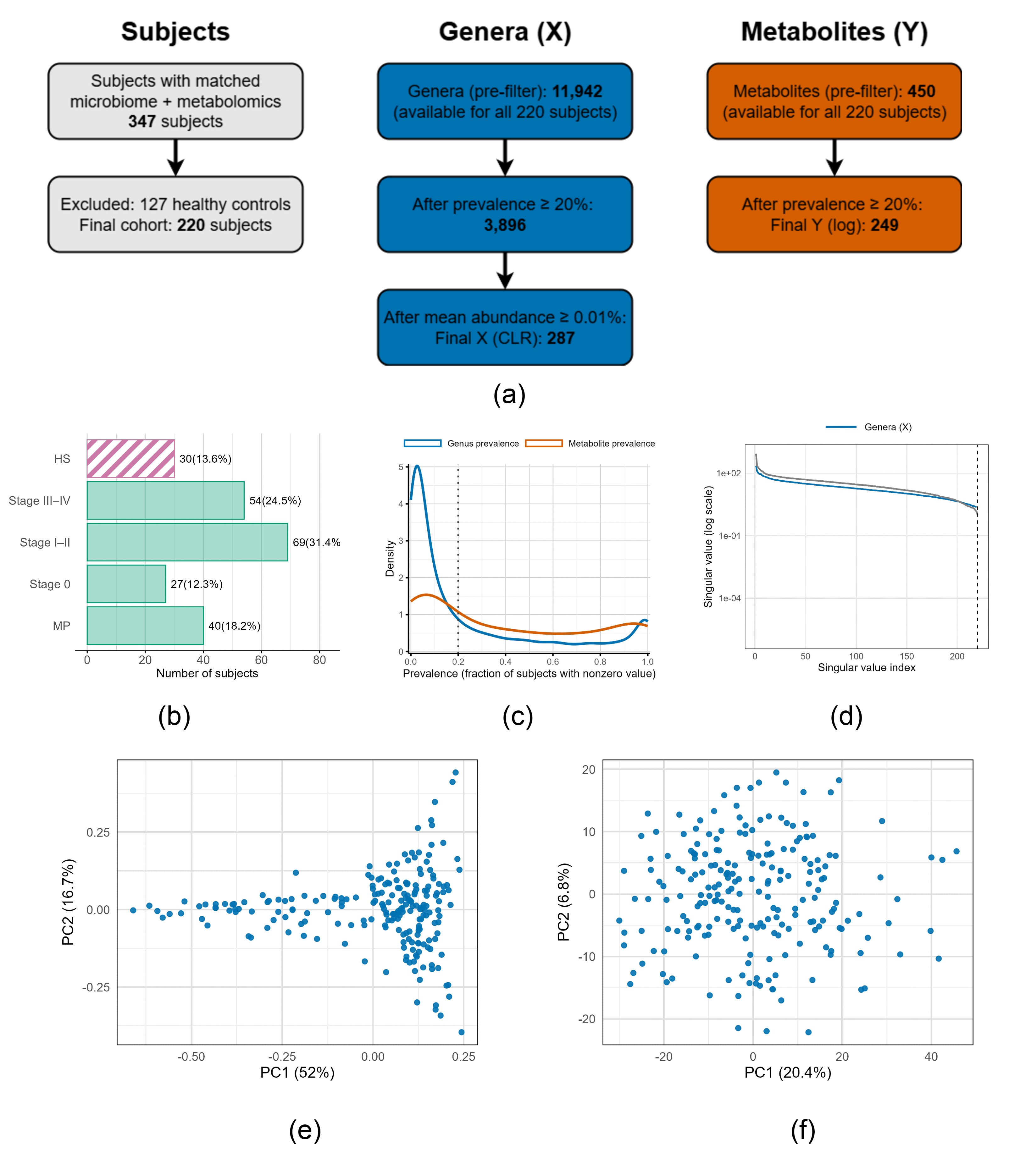}\vspace{-0.6cm}
    \caption{Overview of the CRC microbiome--metabolome dataset.  
    (a) Cohort and filtering workflow. (b) Stage distribution.  
    (c) Density of genus and metabolite prevalence.  
    (d) Scree plot of singular values of $X$ (CLR scale).  
    (e--f) First two PCs of genera before/after CLR transformation.}
    \label{fig:case_study_data}\vspace{-0.3cm}
\end{figure}

We analyze matched microbiome and metabolome data from 347 subjects recruited at the National Cancer Center Hospital (Tokyo) \citep{Muller2022}. After filtering, the final cohort includes 220 CRC subjects, 287 microbial genera, and 249 metabolites (Figure~\ref{fig:case_study_data}). The cohort spans stages 0--IV, precancerous multiple polypoid adenomas, and post-surgical patients. Microbial abundances were provided on a relative-abundance scale and were filtered to retain genera present in at least 20\% of samples with mean abundance exceeding 0.01\%. To address compositionality, we applied a centered log-ratio (CLR) transformation; before taking logarithms, exact zero entries were replaced internally by a pseudocount equal to one-half of the smallest positive relative abundance, following the default implementation of the CLR transformation in the \texttt{microbiome} R package and common practice in compositional microbiome analyses. Metabolites present in fewer than 20\% of samples were removed; remaining zero values were replaced by one-half of the smallest observed positive metabolite value before log transformation. Exploratory analyses reveal typical high-dimensional challenges, including sparsity, non-Gaussianity, compositional dependence in microbiome measurements, and substantial cross-domain correlation. Supplement Figures~S7--S9 further highlight covariate distributions, non-Gaussian metabolite residuals, and cross-domain correlations. To identify key microbiome components influencing metabolites, B-MASTER is fitted with hyper-parameters consistent with the simulation study. Inference is pursued based on 2000 posterior samples, with the first 100 discarded as burn-in.

To provide an interpretable measure of how influence is distributed across metabolites, we introduce a \emph{Fractional Influence Score (FIS)} for each microbial genus identified by B-MASTER. For any metabolite outcome $Y_q$, suppose it is influenced by $h_q$ genera (as determined by the B-MASTER selection step). Instead of attributing a full unit of influence to every selected genus, we allocate equal shares: each contributing genus receives a score of $1/h_q$. Summing these fractional contributions across all metabolites yields the overall score for a genus,
\[
\mathrm{FIS}(g) \;=\; \sum_{q=1}^{Q} \frac{\mathbf{1}\{g \in \mathcal{I}_q\}}{h_q},
\]
where $\mathcal{I}_q$ is the set of influencing genera for metabolite $q$. Because $h_q$ appears in the denominator, this score naturally distinguishes genera that uniquely regulate a metabolite from those that appear only as part of larger sets of co-predictors. Posterior uncertainty for selected genus--metabolite associations is summarized through credible-interval-based selection and Bayesian posterior $p$-values, with additional FIS sensitivity summaries reported in Supplement Section~S5.2. Specifically, the Bayesian posterior $p$-value is computed as $2\min(\hat p^+,\hat p^-)$, where $\hat p^+$ and $\hat p^-$ denote the posterior proportions of samples greater than zero and less than or equal to zero, respectively; details are provided in Supplement Section~S3.1.
\begin{figure}[]
\centering
\includegraphics[width=.98\linewidth]{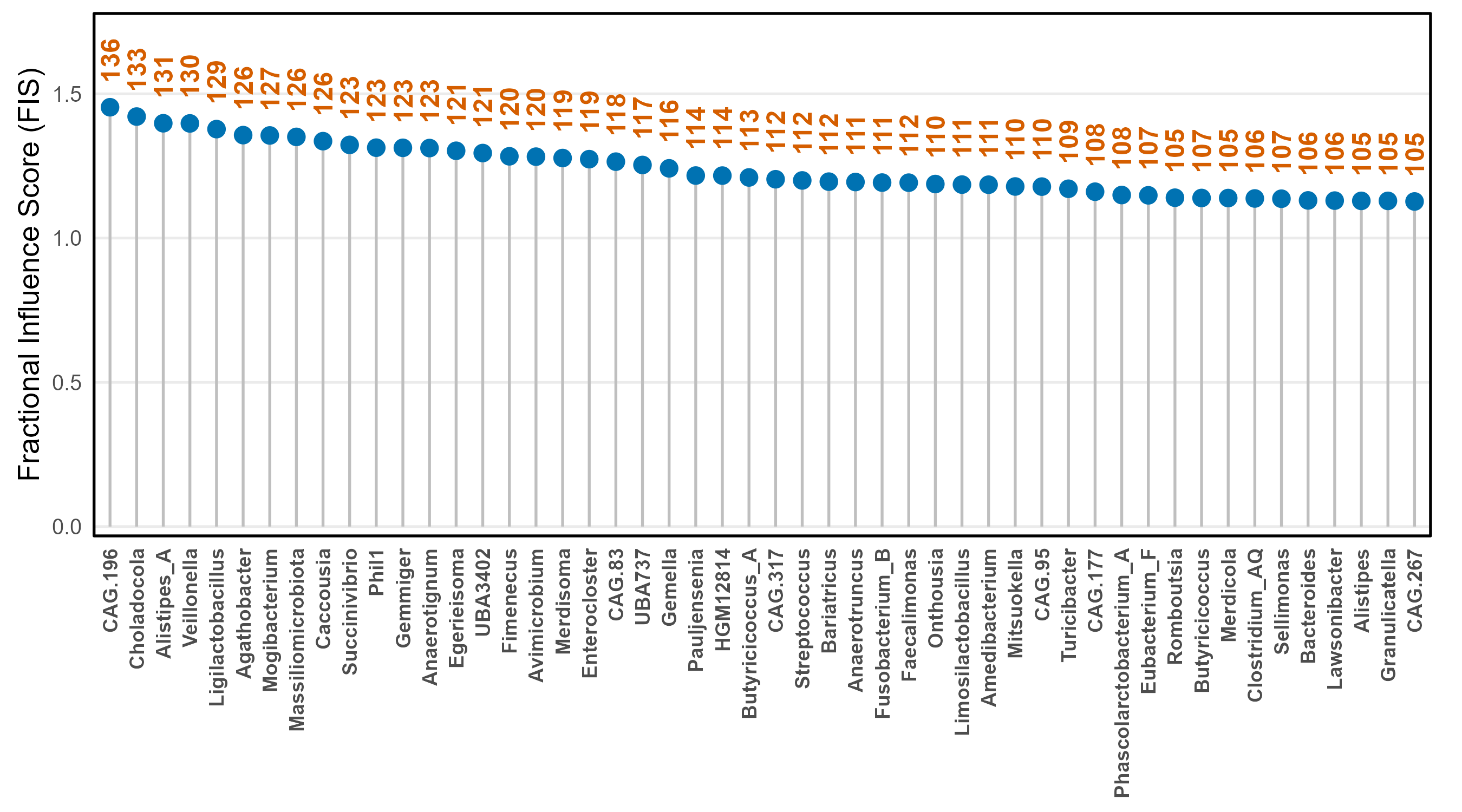} \vspace{-0.3cm}
\caption{Functional Influence Scores (FIS) for the top 50 genera identified by B-MASTER. Bar height reflects each genus’s FIS; the numbers above the bars indicate, for that genus, the number of metabolites it influences.}\vspace{-0.6cm}
\label{Top50genera}
\end{figure}

\begin{figure*}[htbp]
\centering
\includegraphics[width=0.9\textwidth]{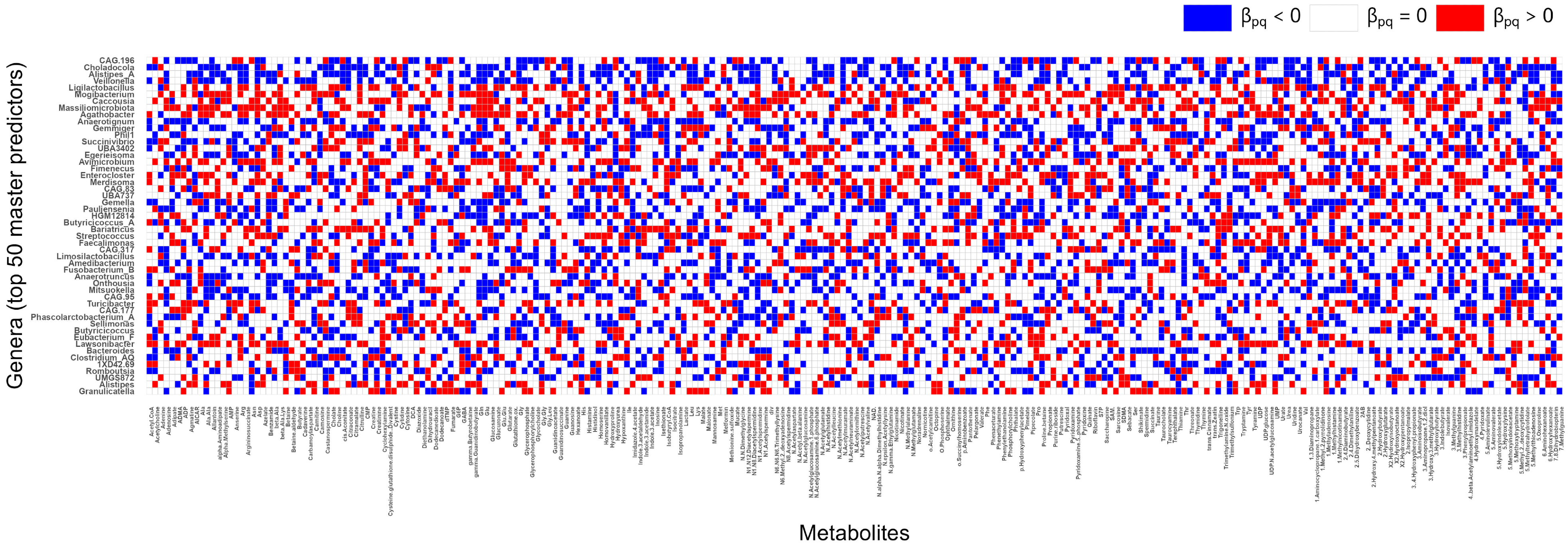} \vspace{-0.45cm}
\caption{Heatmap illustrating the direction of the impact for each of the identified top 50 master predictors on each metabolite.}\vspace{-0.45cm}
\label{Top50genera_heat}
\end{figure*}

Using B-MASTER, we identify the top 50 microbiome genera influencing the largest number of metabolites based on FIS scores (Figure~\ref{Top50genera}), with the number of metabolites regulated by each genus shown above the bars. The top 50 genera were selected as a visualization-focused subset and together account for approximately 25\% of the cumulative FIS across all 287 genera, while the complete ranked FIS distribution is provided in Supplement Section~S5.2, which further reports the full FIS distribution for all 287 genera, together with rankings under 90\% (default), 95\%, and 99\% posterior selection criteria and median Bayesian posterior $p$-values across selected metabolite associations (see Supplementary Table S7-S12). These summaries provide additional uncertainty assessment for the FIS rankings and show that the top-50 threshold is used for visualization and interpretability rather than as a strict biological cutoff. Several master predictors align with prior findings. \cite{Davar2022} reported \textit{CAG.196} and \textit{Choladocola} (ranks 1-2) as key contributors to microbiome-host metabolism, particularly involving bile acids and short-chain fatty acids (SCFAs). \textit{Alistipes} (rank 3) modulates bile acids and SCFAs \citep{Parker2020}, while \textit{Veillonella} (rank 4) and \textit{Ligilactobacillus} (rank 5) participate in lactate, propionate, and bile salt metabolism \citep{Zhang2023b}. \cite{Song2022} linked \textit{Mogibacterium} (rank 7) to increased SCFAs and fat with negative correlations to amino acids. \textit{Agathobacter}, \textit{Gemmiger}, and \textit{Anaerotignum} (ranks 6, 12, 13) contribute to fiber degradation and SCFA production, and \textit{Massiliomicrobiota} (rank 8) has been associated with amino acid and energy metabolism \citep{Lu2024}. In contrast, genera such as \textit{Caccousia}, \textit{Succinivibrio}, and \textit{Phil1} (ranks 9-11) remain sparsely characterized, highlighting potentially novel regulators. Figure~\ref{Top50genera_heat} presents a heatmap of the top 50 predictors, illustrating their metabolite-specific effects and directionality.

Beyond identifying overall master predictors, we examine key genera affecting two biologically important metabolite subsets. Subset~1 comprises the 10 most abundant metabolites \citep{Yachida2019}: \textit{Propionate}, \textit{Butyrate}, \textit{Dihydrouracil}, \textit{Glutamate}, \textit{Urea}, \textit{Succinate}, \textit{5-Aminovalerate}, \textit{Valerate}, \textit{Lysine}, and \textit{Alanine}. Subset~2 includes 11 cancer-associated metabolites \citep{Yachida2019}: \textit{X\_DCA}, \textit{Glycocholate}, \textit{Taurocholate}, \textit{Isovalerate}, \textit{L-Isoleucine}, \textit{L-Leucine}, \textit{L-Valine}, \textit{L-Phenylalanine}, \textit{L-Tyrosine}, \textit{L-Serine}, and \textit{Glycine}. For each subset, the top 15 master predictor genera were identified, and their directional effects are summarized in Figures~\ref{figsubset}(a) and (b), with full results in Supplement Tables~S13-S14. Canonical correlations with cumulatively included predictors are shown in Supplement Figure~S11.

For Subset~1, \textit{Veillonella} (rank 1) emerges as a dominant regulator, negatively associated with multiple SCFAs and amino acids while positively linked to \textit{Dihydrouracil} and \textit{Urea}, suggesting a shift toward nitrogen metabolism \citep{Mashima2021,Zhang2023}. \textit{Oliverpabstia} (rank 2) and \textit{Merdicola} (rank 3) display similar mixed patterns, with reduced SCFAs but increased nitrogen-related metabolites. \textit{Sellimonas} (rank 4) shows broadly positive effects, whereas \textit{Avimicrobium} (rank 5) positively influences amino acids but negatively associates with \textit{Succinate}. \textit{UBA4372} (rank 6) exhibits reduced SCFAs and increased nitrogen metabolites \citep{Xu2024}. \textit{Mogibacterium} and \textit{Lancefieldella} (ranks 8-9) are positively linked to amino acid metabolism, while \textit{Turicibacter}, \textit{Intestinibacter}, and \textit{Anaerotignum} show predominantly negative associations, suggesting suppressive roles in host-associated metabolic pathways.
\begin{figure}[h]
\centering
\includegraphics[width=.9\linewidth]{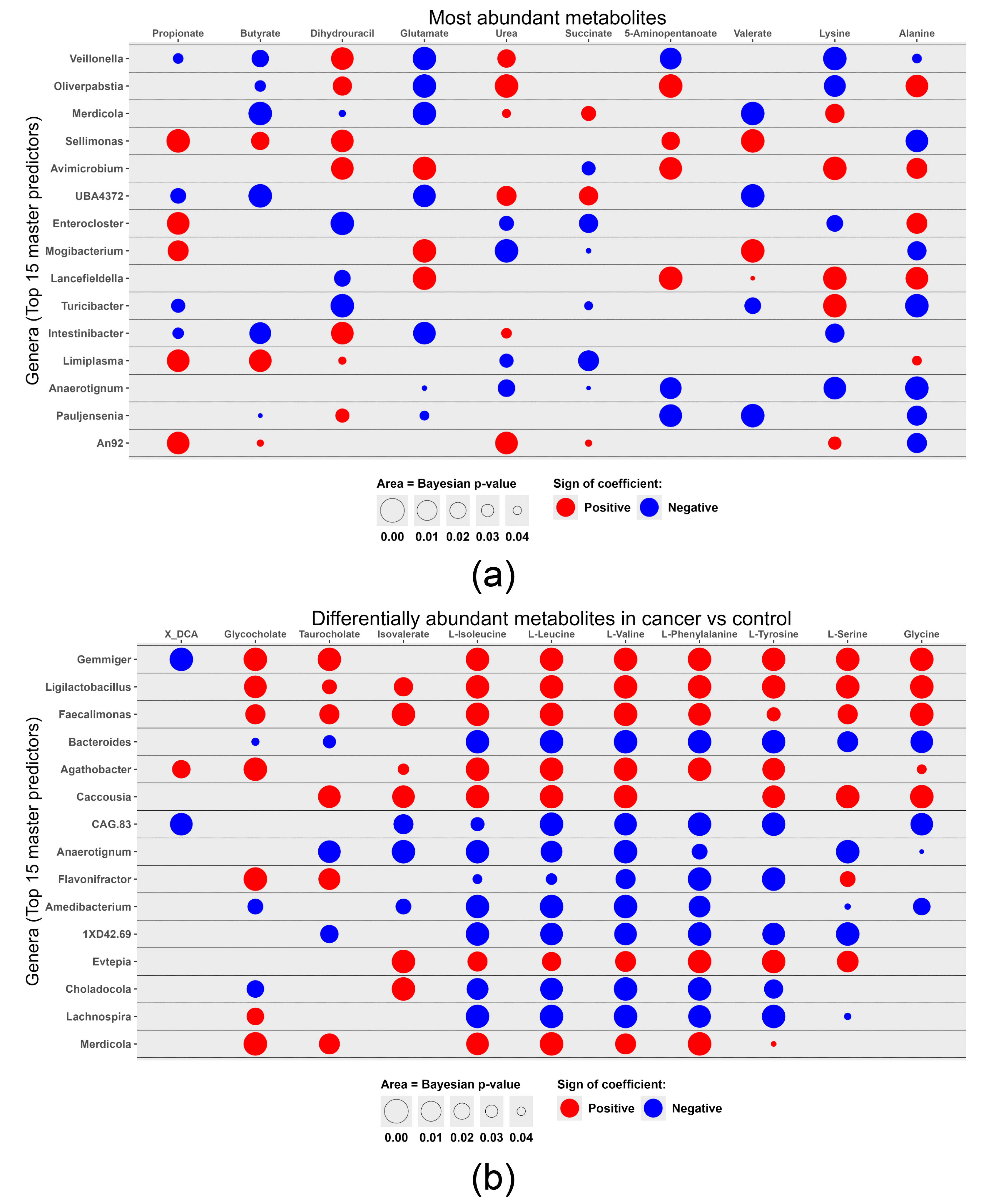} \vspace{-0.4cm}
\caption{The plot demonstrates the direction and statistical significance of the relationships between (a) the most abundant metabolites (Subset 1) on the corresponding top 15 key genera, and (b) differentially abundant metabolites (Subset 2) in the cancer versus control comparison on the corresponding top 15 key genera, identified via B-MASTER analysis. Progressively larger circles are associated with smaller Bayesian p-values.}
\label{figsubset}\vspace{-0.8cm}
\end{figure}

For Subset~2, a more systemic regulatory pattern emerges, with many genera influencing nearly all cancer-associated metabolites. \textit{Gemmiger} (rank 1) acts as a dominant positive regulator across bile acids and amino acids, including \textit{Glycocholate}, \textit{Taurocholate}, \textit{L-Isoleucine}, \textit{L-Leucine}, \textit{L-Valine}, \textit{L-Phenylalanine}, \textit{L-Tyrosine}, \textit{L-Serine}, and \textit{Glycine}, while uniquely showing a negative association with \textit{X\_DCA}. \textit{Ligilactobacillus} (rank 2) and \textit{Faecalimonas} (rank 3) display similarly broad positive effects. In contrast, \textit{Bacteroides} (rank 4) and \textit{CAG.83} (rank 7) exhibit consistent negative associations across bile acids and amino acids. \textit{Agathobacter} (rank 5) and \textit{Caccousia} (rank 6) are uniformly positive, whereas \textit{Anaerotignum} (rank 8) shows widespread negative effects, particularly on branched-chain amino acids. Mixed profiles are observed for \textit{Flavonifractor} (rank 9) and \textit{Lachnospira} (rank 14). Strong positive regulators include \textit{Evtepia} (rank 12) and \textit{Merdicola} (rank 15), while \textit{Amedibacterium}, \textit{1XD42.69}, and \textit{Choladocola} (ranks 10-13) are predominantly negative.

Overall, Subset~1 highlights targeted modulation of SCFA and amino acid pathways, whereas Subset~2 reveals broad regulators of CRC-associated metabolites. To our knowledge, this represents the first genus-level characterization of such coordinated microbiome-metabolite regulation in CRC, extending beyond higher taxonomic analyses \citep{Dueholm2024}.

As an additional disease-stratified analysis, Supplement Section~S5.4 applies B-MASTER to the healthy control samples ($n=127$) using the same posterior selection rule and FIS definition, and compares the resulting master-predictor structures with the CRC cohort. The comparison reveals limited overlap between CRC and healthy top-ranked master predictors, suggesting that the inferred microbiome-metabolite regulatory architecture is largely disease-specific rather than reflecting generic high-abundance microbial effects. In particular, the analysis identifies CRC-enriched and healthy-enriched genera through differential FIS values and percentile rank shifts, while defining target sets consistently as the metabolites selected for each genus under the posterior selection rule (see Supplementary Figure~S12 and Table~S15).

\vspace{-0.5cm}
\section{Discussion}\label{sec:discussion}
In this article, we introduce B-MASTER, a fully Bayesian framework for identifying essential regressors in multivariate regression via an efficient and scalable Gibbs sampler. Empirically, B-MASTER outperforms existing approaches while maintaining appropriate sparsity and strong signal recovery. Most notably, it scales effectively in high-dimensional settings: unlike remMap and mSSL, which failed to converge for $P=287, Q=249$, B-MASTER remained stable even for models with up to 4 million parameters ($P=Q=2000$). Its runtime is nearly invariant to sample size and is observed to increase linearly with the number of parameters. These computational properties, combined with theoretical guarantees of posterior consistency and minimax-rate contraction under sparsity, make B-MASTER well suited for ultra-high-dimensional microbiome-metabolome analysis. While discrete selection priors such as Spike-and-Slab offer explicit posterior inclusion probabilities, they typically lead to substantially more complex posterior exploration and increased computational burden in large-scale multivariate settings. B-MASTER instead employs a continuous shrinkage formulation that preserves strong sparsity-inducing behavior while enabling efficient Gibbs sampling and scalable uncertainty quantification. Future work could investigate Spike-and-Slab \citep{GeorgeMcCulloch1993} extensions of B-MASTER for settings where explicit inclusion probabilities are of primary interest, as well as nonlinear or distributionally robust extensions based on spline, kernel, or non-Gaussian response formulations to accommodate complex predictor--response relationships and metabolite distributions encountered in biological systems.

Using the B-MASTER framework, we identify key components of the microbiome that influence overall metabolite profiles. Specifically, we determine the top 50 ``master predictors'' of gut microbiota in regulating the microbiome-metabolite dependence structure based on colorectal cancer data \citep{Yachida2019}. Furthermore, we identify a set of key genera that influence the most abundant metabolites. Our findings align with existing studies in this area. In addition, we examine the set of genera influencing differentially abundant metabolites in cancer versus control cases. Notably, we observe a distinctive regulatory pattern where key genera consistently influence these differential metabolites either positively or negatively. Aligning our results with the findings of \citet{Yachida2019}, we conclude that elevated levels of \textit{Ligilactobacillus}, \textit{Faecalimonas}, \textit{Agathobacter}, \textit{Caccousia}, \textit{Evtepia}, and \textit{Merdicola} are positively associated with colorectal disease progression. This study represents the first attempt to elucidate the role of key microbiome components at the genus level potentially associated with colorectal cancer by exploring their influence on differentially abundant metabolites of colorectal cancer patients across various stages. We hope that our novel findings provide a foundational basis for further studies on modulating these identified genera as potential targets for colorectal cancer treatment or prevention in the future. 
\vspace{-0.6cm}
\section*{Conflict of interest}\vspace{-0.15cm}
The authors declare that they have no competing interests.
\vspace{-0.5cm}
\section*{Funding}\vspace{-0.15cm}
PD is partially supported by NIH/NCI CCSG P30 CA016059. CBP is partially supported by NIH R01 HL158796, NIH/NCI CCSG P30CA016672, and an Andrew Sabin Family Fellowship.
\vspace{-0.4cm}
\section*{Data availability}\vspace{-0.15cm}
The data analyzed in this study are publicly available and were obtained from the colorectal neoplasia cohort originally reported by \cite{Yachida2019}, comprising subjects undergoing total colonoscopy at the National Cancer Center Hospital, Tokyo, Japan.
\vspace{-0.4cm}
\section*{Acknowledgments}\vspace{-0.15cm}
High Performance Computing resources provided by the High Performance Research Computing (HPRC)
core facility at Virginia Commonwealth University (\url{https://hprc.vcu.edu}) were used for conducting
the research reported in this work.\vspace{-0.4cm}
\bibliographystyle{apalike}
\bibliography{reference}

\end{document}


\def\spacingset#1{\renewcommand{\baselinestretch}%
{#1}\small\normalsize} \spacingset{1}


\if1\blind
{
  \title{\bf Supplementary Material for ``B-MASTER: Scalable Bayesian Multivariate Regression for Master Predictor Discovery in Colorectal Cancer Microbiome-Metabolite Profiles''}
  \author{Priyam Das
    \hspace{.2cm}\\
 {\normalsize \medskip Department of Biostatistics, Virginia Commonwealth University}\\
    Tanujit Dey \\
{\normalsize \medskip Center for Surgery and Public Health, Brigham and Women's Hospital,}\\
  {\normalsize \medskip Harvard Medical School}\\
    Christine B. Peterson \\
    {\normalsize \medskip Department of Statistics, Rice University}\\
    \medskip and \\
     Sounak Chakraborty \\
    {\normalsize \medskip Department of Statistics, University of Missouri}
    }
    \date{}
  \maketitle
} \fi

\if0\blind
{
  \bigskip
  \bigskip
  \bigskip
  \begin{center}
    {\LARGE\bf Supplementary Material for ``B-MASTER: Scalable Bayesian Multivariate Regression Analysis for Selecting Targeted Essential Regressors''}
\end{center}
  \medskip
} \fi

\spacingset{1.45} 
\bigskip
\tableofcontents
\clearpage

\section*{Notation}
For convenience, we summarize here the key notation used throughout the Supplement. 

\begin{table}[h!]
\centering
\begin{tabular}{ll}
\hline
Symbol & Description \\
\hline
$X \in \mathbb{R}^{N \times P}$ & Design matrix with $N$ subjects and $P$ predictors \\
$Y \in \mathbb{R}^{N \times Q}$ & Response matrix with $Q$ outcomes \\
$B = [\beta_{pq}] \in \mathbb{R}^{P \times Q}$ & Regression coefficient matrix \\
$\varepsilon_i \sim \mathcal{N}_Q(0,\Sigma)$ & Error term for subject $i$ \\
$\Sigma \in \mathbb{R}^{Q \times Q}$ & Error covariance matrix \\
$T^2,\, G^2$ & Local shrinkage parameters for rows and groups, respectively \\
$\tau^2_{pq}$ & Local element–specific shrinkage parameter for $\beta_{pq}$ \\
$\gamma^2_p$ & Group–specific shrinkage parameter for predictor $p$ \\
$\lambda_1,\, \lambda_2$ & Global shrinkage hyperparameters \\
$s$ & Sparsity level (number of nonzero entries in $B$) \\
$\|\cdot\|_F$ & Frobenius norm (square root of the sum of squared entries) \\
$\|\cdot\|_{\mathrm{op}}$ & Operator norm (largest singular value) \\
\hline
\end{tabular}
\caption{Key notation used in the Supplement.}
\label{tab:notation}
\end{table}

\section{Model, Full Posterior and Gibbs Sampling} 
\subsection{Comparison with Elastic Net}
At first glance, the conditional prior in equation (3) of the main draft may appear similar to the Elastic Net penalty \citep{Zou2005}, which also combines $l_1$- and $l_2$-type regularization. However, there are key differences. First, the Elastic Net applies an $l_1$ penalty on individual coefficients and an $l_2$ penalty on the entire coefficient vector, whereas our formulation applies the $l_2$ penalty at the level of predictor-specific coefficient vectors $(\beta_{p1},\ldots,\beta_{pQ})$. This design induces selection of predictors with shared effects across multiple responses, a property not inherent to the Elastic Net. Second, our Bayesian formulation introduces hierarchical priors on the tuning parameters $(\lambda_1, \lambda_2)$, allowing them to be adaptively learned from the data rather than fixed by cross-validation. Taken together, these differences demonstrate that the B-MASTER prior is tailored to multivariate response settings with structured sparsity, rather than being a direct analogue of Elastic Net penalization.

\subsection{Scalability: B-MASTER vs remMap}
It is important to note several caveats regarding analogous frequentist approaches for optimizing the coefficient matrix $\bm{B}$ within a penalized likelihood framework, as discussed in (\citealp{Peng2010}; referred to as `remMap'). First, the remMap method uses the ``active-shooting'' algorithm \citep{Friedman2008, Peng2009} to optimize $\bm{B}$ for a given $\lambda_1$, with computational complexity $O(NPQ)$. As sample size increases, this results in substantial computational burden. In contrast, the proposed Gibbs sampler algorithm maintains nearly constant computation time, as only a small portion involves dataset-related operations efficiently handled by MATLAB's multi-threading. As $N$ grows, higher levels of multi-threading mitigate computation time increases \citep{MathWorks2024}. For fixed $P$ and $Q$, the Gibbs sampler's computation time remains largely invariant with sample size. Figure~5 (in the main draft; see Section S3.2 for data generation) shows that remMap's computation time depends strongly on sample size, while B-MASTER exhibits near invariance. Second, selecting an optimal $\lambda_1$ via cross-validation intensifies computational demands. If uncertainty quantification is required, bootstrapping may result in prohibitively long computation times. A detailed examination of remMap's performance limitations in high-dimensional settings is provided in the simulation study section. These challenges can be alleviated using a Bayesian framework, where model estimation occurs once, yielding both optimal coefficients and their uncertainty measures. Lastly, the computational time per iteration of the proposed Gibbs sampler is independent of dataset characteristics, assuming $P$ and $Q$ are fixed. In contrast, remMap's ``active-shooting'' algorithm exhibits strong data dependence, with convergence time varying significantly across datasets of the same dimensionality. For instance, as shown in Figure~5 (left) (main draft), with $P=Q=40$, the computation time for $N=40$ exceeded that for $N=200$, and for $P,Q=50$, $N=50$. B-MASTER, in contrast, shows predictable scaling (Figure~5 (right) in the main draft). It is observed that B-MASTER takes slightly more time for $N=P$ scenarios compared to $N=5P$ and $N=10P$ scenarios, likely due to increased multi-threading activation in MATLAB for high-dimensional matrix operations \citep{MathWorks2024}.

\subsection{Posterior Distribution}
We introduce latent parameters $\bm{T}^2 = (\bm{\tau}_1^2, \ldots,  \bm{\tau}_P^2)^T = \big(\tau_{pq}^2\big)_{P \times Q}$ and $\bm{G}^2 = \big(\gamma_{p}^2\big)_{P \times 1}$. This formulation yields the following hierarchical representation of the B-MASTER prior \citep{Xu2015}:
\vspace{-0.4cm}
\begin{align}
    &\bm{\beta}^{(p)}|\bm{\tau}_p^2 ,\gamma_p^2,\sigma_1^2,\ldots,\sigma_Q^2 \sim 
	N_Q(\bm{0}, \bm{V}_p),\; \bm{V}_p = diag\left\{\sigma^2_q\left(\frac{1}{\tau_{pq}^2} + \frac{1}{\gamma^2_p}\right)^{-1}\right\} \; \; \text{for} \; \; q = 1,\ldots,Q, \nonumber\\
    &\pi(\bm{\tau}_p^2,\gamma^2_p) \propto C_p^*(\lambda_1^2,\lambda_2^2)\prod_{q=1}^Q \left[ (\tau_{pq}^2)^{-1/2}\left(\frac{1}{\tau_{pq}^2} + \frac{1}{\gamma^2_p}\right)^{-1/2}\right] (\gamma^2_p)^{-1/2} \exp\left\{-\frac{\lambda_1^2}{2}\sum_{q=1}^Q\tau_{pq}^2 - \frac{\lambda_2^2}{2}\gamma^2_p\right\}
    \label{tau_pq}
\end{align}
\noindent where $C_p^*(\lambda_1^2,\lambda_2^2)$ is a normalizing constant involving both penalty terms. It is introduced solely to ensure propriety of the hierarchical prior representation. The propriety of the prior distribution introduced in \eqref{tau_pq} is established in Lemma \ref{lem:propriety}. We put a non-informative prior on $\sigma_q^2$, $\pi(\sigma_q^2)\propto \frac{1}{\sigma_q^2}, \mbox{ for } q=1,\ldots,Q$.
Hereafter, combining the likelihood from the model (1) of the main paper along with the aforementioned prior setup, we obtain the joint posterior distribution:
\vspace{-0.6\baselineskip}
\begin{align}
    \pi(\bm{B},\bm{T}^2,\bm{G}^2,\bm{\sigma}^2,  \mid \bm{X},\bm{Y}) \propto
     & \prod_{q=1}^{Q}\left[(\sigma_q^2)^{-\frac{(N+P)}{2}-1}
	\exp\left\{-\frac{(\bm{y}_q-\bm{X}\bm{\beta}_q)'(\bm{y}_q-\bm{X}\bm{\beta}_q)+
	\bm{\beta}_q'D_{\tau_q}\bm{\beta}_q}{2\sigma_q^2}
	\right\}\right]. \nonumber\\
   & \prod_{p=1}^P \left[C_p^*(\lambda_1^2,\lambda_2^2)\prod_{q=1}^Q \left\{ (\tau_{pq}^2)^{-1/2}\left(\frac{1}{\tau_{pq}^2} + \frac{1}{\gamma^2_p}\right)^{-1/2}\right\} (\gamma^2_p)^{-1/2}\right]. \nonumber\\
   & \exp\left\{-\frac{\lambda_1^2}{2}\sum_{q=1}^Q\tau_{pq}^2 - \frac{\lambda_2^2}{2}\gamma^2_p\right\},
   \label{post_temp}
\end{align}
where $\bm{y}_q = (y^1_q,y^2_q,\ldots, y^N_q)$ denotes the $q$-th component (column) of the response matrix $\bm{Y}$ ($N\times Q$ matrix of response); $\bm{x}^i = (x^i_1,x^i_2,\ldots, x^i_P)$ denotes the $i$-th row of the covariate matrix $\bm{X}$ ($N\times P$ matrix of covariates); and $D_{\tau_q}$ is a $P\times P$ diagonal matrix such that $D_{\tau_q}(p,p) = \frac{1}{\tau_{pq}^2}+\frac{1}{\gamma^{2}_p}$ for $p=1,\ldots,P$.
\vspace{-0.4cm}
\subsection{Gibbs Sampling}
\label{gibbs}

From the joint posterior distribution (\ref{post_temp}), we derive a Gibbs sampling algorithm to generate posterior samples of $\bm{B} = \big(\beta_{pq}\big)_{P \times Q}$, $\bm{T}^2 = (\bm{\tau}_1^2, \ldots,  \bm{\tau}_P^2)^T = \big(\tau_{pq}^2\big)_{P \times Q}$, $\bm{G}^2 = \big(\gamma_{p}^2\big)_{P \times 1}$, and $\bm{\sigma}^2 = (\sigma_1^2,\sigma_2^2,\ldots,\sigma_Q^2)$ from their full conditional posteriors as follows, 
\begin{itemize}
\item Sample $\bm{\beta}_q = (\beta_{1q},\ldots,\beta_{Pq})^T$ for $q=1,\ldots,Q$, from
\begin{align*}
\bm{\beta}_q|\bm{T}^2,\bm{G}^2,\sigma_q^2 \sim N_p
	\left((\bm{X}'\bm{X}+D_{\tau_q})^{-1}\bm{X}'\bm{y}_q,
	\sigma_q^2(\bm{X}'\bm{X}+D_{\tau_q})^{-1}\right).
\end{align*}
\item Sample $\tau_{pq}^2$, for $p = 1,\ldots,P;q = 1,\ldots,Q$, such that $\tau_{pq}^2 = \frac{1}{\delta_{pq}^{(1)}}$, where
\begin{align*}
 \delta_{pq}^{(1)}|\beta_{pq},\sigma_q^2, \lambda_1^2  \sim \mbox{Inv-Gaussian}\left(\sqrt{\frac{\lambda_1^2\sigma_q^2}{\beta_{pq}^2}},\lambda_1^2\right).
\end{align*}
\item Sample $\gamma_{p}^2$ for $p = 1,\ldots,P$, such that $\gamma_{p}^2 = \frac{1}{\delta_{p}^{(2)}}$, where
\begin{align*}
	\delta_{p}^{(2)}|\bm{B}, \bm{\sigma}^2,\lambda_2^2 \sim
	\mbox{Inv-Gaussian}\left(\sqrt{\frac{\lambda_2^2}{\sum_{q=1}^Q(\beta_{pq}^2/\sigma_q^2)}},\lambda_2^2\right).
\end{align*}
\item Sample $\sigma_q^2$ for $q=1,\ldots,Q$, from
\begin{align*}
 \sigma_q^2|\bm{\beta}_q,\bm{T}^2,\bm{G}^2 \sim \mbox{Inv-Gamma}\bigg(\frac{N+P}{2},\frac{(\bm{y}_q-\bm{X}\bm{\beta}_q)'(\bm{y}_q-\bm{X}\bm{\beta}_q)+
	\bm{\beta}_q'D_{\tau_q}\bm{\beta}_q}{2}\bigg).
\end{align*}
\end{itemize}
Note that the tuning parameters in \eqref{post_temp} are incorporated as fixed values. Thus, this proposed sampling scheme can be used to perform posterior inference of B-MASTER parameters for given fixed values of the tuning parameters $\lambda_1$ and $\lambda_2$. We further delve into putting appropriate hyper-prior on the prior parameters in the following sub-section. 
\subsection{Prior on Tuning Parameters and Gibbs Sampling}
The Gibbs sampling scheme outlined in the previous subsection is conceptually analogous to the sparse model estimation process within the frequentist framework. In the frequentist setting, for fixed values of the tuning parameters, the model is fitted, and optimal tuning parameter values are subsequently selected through methods such as cross-validation or approaches based on Stein's unbiased risk estimate \citep{Tibshirani1996}.
In contrast, the Bayesian framework circumvents this extended cross-validation process by assigning suitable hyper-priors to $\lambda_1$ and $\lambda_2$ \citep{Park2008, Kyung2010, Roy2017}. This facilitates a fast and efficient mechanism for sampling $\lambda_1$ and $\lambda_2$ simultaneously with other model parameters. Following the recommendations of \cite{Kyung2010} and \cite{Xu2015}, we assign the following joint prior distribution to $\lambda_1^2$ and $\lambda_2^2$:
\begin{align}
\pi(\lambda_1^2, \lambda_2^2) \propto  C(\lambda_1^2,\lambda_2^2) (\lambda_1^2)^{a_1} \exp{(-b_1\lambda_1^2)}(\lambda_2^2)^{a_2} \exp{-(b_2\lambda_2^2)},
\label{lambda_prior}
\end{align}
where $C(\lambda_1^2,\lambda_2^2) = \left(\prod_{p=1}^P C_p^*(\lambda_1^2,\lambda_2^2)\right)^{-1}$ is the induced normalizing constant from the hierarchical prior construction, and $a_1,b_1,a_2,b_2$ are some positive constants.

Such a prior structure is well-studied in the context of Bayesian group lasso and sparse graphical models \citep{Xu2015, Das2020}. Following their recommendation, to impose a stronger penalization structure as the total number of regression coefficients increases, hyper-prior parameter values are chosen as $a_1 = PQ + r_1$, $a_2 = \frac{PQ}{2} + r_2$, $b_1 = \frac12 \sum_{p=1}^P \sum_{q=1}^Q \tau_{pq}^2 + \delta_1$ and $b_2 = \frac{1}{2}\sum_{p=1}^P\gamma_p^2+\delta_2,$  where $r_1,r_2,\delta_1,\delta_2$ are positive constants. In B-MASTER analyses, we set $r_1=r_2=1$ and $\delta_1=\delta_2=0.1$. This prior structure is found to perform well across a broad range of $P$ and $Q$ values, maintaining excellent control over the true positive rate (TPR) and false positive rate (FPR), without issues in both low and high-dimensional scenarios, as detailed in the simulation study section of the main paper.  The propriety of the prior introduced in \eqref{lambda_prior} is discussed in detail in Lemma \ref{lem:propriety}.  The conditional posterior distributions of the tuning parameters are given as follows:

\begin{align}
\lambda_1^2|\bm{T}^2 \sim & \mbox{ Gamma}\bigg(PQ+r_1, 
	\frac{1}{2}\sum_{p,q}\tau_{pq}^2+\delta_1 \bigg), \\
\lambda_2^2|\bm{G}^2 \sim &  \mbox{ Gamma}\bigg(\frac{PQ}{2}+r_2,
	\frac{1}{2}\sum_{p=1}^P\gamma_p^2+\delta_2 \bigg). 
\end{align}
Note that the full conditional posterior distributions of $\lambda_1^2$ and $\lambda_2^2$ do not involve any computation of $C(\lambda_1^2, \lambda_2^2)$ resulting in substantial improvement in computational efficiency. By integrating the sampling steps for $\lambda_1^2$ and $\lambda_2^2$ outlined above with the Gibbs sampling procedure detailed in Section \ref{gibbs}, the complete Gibbs sampling framework for B-MASTER is established.


\section{Theoretical results and proofs}\label{sec:supp-theory}

Throughout, let $X\in\mathbb{R}^{N\times P}$ collect the row vectors $x_i^\top$, $Y\in\mathbb{R}^{N\times Q}$ collect $y_i^\top$, and consider
\[
y_i \;=\; x_i B_0 + \varepsilon_i,\qquad \varepsilon_i\stackrel{\mathrm{i.i.d.}}{\sim}\mathcal{N}_Q(0,\Sigma_0),\quad i=1,\ldots,N,
\]
where $B_0\in\mathbb{R}^{P\times Q}$ is the true coefficient matrix and $\Sigma_0\in\mathbb{R}^{Q\times Q}$ the true error covariance. Denote by $\|\cdot\|_F$ the Frobenius norm and by $\|\cdot\|_{\mathrm{op}}$ the operator norm. For a matrix $B=[\beta_{pq}]$, define its sparsity level
\[
s\;:=\; \#\{(p,q): \beta_{pq}\neq 0\}.
\]
We adopt the continuous shrinkage prior of the main text together with Gamma hyperpriors on $(\lambda_1^2,\lambda_2^2)$; for the error scales we keep the implementation’s prior (see (A5)(ii)).

\paragraph{Status of assumptions in \textsc{B-MASTER}.}
The implementation fits the Gaussian multivariate regression with a \emph{diagonal} working covariance $\Sigma=\mathrm{diag}(\sigma_1^2,\ldots,\sigma_Q^2)$, Jeffreys priors $\pi(\sigma_q^2)\propto 1/\sigma_q^2$, and the continuous shrinkage prior on $B$ from (A5)(i). Under this setup:
\begin{itemize}
\item \emph{Directly applicable to the implementation.} 
Lemma~\ref{lem:propriety} (posterior propriety and $N^{-1/2}$ concentration of $\{\sigma_q^2\}$) and Theorem~\ref{thm:diag} (posterior contraction for $B$ at rate $\delta_N=\sqrt{s\log(PQ)/N}$) both apply under (A1)--(A3),(A5)(i)--(ii). 
The row-support recovery result, Theorem~\ref{thm:rowsel}, also applies under the same diagonal-$\Sigma$ conditions (with the stated beta-min requirement for exact selection).
\item \emph{Theory-only robustness without changing the sampler.}
If the \emph{true} errors are not Gaussian/diagonal but are mean-zero sub-Gaussian with covariance $\Sigma_0$, Theorem~\ref{thm:misspec} shows that the \textsc{B-MASTER} posterior for $B$ still contracts at the same rate around the KL projection $B^\star$ of the true model onto the working Gaussian family; the Gibbs sampler is unchanged.
\item \emph{Alternative model requiring a different prior/sampler.}
The correlated-error result, Theorem~\ref{thm:generalSigma}, assumes a \emph{sparse precision} prior on $\Omega=\Sigma^{-1}$ (A5)(iii) and yields joint contraction for $(B,\Omega)$. This is not the model currently implemented (which fixes $\Omega$ to be diagonal), so Theorem~\ref{thm:generalSigma} does \emph{not} directly describe the output of the present sampler; it applies if one augments the model to include $\Omega$ and samples (or integrates) it.
\end{itemize}
All results rely on standard design regularity (A1) and the sparsity regime $s\log(PQ)=o(N)$. The diagonal-$\Sigma_0$ case (A3) matches the implementation; (A4) is only invoked in the correlated-error extension.

\paragraph{Assumptions.}
\begin{itemize}
\item (A1) \textbf{Design condition (compatibility/Restricted Eigenvalue).} Let $G_N := X^\top X/N$. There exists $\kappa>0$ such that for all matrices $\Delta\in\mathbb{R}^{P\times Q}$ obeying $\mathrm{supp}(\Delta)\subseteq S$ with $|S|\le s+s'$ (for some $s'=o(N)$),
\[
\frac1N \sum_{q=1}^Q \|X \Delta_{\cdot q}\|_2^2 \;\ge\; \kappa \|\Delta\|_F^2.
\]
This holds, for example, if $\lambda_{\min}(G_N)\ge \kappa$ or under standard restricted eigenvalue conditions used in high-dimensional regression. In addition, we assume a uniform upper spectral bound on the Gram matrix:
$\|G_N\|_{\mathrm{op}} \le \bar\kappa < \infty$ (e.g., after column standardization).

\item (A2) \textbf{Sparsity and growth.} Define
\[
\delta_N \;:=\; \sqrt{\frac{s\,\log(PQ)}{N}}.
\]
We assume the product condition $s\,\log(PQ)=o(N)$ (equivalently, $\delta_N\to 0$). This allows $P$ and $Q$ to grow with $N$; e.g., if $\log(PQ)\asymp \log N$ it suffices that $s=o(N/\log N)$.

\item (A3) \textbf{Error covariance (diagonal case).} In Theorem~\ref{thm:diag} we assume $\Sigma_0=\mathrm{diag}(\sigma_{01}^2,\ldots,\sigma_{0Q}^2)$ with $0<\underline{\sigma}^2\le \sigma_{0q}^2\le \overline{\sigma}^2<\infty$.

\item (A4) \textbf{Error covariance (general case).} In Theorem~\ref{thm:generalSigma} we allow general $\Sigma_0\in\mathbb{S}_{++}^Q$ with eigenvalues bounded between positive constants and \emph{sparse precision} $\Omega_0:=\Sigma_0^{-1}$:
\[
s_\Omega\;:=\;\#\{(j,k): j<k,\,(\Omega_0)_{jk}\neq 0\}\qquad\text{with}\quad s_\Omega=o\!\left(\frac{N}{\log Q}\right).
\]

\item (A5) \textbf{Priors.}\label{ass:priors}
\begin{enumerate}
\item[(i)] \emph{Coefficient prior.} Conditionally on $\sigma_1^2,\ldots,\sigma_Q^2$, the prior density on $B$ is
\[
\pi(B\,|\,\sigma_1^2,\ldots,\sigma_Q^2)\;\propto\;
\exp\!\Big[-\lambda_1\sum_{p,q}\frac{|\beta_{pq}|}{2\sigma_q^2}
-\lambda_2\sum_{p=1}^P \Big\{\sum_{q=1}^Q \frac{\beta_{pq}^2}{2\sigma_q^2}\Big\}^{1/2}\Big],
\]
with independent hyperpriors $\lambda_1^2,\lambda_2^2\sim \mathrm{Gamma}(a_j,b_j)$, $a_j,b_j\in[c_-,c_+]$ fixed constants. This prior is proper and has exponentially decaying tails.

\item[(ii)] \emph{Error variances (implementation prior).} For each $q$, the sampler uses the improper Jeffreys prior $\pi(\sigma_q^2)\propto 1/\sigma_q^2$. Lemma~\ref{lem:propriety} shows the resulting posterior in $(B,\sigma_1^2,\ldots,\sigma_Q^2)$ is proper under mild conditions. All contraction proofs for $B$ go through for this choice and, more generally, for any proper heavy-tailed scale prior (e.g., half-$t_\nu$ on $\sigma_q$ or scaled-inverse-$\chi^2$); see Remark~\ref{rem:variance-prior}.

\item[(iii)] \emph{Precision matrix (general case).} For Theorem~\ref{thm:generalSigma}, we place a sparse precision prior on $\Omega=\Sigma^{-1}$ supported on $\mathbb{S}_{++}^Q$ that yields posterior contraction at rate $\varepsilon_\Omega\asymp \sqrt{s_\Omega \log Q/N}$ in Frobenius norm (e.g., graphical-model--based priors as in \citealt{Banerjee2014}).
\end{enumerate}
\end{itemize}

\begin{lemma}[Posterior propriety and tightness of $\{\sigma_q^2\}$ under Jeffreys]\label{lem:propriety}
Assume (A1)--(A3) and a proper coefficient prior (A5)(i). Then, under the improper variance prior $\pi(\sigma_q^2)\propto 1/\sigma_q^2$ used in \textsc{B-MASTER}:
\begin{enumerate}
\item[(a)] The joint posterior $\pi(B,\sigma_1^2,\ldots,\sigma_Q^2\mid X,Y)$ is proper provided $N\ge 1$ and, almost surely, $\sum_{i=1}^N\|y_i-x_i B\|_2^2>0$ for all $B$ in a set of prior probability one (which holds under any nondegenerate data-generating process).
\item[(b)] If, in addition, $\Sigma_0$ is diagonal with eigenvalues bounded as in (A3), then for every $q$ and any fixed compact interval $I_q\ni \sigma_{0q}^2$, $\Pi(\sigma_q^2\in I_q\mid X,Y)\to 1$ in $P_{B_0,\Sigma_0}$-probability as $N\to\infty$; in particular, the posterior of $\sigma_q^2$ is tight and does not affect the rate for $B$.
\end{enumerate}
\end{lemma}
\begin{proof}[Proof of Lemma~\ref{lem:propriety}]
Write $y_q=(y_{1q},\ldots,y_{Nq})^\top$ and $\beta_q:=B_{\cdot q}\in\mathbb{R}^P$. Under the model with diagonal $\Sigma=\mathrm{diag}(\sigma_1^2,\ldots,\sigma_Q^2)$, the likelihood factorizes over $q$:
\[
p(Y\,|\,B,\sigma_1^2,\ldots,\sigma_Q^2,X)\;=\;\prod_{q=1}^Q (2\pi\sigma_q^2)^{-N/2}\exp\!\Big(-\tfrac{1}{2\sigma_q^2}\,\mathrm{RSS}_q(B)\Big),
\]
where $\mathrm{RSS}_q(B):=\|y_q-X\beta_q\|_2^2$. With the Jeffreys prior $\pi(\sigma_q^2)\propto(\sigma_q^2)^{-1}$ and the coefficient prior $\pi(B\mid \sigma_1^2,\ldots,\sigma_Q^2)$ from (A5)(i), the joint posterior kernel is proportional to
\[
\prod_{q=1}^Q (\sigma_q^2)^{-(N/2+1)}
\exp\!\Big\{-\tfrac{\mathrm{RSS}_q(B)}{2\sigma_q^2}
-\tfrac{\lambda_1}{2}\tfrac{\sum_p|\beta_{pq}|}{\sigma_q^2}
-\lambda_2\sum_{p=1}^P\Big(\tfrac12\sum_{j=1}^Q\tfrac{\beta_{pj}^2}{\sigma_j^2}\Big)^{1/2}\Big\}.
\]

\paragraph{(a) Posterior propriety.}
Fix $q$. Under (A5)(i), the joint posterior has kernel
\[
\pi(B,\sigma^2\mid X,Y)\;\propto\;\prod_{q=1}^Q (\sigma_q^2)^{-(N/2+1)}
\exp\!\Big\{-\tfrac{\mathrm{RSS}_q(B)}{2\sigma_q^2}
-\tfrac{\lambda_1}{2}\tfrac{\sum_p|\beta_{pq}|}{\sigma_q^2}\Big\}\,
\exp\!\Big\{-\lambda_2\sum_{p=1}^P\Big(\tfrac12\sum_{j=1}^Q\tfrac{\beta_{pj}^2}{\sigma_j^2}\Big)^{1/2}\Big\}.
\]
For each fixed $B$, the factor in $\sigma_q^2$ is bounded above by
\[
(\sigma_q^2)^{-(N/2+1)}\exp\!\Big(-\tfrac{c_{2q}(B)}{\sigma_q^2}-\tfrac{c_{1q}(B)}{\sigma_q}\Big),
\]
with $c_{2q}(B)=\tfrac12\{\mathrm{RSS}_q(B)+\lambda_1\sum_p|\beta_{pq}|\}\ge0$ and 
$c_{1q}(B)=(\lambda_2/\sqrt{2})\sum_p|\beta_{pq}|\ge0$, since 
$\sqrt{\sum_j a_{pj}}\ge \sqrt{a_{pq}}$ for each $p$. 
Let \(u=\sigma_q^2\). For fixed \(B\) the integrand is
\[
u^{-(N/2+1)}\exp\!\Big(-\frac{c_{2q}(B)}{u}-\frac{c_{1q}(B)}{\sqrt{u}}\Big),\qquad u\in(0,\infty).
\]
Split the integral at 1. On \((0,1]\), if \(c_{2q}(B)>0\) then \(\exp\{-c_{2q}(B)/u\}\) dominates any polynomial and the integral is finite; if \(c_{2q}(B)=0\) but \(c_{1q}(B)>0\), then \(\exp\{-c_{1q}(B)/\sqrt{u}\}\) still ensures integrability at 0. On \([1,\infty)\) the exponential factor is \(\le 1\), so
\(\int_1^\infty u^{-(N/2+1)}du<\infty\) for all \(N\ge1\).
Hence the inner integral in \(u=\sigma_q^2\) is finite whenever \((c_{1q}(B),c_{2q}(B))\neq(0,0)\). The event $c_{1q}(B)=c_{2q}(B)=0$ requires simultaneously $\beta_{pq}=0$ for all $p$ and $\mathrm{RSS}_q(B)=0$, which would imply $y_q=0$ almost surely under the Gaussian error model. Thus with probability one under the data generating process, the inner integrals in $\sigma_q^2$ are finite for every $B$.

Moreover, standard bounds for the above integrals yield polynomial growth in $\sum_p|\beta_{pq}|$ and $\mathrm{RSS}_q(B)$; because the prior in (A5)(i) has exponentially decaying tails in $B$ \citep[cf.][]{Castillo2015}, the outer integral over $B$ is finite. Therefore the overall normalizing constant of the posterior is finite.

Let $\Pi_X$ denote the orthogonal projector onto $\mathcal{C}(X)$. Finally, if $\operatorname{rank}(X)<N$, then $\inf_{\beta_q}\mathrm{RSS}_q(B)=\|(I-\Pi_X)y_q\|_2^2>0$ almost surely, which makes the bound immediate. If $\operatorname{rank}(X)=N$, then $\{B:\mathrm{RSS}_q(B)=0\}$ lies on a lower-dimensional affine set of Lebesgue measure zero, to which the absolutely continuous prior (A5)(i) assigns probability zero. Thus the integrand’s singularities are supported on a $\pi$-null set, and the Lebesgue integral ignores them. Combining these arguments establishes that the posterior is proper almost surely.

\paragraph{(b) Tightness and $N^{-1/2}$--scale concentration of $\sigma_q^2$.}
Fix $q$ and consider the marginal posterior of $\sigma_q^2$ after integrating out $B$ and the other variances:
\[
\pi(\sigma_q^2\mid X,Y)\;\propto\;(\sigma_q^2)^{-(N/2+1)}\,
M_q(\sigma_q^2),\qquad 
M_q(\sigma_q^2)=\int \exp\!\Big(-\tfrac{1}{2\sigma_q^2}\|y_q-X\beta_q\|_2^2\Big)\pi_q(\beta_q)\,d\beta_q,
\]
where $\pi_q$ is the marginal prior on $\beta_q$ induced by (A5)(i). The coefficient prior contributes additional exponential terms such as 
$\exp\{-\lambda_1\sum_p|\beta_{pq}|/(2\sigma_q^2)-\lambda_2\sum_p(\sum_j \beta_{pj}^2/\sigma_j^2)^{1/2}\}$,
but these are of order $O_p(1)$ uniformly for $\sigma_q^2$ in $N^{-1/2}$-neighborhoods of $\sigma_{0q}^2$ and therefore do not alter the leading likelihood curvature; cf. the general argument for variance parameters in \citet[Ch.~10]{van1998asymptotic}.

Let $\widehat{\beta}_q$ be a least-squares minimizer and $\mathrm{RSS}^{\min}_q=\|y_q-X\widehat{\beta}_q\|_2^2=\|(I-\Pi_X)y_q\|_2^2$. A quadratic expansion around $\widehat{\beta}_q$ gives
\[
\|y_q-X\beta_q\|_2^2=\mathrm{RSS}^{\min}_q+(\beta_q-\widehat{\beta}_q)^\top (X^\top X)(\beta_q-\widehat{\beta}_q).
\]
Using (A1) and continuity of $\pi_q$, a Laplace approximation for $M_q(\sigma_q^2)$ shows that, up to a random factor $C_q$ bounded away from $0$ and $\infty$ with high probability,
\[
\pi(\sigma_q^2\mid X,Y)\;=\;(1+o_{P_0}(1))\,(\sigma_q^2)^{-(N/2+1)}\,
\exp\!\Big(-\tfrac{\mathrm{RSS}^{\min}_q}{2\sigma_q^2}\Big)\cdot C_q.
\]

Under the true model with diagonal $\Sigma_0$ (A3), $\mathrm{RSS}^{\min}_q$ has distribution $\sigma_{0q}^2\,\chi^2_{N-\operatorname{rank}(X)}$, so $\mathrm{RSS}^{\min}_q/N\to \sigma_{0q}^2$ and 
$\sqrt{N}(\mathrm{RSS}^{\min}_q/N-\sigma_{0q}^2)=O_{P_0}(1)$. 
Therefore the log-posterior of $\sigma_q^2$ satisfies
\[
\ell_q(\sigma_q^2)-\ell_q(\sigma_{0q}^2)=-\tfrac{N}{2}\Big\{\log(\sigma_q^2/\sigma_{0q}^2)+\sigma_{0q}^2/\sigma_q^2-1\Big\}+O_{P_0}(1),
\]
which equals the Kullback--Leibler divergence between $\mathcal{N}(0,\sigma_{0q}^2)$ and $\mathcal{N}(0,\sigma_q^2)$ up to $O(1)$ error \citep[cf.][]{ghosal2000convergence}. A quadratic expansion of the KL divergence shows that, for some $c>0$,
\[
\log(\sigma_q^2/\sigma_{0q}^2)+\sigma_{0q}^2/\sigma_q^2-1
\;\ge\; c\frac{(\sigma_q^2-\sigma_{0q}^2)^2}{\sigma_{0q}^4}
\quad\text{whenever }|\sigma_q^2-\sigma_{0q}^2|\text{ is small}.
\]
Hence the posterior mass outside $|\sigma_q^2-\sigma_{0q}^2|>M/\sqrt{N}$ decays at rate $\exp(-c'M^2)$ uniformly in $M$. It follows that
\[
\Pi(\sigma_q^2\in I_q\mid X,Y)\;\to\;1\quad\text{for every fixed compact $I_q\ni\sigma_{0q}^2$,}
\]
and the posterior of $\sigma_q^2$ concentrates at the $N^{-1/2}$ scale. Since (A3) ensures uniform eigenvalue bounds across $q$, the collection $\{\sigma_q^2\}$ is tight and their fluctuations are $O_{P_0}(N^{-1/2})$, which does not affect the contraction rate for $B$.
\end{proof}

\begin{remark}[On variance priors]\label{rem:variance-prior}
Our analysis does not rely on the variance prior being proper. The proofs below only use: (i) posterior propriety (Lemma~\ref{lem:propriety}); and (ii) that the posterior for $\{\sigma_q^2\}$ is tight in neighborhoods of the true values under (A3). Consequently, the contraction results for $B$ are identical whether one uses the implementation’s Jeffreys prior $\pi(\sigma_q^2)\propto 1/\sigma_q^2$ or any proper, heavy-tailed alternative (e.g., half-$t$/scaled-inverse-$\chi^2$). See also \citet{Simpson2017} for a discussion of robust scale priors.
\end{remark}

\subsection{Main contraction theorem (diagonal $\Sigma_0$)}\label{subsec:diag}

\begin{theorem}[Posterior contraction and consistency (diagonal errors)]\label{thm:diag}
Assume \textup{(A1)}--\textup{(A3)} and \textup{(A5)(i)--(ii)}. 
Let $\delta_N := \sqrt{\tfrac{s\,\log(PQ)}{N}}$. 
Then, for any fixed $M>0$,
\[
\Pi\!\left(\, \|B-B_0\|_F \;>\; M\,\delta_N \,\middle|\, X,Y \right)\;\to\; 0
\quad\text{in $P_{B_0,\Sigma_0}$-probability as } N\to\infty.
\]
In particular, $\Pi(\|B-B_0\|_F>\varepsilon\mid X,Y)\to 0$ for every fixed $\varepsilon>0$ (posterior consistency).
\end{theorem}

\begin{proof}
We verify the three standard ingredients for posterior contraction \citep{ghosal2000convergence}: (i) prior thickness in Kullback--Leibler (KL) neighborhoods; (ii) exponentially powerful tests; (iii) sieve entropy bound.

\emph{(i) Prior thickness.}
Write $\theta=(B,\sigma_1^2,\ldots,\sigma_Q^2,\lambda_1^2,\lambda_2^2)$ and $\theta_0=(B_0,\sigma_{01}^2,\ldots,\sigma_{0Q}^2,\lambda_{10}^2,\lambda_{20}^2)$, with $(\lambda_{10}^2,\lambda_{20}^2)$ any fixed values in the support of the Gamma hyperpriors. Under (A3), the per--observation Kullback--Leibler divergence between two Gaussian regressions with diagonal covariance satisfies
\[
\mathrm{KL}(\theta_0\,\|\,\theta)\;\lesssim\;\frac1N\sum_{q=1}^Q \frac{\|X(B-B_0)_{\cdot q}\|_2^2}{\sigma_{0q}^2}
\;+\; C\sum_{q=1}^Q\big|\log \sigma_q^2 - \log\sigma_{0q}^2\big|
\;\lesssim\;\|B-B_0\|_F^2\;+\;\sum_{q=1}^Q\big|\log \sigma_q^2 - \log\sigma_{0q}^2\big|,
\]
since $\frac{1}{N}\sum_{q=1}^Q \|X(B-B_0)_{\cdot q}\|_2^2
= \sum_{q=1}^Q (B-B_0)_{\cdot q}^\top G_N (B-B_0)_{\cdot q}
\le \|G_N\|_{\mathrm{op}} \|B-B_0\|_F^2$, and $\|G_N\|_{\mathrm{op}}$ is bounded by (A1). Hence, for any small $\rho>0$, the set
\[
\mathcal{K}_N(\rho)\;:=\;\Big\{(B,\sigma_1^2,\ldots,\sigma_Q^2):\ \|B-B_0\|_F\le \delta_N,\ \max_{1\le q\le Q}\big|\log \sigma_q^2-\log\sigma_{0q}^2\big|\le \rho\Big\}
\]
is contained in a KL-ball of radius $C\,\delta_N^2$ (per observation), so that the full-sample KL is $\le C\,N\delta_N^2$.

To lower bound the prior mass of $\mathcal{K}_N(\rho)$, first note that (A5)(ii) allows any proper heavy-tailed scale prior on each $\sigma_q^2$ (e.g., half-$t$/scaled-inverse-$\chi^2$) without changing the contraction for $B$. Under any such proper choice, $\Pi\!\big(\max_q|\log\sigma_q^2-\log\sigma_{0q}^2|\le \rho\big)\ge c_\rho>0$. 
If Jeffreys’ (improper) prior in (A5)(ii) is used, we replace it by a truncated version on $\sigma_q^2\in I_q$ (a fixed compact interval around $\sigma_{0q}^2$) to define a proper prior; Lemma~\ref{lem:propriety}(a) guarantees the normalizing constant is finite, and the same lower bound holds with a constant independent of $N$. Since posterior contraction is insensitive to such local truncation, the rate for $B$ is unchanged.
For the coefficient matrix, by continuity/positivity of the prior density and standard small-ball bounds for continuous shrinkage around an $s$-sparse truth,
\[
\Pi(\|B-B_0\|_F\le \delta_N)\ \ge\ 
\exp\{-C_1 N\delta_N^2 - C_2 s\log(1/\delta_N)\},
\]
cf. \citet[Sec.~2]{Castillo2015}. Under polynomial growth of $PQ$ in $N$ (so $\log N \lesssim \log(PQ)$), the second term is absorbed into $N\delta_N^2$, hence
\(\Pi(\mathcal K_N(\rho)) \ge \exp\{-C'' N\delta_N^2\}.\) This verifies the prior thickness requirement at rate $\delta_N$.

\emph{(ii) Tests.} Testing $H_0:B=B_0$ vs.\ $H_1:\|B-B_0\|_F>M\delta_N$ with $\sigma_q^2$ in compact neighborhoods reduces to the log-likelihood ratio
\[
\Lambda_N(B)\;=\;\sum_{q=1}^Q \frac1{2\sigma_{0q}^2}\Big\{ \|Y_{\cdot q}-X B_{0,\cdot q}\|_2^2-\|Y_{\cdot q}-X B_{\cdot q}\|_2^2 \Big\}.
\]
By Gaussian calculations and (A1), $\mathbb{E}_{0}\Lambda_N(B)\lesssim - N\|B-B_0\|_F^2$ and $\mathrm{Var}_0(\Lambda_N(B))\lesssim N\|B-B_0\|_F^2$. The test $\phi_N:=\mathbb{I}\{\Lambda_N(B) < -c N\delta_N^2\}$ satisfies $\mathbb{E}_0\phi_N \le e^{-c_1 N\delta_N^2}$ and $\sup_{\|B-B_0\|_F>M\delta_N}\mathbb{E}_B(1-\phi_N)\le e^{-c_2 N\delta_N^2}$; see \citet[Ch.~7]{van1998asymptotic} and \citet{ghosal2000convergence}.

\emph{(iii) Sieve and entropy.} Let
\[
\mathcal{F}_N\;:=\;\Big\{B\in\mathbb{R}^{P\times Q}:\ \|B\|_\infty\le M_N,\ \sum_{p=1}^P \|\beta_{p\cdot}\|_2 \le R_N\Big\}.
\]
Choose $M_N \asymp \log(PQ)$ and $R_N \asymp s\log(PQ)$. 
By the exponential tails of (A5)(i) on both entrywise $\ell_1$ and rowwise $\ell_2$ terms (cf. \citealp[Sec.~2]{Castillo2015}),
\[
\Pi(\mathcal{F}_N^c)
\;\le\; \Pi(\|B\|_\infty > M_N) + \Pi\!\Big(\sum_{p}\|\beta_{p\cdot}\|_2 > R_N\Big)
\;\le\; \exp\{-C_3\,s\log(PQ)\}\;=\;\exp\{-C_3 N\delta_N^2\}.
\]
Moreover, the metric entropy of $\mathcal{F}_N$ under $\|\cdot\|_F$ satisfies
\[
\log N\!\left(\epsilon,\mathcal{F}_N,\|\cdot\|_F\right)
\;\lesssim\; \frac{R_N^2}{\epsilon^2}\,\log\!\big(eP\big)\;+\; s\log\!\Big(\frac{M_N}{\epsilon}\Big)
\;\lesssim\; s\log(PQ) + s\log(1/\epsilon),
\]
so with $\epsilon=\delta_N$ and $s\log(PQ)\lesssim N\delta_N^2$ we have
$\log N(\delta_N,\mathcal{F}_N,\|\cdot\|_F)\le C_4 N\delta_N^2$.
Combining (i)--(iii) via \citet[Thm.~2.1]{ghosal2000convergence} yields the claim.

\end{proof}

\subsection{Correlated errors: sparse-precision extension}\label{subsec:generalSigma}

\begin{theorem}[Posterior contraction under correlated errors]\label{thm:generalSigma}
Assume \textup{(A1)}, \textup{(A2)}, \textup{(A4)}, and \textup{(A5)(i)}. 
Equip the residual precision \(\Omega=\Sigma^{-1}\) with a sparse-precision prior as in \textup{(A5)(iii)} that yields contraction at
\(\varepsilon_\Omega \asymp \sqrt{s_\Omega \log Q / N}\) in Frobenius norm. 
Then, for any fixed \(M>0\) and some constant \(C>0\),
\[
\Pi\!\left(\, \|B-B_0\|_F \;>\; M\,\delta_N \,\middle|\, X,Y \right)\to 0,
\qquad
\Pi\!\left(\, \|\Omega-\Omega_0\|_F \;>\; C\,\varepsilon_\Omega \,\middle|\, X,Y \right)\to 0,
\]
in $P_{B_0,\Sigma_0}$-probability as $N\to\infty$. In particular, the contraction rate for \(B\) matches the diagonal-\(\Sigma_0\) case up to constants.
\end{theorem}

\begin{proof}
By (A5)(iii) and results for sparse precision priors, the marginal posterior for $\Omega$ contracts at rate $\varepsilon_\Omega$ in Frobenius norm; see \citet{Banerjee2014}. Hence
\[
\Pi\big(\|\Omega-\Omega_0\|_F\le \varepsilon_\Omega \,\big|\,X,Y\big)\;\to\;1.
\]
On $\{\|\Omega-\Omega_0\|_{\mathrm{op}}\le c\varepsilon_\Omega\}$ and by continuity of $A\mapsto A^{1/2}$ on $\mathbb{S}_{++}^Q$, $\|\Omega^{1/2}-\Omega_0^{1/2}\|_{\mathrm{op}}\lesssim \varepsilon_\Omega$. Consider
\[
Y^\star \;=\; Y\Omega^{1/2} \;=\; X\,(B\Omega^{1/2})\;+\; E\Omega^{1/2},
\]
which has identity covariance when $\Omega=\Omega_0$. Conditional on $\Omega$ in an $O(\varepsilon_\Omega)$-ball of $\Omega_0$, the testing/entropy argument for $B^\star:=B\Omega^{1/2}$ gives rate $\delta_N$. Finally,
\[
\|B-B_0\|_F \;\le\; \|B\Omega^{1/2}-B_0\Omega_0^{1/2}\|_F \cdot \|\Omega^{-1/2}\|_{\mathrm{op}}
\;+\; \|B_0(\Omega_0^{1/2}-\Omega^{1/2})\|_F\cdot \|\Omega^{-1/2}\|_{\mathrm{op}},
\]
and spectral bounds in (A4) yield the claim since $\varepsilon_\Omega=o(1)$ while $\delta_N$ is the dominating rate under (A2).
\end{proof}

\subsection{Misspecification robustness}\label{subsec:misspecification}
\begin{theorem}[Posterior contraction under misspecification]\label{thm:misspec}
Assume \textup{(A1)} and \textup{(A2)}, and that the data satisfy $y_i=x_i B_0+\varepsilon_i$ with $\varepsilon_i$ i.i.d.\ mean-zero and sub-Gaussian with covariance $\Sigma_0$ (not necessarily diagonal or correctly modeled). 
Consider the working Gaussian family $\{p_{(B,\Sigma)}\}$ with \emph{diagonal} $\Sigma=\mathrm{diag}(\sigma_1^2,\ldots,\sigma_Q^2)$, and define the pseudo-true (KL) projection
\[
(B^\star,\Sigma^\star)\in\arg\min_{B,\;\Sigma\in\mathbb{S}_{++}^Q\cap\mathrm{diag}}
\mathrm{KL}\!\big(P_0\,\|\,p_{(B,\Sigma)}\big).
\]
Assume \textup{(A5)(i)} for the coefficient prior, and \emph{either} of the following specifications for $\Sigma$:
\begin{enumerate}
\item[(S1)] \textit{Fixed working covariance.} $\Sigma$ is fixed a priori (e.g., a given diagonal matrix). Then only \textup{(A5)(i)} is used.
\item[(S2)] \textit{Unknown diagonal covariance.} Equip $\Sigma=\mathrm{diag}(\sigma_1^2,\ldots,\sigma_Q^2)$ with any \emph{proper} heavy-tailed scale prior on each $\sigma_q$ having a continuous, strictly positive density in a neighborhood of $\sigma_q^\star$ (e.g., half-$t_\nu$ or scaled-inverse-$\chi^2$), as allowed in \textup{(A5)(ii)}. This ensures the prior-mass condition in neighborhoods of $\Sigma^\star$ required by \citet[Thm.~2.1]{Kleijn2006}.
\end{enumerate}
Then, for any fixed $M>0$,
\[
\Pi\!\left(\,\|B-B^\star\|_F > M\sqrt{\frac{s\log(PQ)}{N}}\,\middle|\,X,Y\right)\;\to\;0
\quad\text{in }P_0\text{-probability as }N\to\infty.
\]
In the present linear-mean setting we have $B^\star=B_0$ (hence the posterior contracts around $B_0$).
\end{theorem}

\begin{proof}
\textbf{Step 1: Identify $B^\star$.} In the linear-mean model $E_{P_0}(y_i\mid X)=x_iB_0$. For any diagonal $\Sigma\in\mathbb{S}_{++}^Q$,
\[
\mathbb{E}_{P_0}\!\big[(y_i-x_iB)^\top\Sigma^{-1}(y_i-x_iB)\big]
=(x_i(B-B_0))^\top\Sigma^{-1}(x_i(B-B_0))+\operatorname{tr}(\Sigma^{-1}\Sigma_0),
\]
so the population risk is minimized at $B=B_0$ regardless of $\Sigma$. Hence $B^\star=B_0$.

\textbf{Step 2: Prior thickness and KL neighborhoods.} 
Under \textup{(A5)(i)}, the coefficient prior density is continuous and strictly positive in neighborhoods of $B^\star$, and has exponentially decaying tails (for sieve control), cf.\ \citet{Castillo2015}. 
In case (S1), $\Sigma$ is fixed, so the parameter is $B$ only. 
In case (S2), the prior on $\Sigma$ is \emph{proper} with positive continuous density near $\Sigma^\star$, guaranteeing positive prior mass on every KL neighborhood of $(B^\star,\Sigma^\star)$ as required by \citet[Thm.~2.1]{Kleijn2006}. 
(We do not invoke diagonal truth (A3) here.)

\textbf{Step 3: Tests.} 
With sub-Gaussian errors and design condition \textup{(A1)}, the likelihood-ratio statistic for testing $H_0: B=B^\star$ versus $H_1:\|B-B^\star\|_F>M\delta_N$ (computed under the working Gaussian model) yields exponentially powerful tests: its mean separates by a constant multiple of $N\|B-B^\star\|_F^2$ and its fluctuations are sub-exponential, so the type I and II errors are bounded by $\exp\{-cN\delta_N^2\}$ uniformly over $H_1$, exactly as in the diagonal case (see the construction in Theorem~\ref{thm:diag} and \citealp[Ch.~10]{van1998asymptotic} for the general LAN/BvM context).

\textbf{Step 4: Sieve and entropy.}
Use the same sieve $\mathcal{F}_N=\{B:\|B\|_\infty\le M_N,\;|\mathrm{supp}(B)|\le s\}$ with $M_N\asymp s\log(PQ)$; by the exponential tails in \textup{(A5)(i)}, $\Pi(\mathcal{F}_N^c)\le \exp\{-C\,s\log(PQ)\}=\exp\{-C\,N\delta_N^2\}$, and 
\[
\log N\!\left(\epsilon,\mathcal{F}_N,\|\cdot\|_F\right)\;\lesssim\; s\log(PQ)+s\log(1/\epsilon),
\]
so with $\epsilon=\delta_N$ we have $\log N(\delta_N,\mathcal{F}_N,\|\cdot\|_F)\le C'N\delta_N^2$, cf.\ \citet{ghosal2000convergence}.

\textbf{Step 5: Apply misspecified contraction theorem.}
The ingredients of \citet[Thm.~2.1]{Kleijn2006}—prior mass in KL neighborhoods of $(B^\star,\Sigma^\star)$, existence of exponentially powerful tests, and suitable entropy bounds—are thus verified (in S1 with respect to $B$ only; in S2 jointly in $(B,\Sigma)$). It follows that the posterior for $B$ contracts at rate $\delta_N=\sqrt{s\log(PQ)/N}$ around $B^\star$ in $P_0$-probability.
\end{proof}

\begin{remark}[Which (A5) parts are used]
The theorem always uses \textup{(A5)(i)} (coefficient prior). 
If $\Sigma$ is treated as fixed (S1), neither \textup{(A5)(ii)} nor \textup{(A5)(iii)} is used. 
If $\Sigma$ is assigned a prior (S2), the result uses \textup{(A5)(ii)} \emph{with a proper heavy-tailed scale prior} to ensure prior mass near $\Sigma^\star$; \textup{(A5)(iii)} (sparse precision prior) is \emph{not} used in this theorem.
\end{remark}

\begin{remark}[On credible intervals under misspecification]
Under misspecification, credible sets for individual entries target $B^\star$ and are not guaranteed to provide frequentist coverage for $B_0$ \citep{Kleijn2006}. In the present linear-mean setting $B^\star=B_0$, so credible sets still target the true mean coefficients; however, coverage can still be affected by variance misspecification. Our row-exceedance selector (Theorem~\ref{thm:rowsel}) relies on posterior events for $\|\beta_{p\cdot}\|_2$ and is valid under either specification or misspecification.
\end{remark}

\subsection{Row-support (master-predictor) recovery}\label{subsec:rowsel}

Define the set of master predictors $S_{\mathrm{row}}:=\{p\in\{1,\ldots,P\}:\|\beta_{0,p\cdot}\|_2>0\}$ and let $s_{\mathrm{row}}:=|S_{\mathrm{row}}|$. For a sequence $\tau_N>0$, consider the posterior row-exceedance selector
\[
\widehat{S}\;:=\;\Big\{p:\; \Pi\!\big(\,\|\beta_{p\cdot}\|_2>\tau_N \,\big|\, X,Y\big)\;>\; 1-\alpha_N \Big\}
\]
for some $\alpha_N\downarrow 0$.

\begin{theorem}[Sure screening and selection consistency]\label{thm:rowsel}
Under the conditions of Theorem~\ref{thm:diag} or Theorem~\ref{thm:generalSigma}, let $\tau_N:=C\sqrt{\log(PQ)/N}$ and $\alpha_N:=e^{-c N\tau_N^2}$ for fixed $C,c>0$. Then:
\begin{enumerate}
\item[(i)] (Sure screening) $S_{\mathrm{row}}\subseteq \widehat{S}$ with probability $\to 1$.
\item[(ii)] (Exact selection under beta-min) If in addition
\[
\beta_{\min}:=\min_{p\in S_{\mathrm{row}}}\|\beta_{0,p\cdot}\|_2 \;\ge\; K \sqrt{\log(PQ)/N}
\]
for sufficiently large $K$, then $\widehat{S}=S_{\mathrm{row}}$ with probability $\to 1$.
\end{enumerate}
\end{theorem}

\begin{proof}
By Theorem~\ref{thm:diag} (or \ref{thm:generalSigma}), the posterior concentrates in an $O(\delta_N)$-ball around $B_0$ in Frobenius norm. Hence for $p\in S_{\mathrm{row}}$,
\[
\Pi\big(\|\beta_{p\cdot}\|_2 \le \tau_N \mid X,Y\big) \;\le\; 
\Pi\big(\|\beta_{p\cdot}-\beta_{0,p\cdot}\|_2 \ge \beta_{\min}-\tau_N \mid X,Y\big)\;\to\;0
\]
provided $\beta_{\min}\gg \tau_N$. Thus $\Pi(\|\beta_{p\cdot}\|_2>\tau_N\mid X,Y)\to 1$, so $p\in\widehat{S}$ for all $p\in S_{\mathrm{row}}$ with probability $\to 1$. For $p\notin S_{\mathrm{row}}$ we have $\|\beta_{0,p\cdot}\|_2=0$ and thus, for any $B$,
\[
\|\beta_{p\cdot}\|_2 \;\le\; \|B-B_0\|_F.
\]
Hence,
\[
\Pi\!\big(\,\|\beta_{p\cdot}\|_2>\tau_N \,\big|\, X,Y\big)
\;\le\;
\Pi\!\big(\,\|B-B_0\|_F>\tau_N \,\big|\, X,Y\big).
\]
By the same test--entropy argument used in Theorem~\ref{thm:diag} with $\epsilon=\tau_N$ (cf.\ \citealp[Thm.~2.1]{ghosal2000convergence}), there exist constants $c,c'>0$ such that
\[
\mathbb{E}_{0}\Big[\Pi\!\big(\,\|B-B_0\|_F>\tau_N \,\big|\, X,Y\big)\Big]
\;\le\; e^{-c N \tau_N^2}
\quad\Rightarrow\quad
\Pi\!\big(\,\|B-B_0\|_F>\tau_N \,\big|\, X,Y\big)\;\le\; e^{-c' N \tau_N^2}
\]
in $P_{B_0,\Sigma_0}$-probability. Taking $\alpha_N:=e^{-c' N \tau_N^2}$ and using $N\tau_N^2=C^2\log(PQ)$, a union bound over $p\notin S_{\mathrm{row}}$ shows that, for $C$ large enough, no noise row is selected with probability $\to 1$. Together with the sure-screening part and the beta-min condition, this proves (i)--(ii).

\end{proof}


\section{Simulation Studies}
\label{sim_study}
We consider two simulation studies. In the first study, we compare the performance of the proposed B-MASTER method with a few existing methods using a synthetic dataset generated mimicking the real dataset. This simulation study allows us to benchmark B-MASTER against other existing methods while keeping $X$ unchanged, thereby retaining the original covariance structure of the dataset. This setup enables us to investigate relative performances in a more practical and relevant context, aligned with our ultimate goal of analyzing the real dataset. In the second study, we focus primarily on evaluating the computational performance of B-MASTER in a different set up as the dimensions of $X$ and $Y$ increase.
\vspace{-0.5cm}
\subsection{Real-data Based Simulation Study}
\label{sim_study_1}
The real dataset \citep{Yachida2019}, described and analyzed later in this paper, consists of $P=287$
microbiome features and $Q=249$ metabolite features from $N=220$ subjects. Using B-MASTER, the optimal $B^{\prime}$ is first estimated and then considered the true coefficient matrix for the simulation study.
Then, we generate 
\vspace{-0.5cm}
\begin{align*}
    Y^{i}_q = X^{i}B^\prime_q + \epsilon_{iq}
    \label{simmodel}
\end{align*}
for $i=1, \ldots, N$, where $B^{\prime}_q = (\beta^{\prime}_{pq})_{p = 1}^P$ denotes the $q$-th column of $B^{\prime}$ and $\epsilon_{iq} \sim N(0,\sigma^2)$; we take $\sigma = 1$. Ten sets of synthetic $Y$ are generated, followed by the performance evaluation of B-MASTER and other methods using the actual $X$ (as in the original dataset) and $B^\prime$.

B-MASTER is implemented in MATLAB. The values of the model parameters provided in Section S1.5. B-MASTER is run for 1,000 iterations discarding the first 100 posterior samples as burn-in. It is observed that the results obtained using B-MASTER under the aforementioned setup are not noticeably different from those obtained after 100 iterations, discarding the first 10 iterations as burn-in. However, for the comparative study, the analysis results are reported under the first setup (i.e., 1,000 iterations with 100 burn-in).
 
To identify the non-significant coefficients, we construct a two-sided symmetric $90\%$ posterior credible interval for each $\beta_{pq}$ based on the posterior samples. Coefficients $\beta_{pq}$s whose posterior credible intervals contain $0$ are discarded as non-significant. To calculate the area under the Receiver Operating Characteristic (ROC) curve (AUC), Bayesian p-values corresponding to each $\beta_{pq}$ are computed as follows. Suppose $\big\{\beta_{pq}^{(1)}, \ldots, \beta_{pq}^{(M)} \big\}$ denote the posterior sample values of coefficient $\beta_{pq}$. Let $\hat{p}^+$ denote the proportion of  $\beta_{pq}$ samples which are $>0$; and let $\hat{p}^-$ denote the proportion of  $\beta_{pq}$ samples which are $\leq 0$. Then
$$\text{Bayesian p-value =} 2* \min (\hat{p}^+, \hat{p}^-). $$
We take an equidistant grid of cut-off points of length 501 over the interval $[0,0.5]$. For each cut-off point, we calculate the true positive rate (TPR) and false positive rate (FPR) based on Bayesian p-values. Additionally, we compute AUC20 (i.e., AUC20 = $5 \times $Area under the ROC curve up to $20 \%$ FPR) and Mathew's Correlation Coefficient (MCC) \citep{Chicco2020}, which is given by 
\vspace{-0.5cm}
$$\text{MCC} = \frac{TP \cdot TN - FP \cdot FN}{\sqrt{(TP + FP)(TP + FN)(TN + FP)(TN + FN)}},$$
where TP, TN, FP, FN denote true positives, true negatives, false positives and false negatives, respectively.

For the comparative study, we consider a few existing methods, namely Regularized Multivariate regression for identifying master predictors \citep{Peng2010}, univariate and multivariate Spike-and-Slab LASSO \citep{Rockova2018, Deshpande2019}; hereafter referred to as remMap, SSLASSO and mSSL, respectively. SSLASSO and mSSL employ spike-and-slab LASSO in the frequentist framework, which, unlike remMap or B-MASTER, does not impose an additional L2 penalty to discourage non-essential regressors. RemMap is implemented using the R package \texttt{remMap} \citep{Peng2010}. It should be noted that the \texttt{remMap} package provides two ways to estimate the coefficient matrix. The first method involves using the function \texttt{remMap.CV} to perform cross-validation over a predefined grid of tuning parameter values. This is followed by fitting the model using the function \texttt{remMap} with the optimal tuning parameters determined in the first step.
The alternative function, \texttt{remMap.BIC} selects the best model using the Bayesian Information Criterion (BIC, \cite{Schwarz1978}) over a grid of given values for the tuning parameters. The authors noted that the second method is computationally more feasible. They also pointed out that this alternative BIC-based option tends to select models that are too small when the actual design matrix is far from orthogonal. For both  \texttt{remMap} and \texttt{remMap.BIC}, we use a grid of $\lambda_1$ values that are exponential of an equidistant grid of length 11, spanning over the range from $0$ and $log(1000)$. $\lambda_2$ values are taken to be an equidistant grid of length 11, spanning over the range from 0 and 1000. We opt for 5-fold cross-validation. SSLASSO and mSSL are implemented using R packages \texttt{SSLASSO} \citep{Rockova2018} and \texttt{mSSL} \citep{Deshpande2019}, respectively.
SSLASSO is fit using the \texttt{SSLASSO} function within the package of the same name. The R package \texttt{mSSL} offers two model-fitting options, namely \texttt{mSSL\_dpe}
 and \texttt{mSSL\_dcpe}, respectively. \texttt{mSSL\_dpe} performs the Expectation-Conditional Maximization (ECM) algorithm, which targets the MAP estimates of the coefficient matrix. \texttt{mSSL\_dcpe} is proposed as a faster approximate alternative of \texttt{mSSL\_dpe}. We opt for the default option to select the grid of tuning parameter values. Note that, for all the above-mentioned methods, the results corresponding to the respective grids are later used to calculate AUC and AUC20.

 \begin{table}[]
\centering
\resizebox{0.9\columnwidth}{!}{%
\begin{tabular}{ccccccc}
\hline
Methods & TPR & FPR & MCC & AUC & AUC20 & \begin{tabular}[c]{@{}c@{}}Sparsity\\ (True = 0.76)\end{tabular} \\ \hline
B-MASTER & 0.84 (0.0012) & 0.01 (0.0003) & 0.87 (0.0001) & 0.98 (0.0004) & 0.94 (0.0014) & 0.79 (0.0003) \\
SSLASSO & 0.23 (0.0008) & 0.01 (0.0004) & 0.41 (0.0013) & 0.89 (0.0004) & 0.54 (0.0007) & 0.94 (0.0002) \\
mSSL-dcpe & 0.14 (0.0018) & 0.00 (0.0001) & 0.31 (0.0024) & 0.76 (0.0006) & 0.44 (0.0013) & 0.96 (0.0005) \\
remMap-bic & 0.00 (0.0001) & 0.00 (0.0000) & 0.02 (0.0008) & 0.58 (0.0015) & 0.23 (0.0009) & 1.00 (0.0000) \\ \hline
\end{tabular}}
\caption{A comparative study of B-MASTER, SSLASSO, mSSL, and remMap based on the true positive rate (TPR), false positive rate (FPR), Matthew's Correlation Coefficient (MCC), Area under the ROC curve (AUC), and AUC20 from 10 simulation experiments is presented. Mean measure values are provided, with standard errors in parentheses.} 
\label{simulation_table} 
\end{table}

All simulation experiments are performed on a desktop running the Windows 10 Enterprise operating system desktop with 32 GB RAM and the following processor characteristics:  12th Gen Intel(R) Core(TM) i7-12700, 2100 Mhz, 12 Cores(s), 20 Logical Processor(s). We observe that \texttt{mSSL\_dpe} and \texttt{remMap.CV} did not terminate or produce results within 48 hours, possibly due to the high number of parameters considered  (P*Q = $249 \times 287 = 71463$) in the simulation settings. We report only the results corresponding to the methods mSSL and remMap, using the functions \texttt{mSSL\_dcpe} (from the R package \texttt{mSSL})  and \texttt{remMap.BIC} (from the R package \texttt{remMap}). The calculated TPR, FPR, MCC, AUC, AUC20, and sparsity proportion of the estimated coefficient matrices under different methods are provided in Table \ref{simulation_table}. The standard errors for all considered measures are noted in parentheses. We observe that B-MASTER outperforms SSLASSO, mSSL, and RemMap by a large margin. It is noted that, except for B-MASTER, the other methods overly shrink the coefficients, resulting in a very low TPR. In terms of AUC, SSLASSO performed the best among the other methods. We would like to point out to the readers that, since the original functions for mSSL and RemMap did not converge within 48 hours, we were only able to use their approximate, faster alternative versions, which are susceptible to the over-shrinking issue, as mentioned by the corresponding authors.

\begin{figure}[h!]
\centering
\includegraphics[width=.95\linewidth]{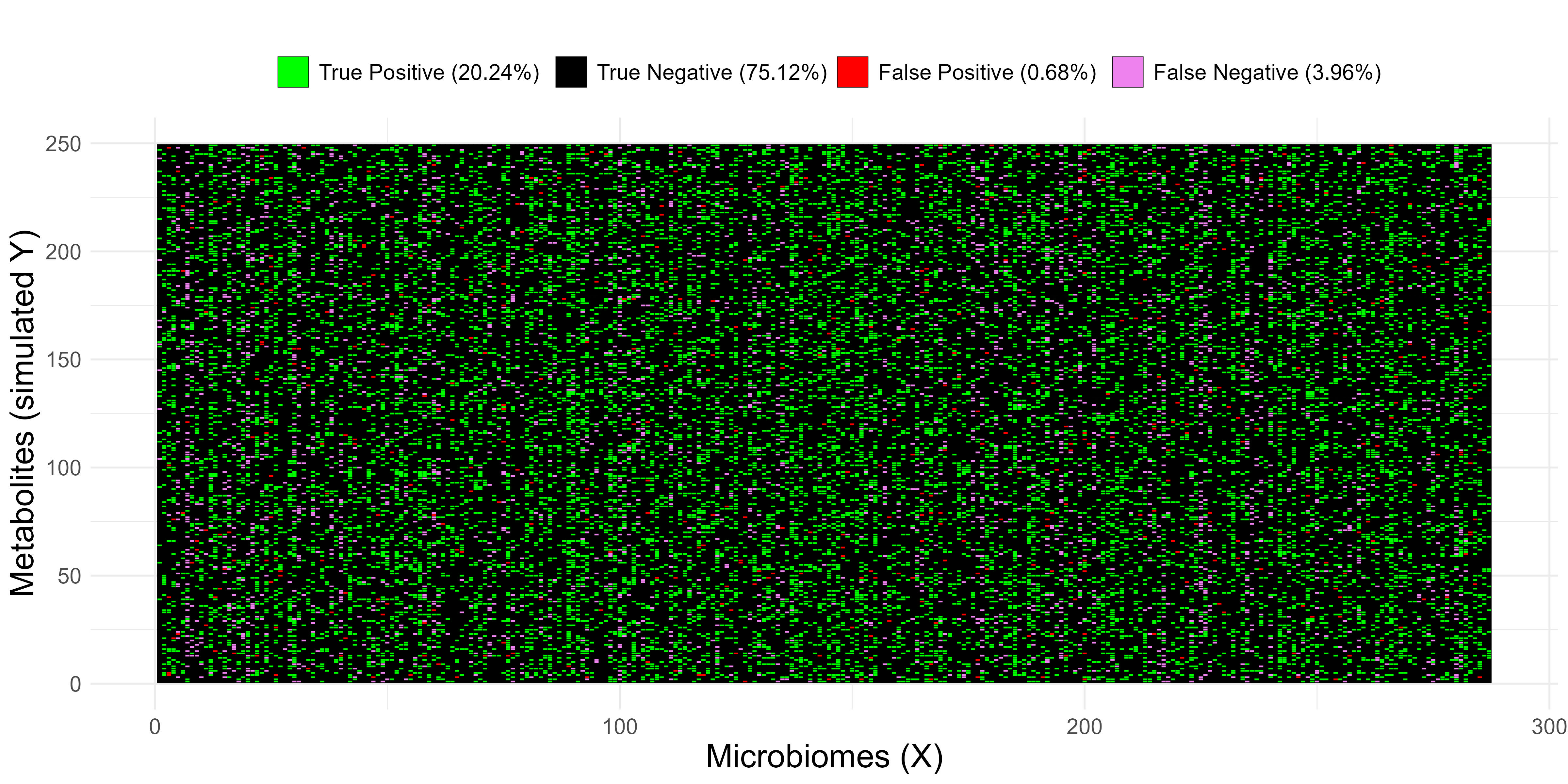} \\
\includegraphics[width=.45\linewidth]{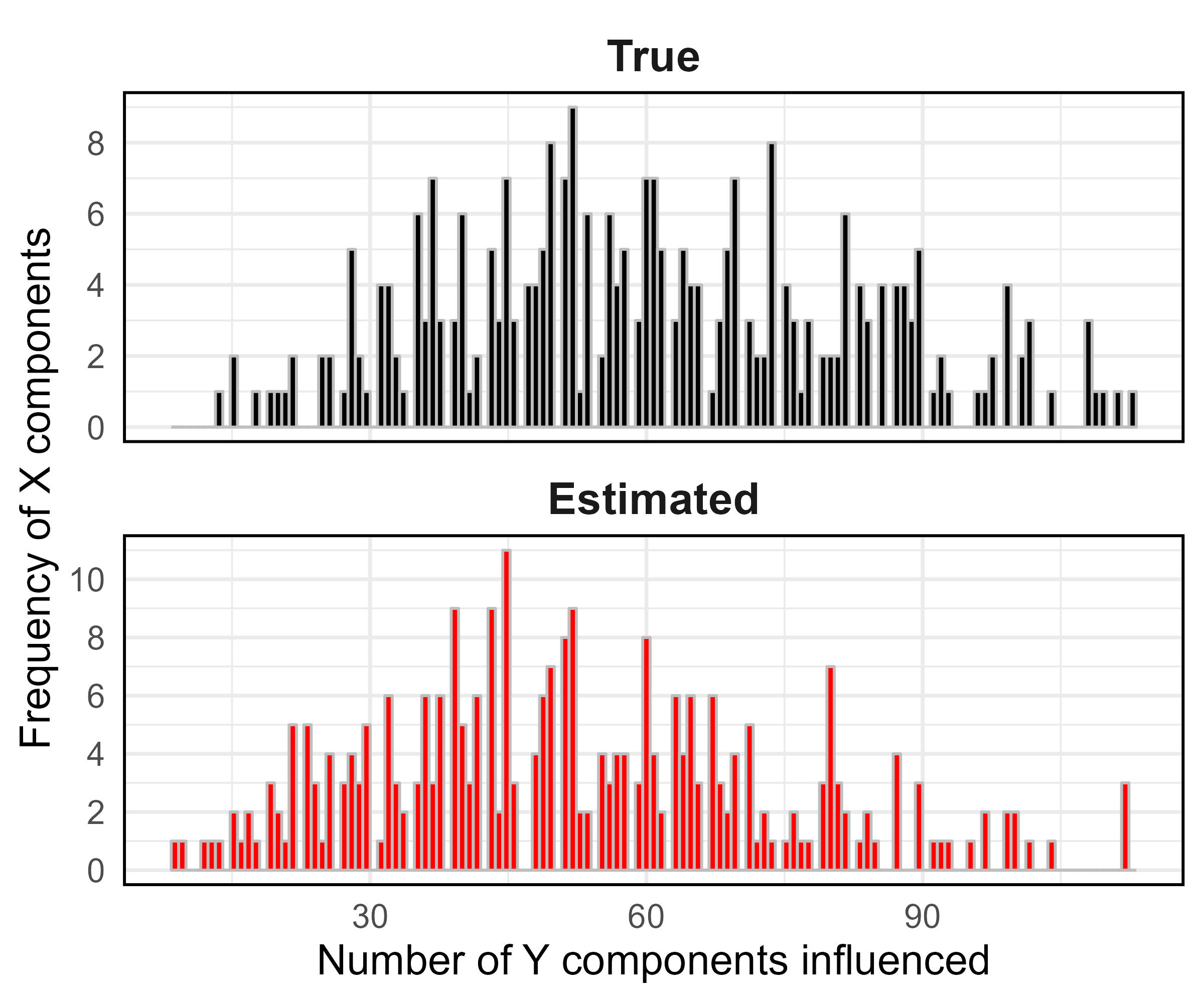} 
\hspace{0.04\linewidth}
\includegraphics[width=.45\linewidth]{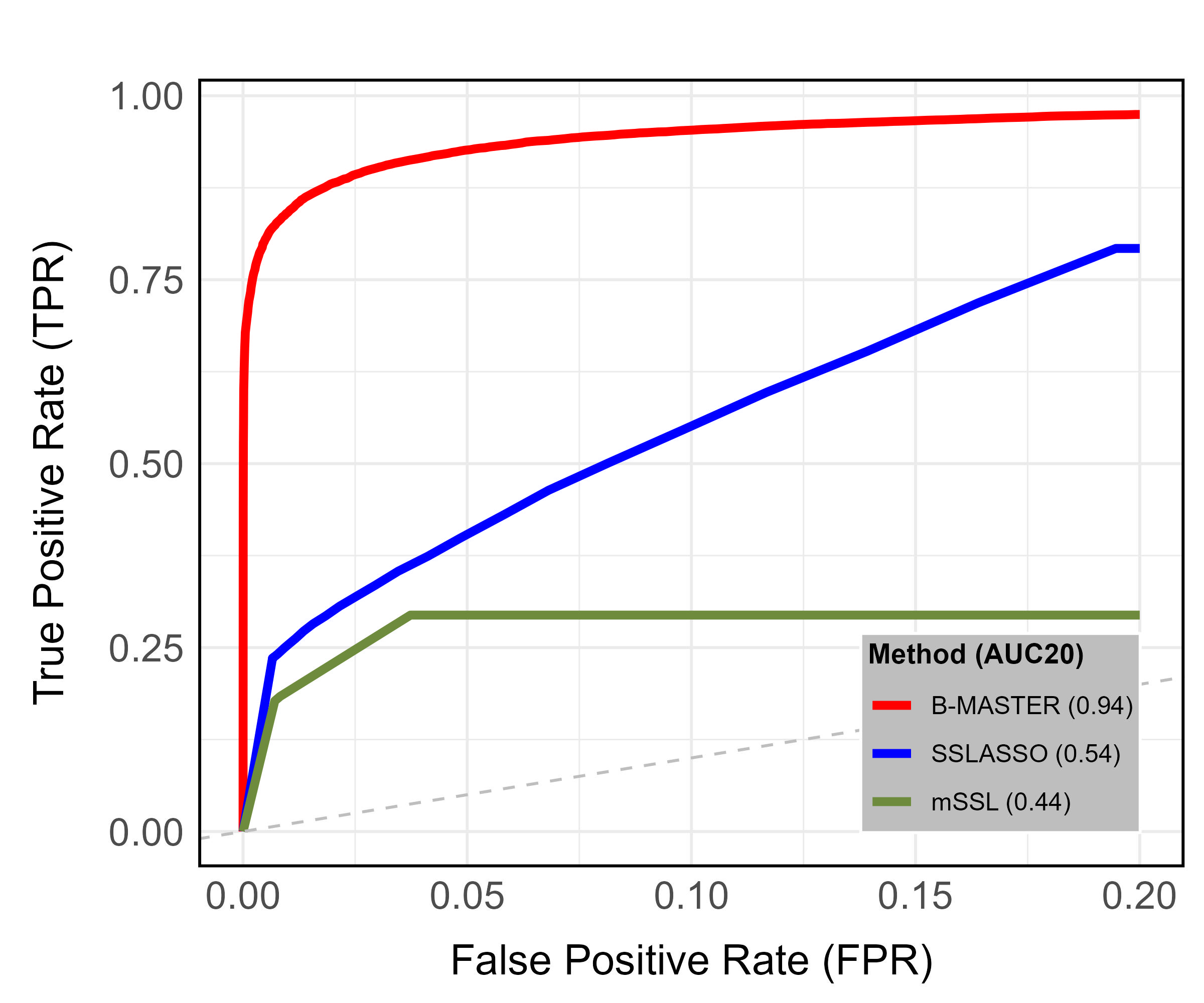} 
\caption{\textit{Up:} Signal detection performance of B-MASTER is evaluated based on Simulation Study 1. True and B-MASTER estimated coefficient signals are depicted, divided into categories: true positives (TP), true negatives (TN), false positives (FP), and false negatives (FN). \textit{Bottom-left:} True and B-MASTER estimated numbers of $Y$ components influenced by $X$ components are depicted. \textit{Bottom-right:} The AUC curve is plotted up to a false positive rate of $20\%$ for the B-MASTER, SSLASSO, and mSSL methods. AUC20 values are provided in parentheses within the legend.}
\label{BMASTER_sim}
\end{figure}

In Figure \ref{BMASTER_sim} (\textit{up}), the true and B-MASTER estimated coefficient signals are depicted, divided into four categories: TP, TN, FP, and FN. The number of $Y$ components influenced by $X$ components under both true and B-MASTER estimated scenarios is shown as a histogram in Figure \ref{BMASTER_sim} (\textit{bottom-left}). It appears that B-MASTER is slightly conservative in selecting variables, which creates a slight leftward shift in the histograms under the B-MASTER estimated scenario. Figure \ref{BMASTER_sim} (\textit{bottom-right}) shows that the AUC curve of B-MASTER dominates over SSLASSO and mSSL across the FPR range from 0\% to 20\%, by a large margin. We chose to display the AUC curve up to the 20\% FPR zone because the grid points of FPR and TPR obtained from the SSLASSO and mSSL model fittings (with grids taken as default options) were very sparse for FPR values greater than 20\%. We refrain from plotting the AUC curve for remMap, as Table \ref{simulation_table} already indicates its under-performance in the considered experiment.

\subsection{Evaluation of B-MASTER in Higher Dimensions}
\label{sim_study_2}
\begin{figure}[h]
\centering
\includegraphics[width=.45\linewidth]{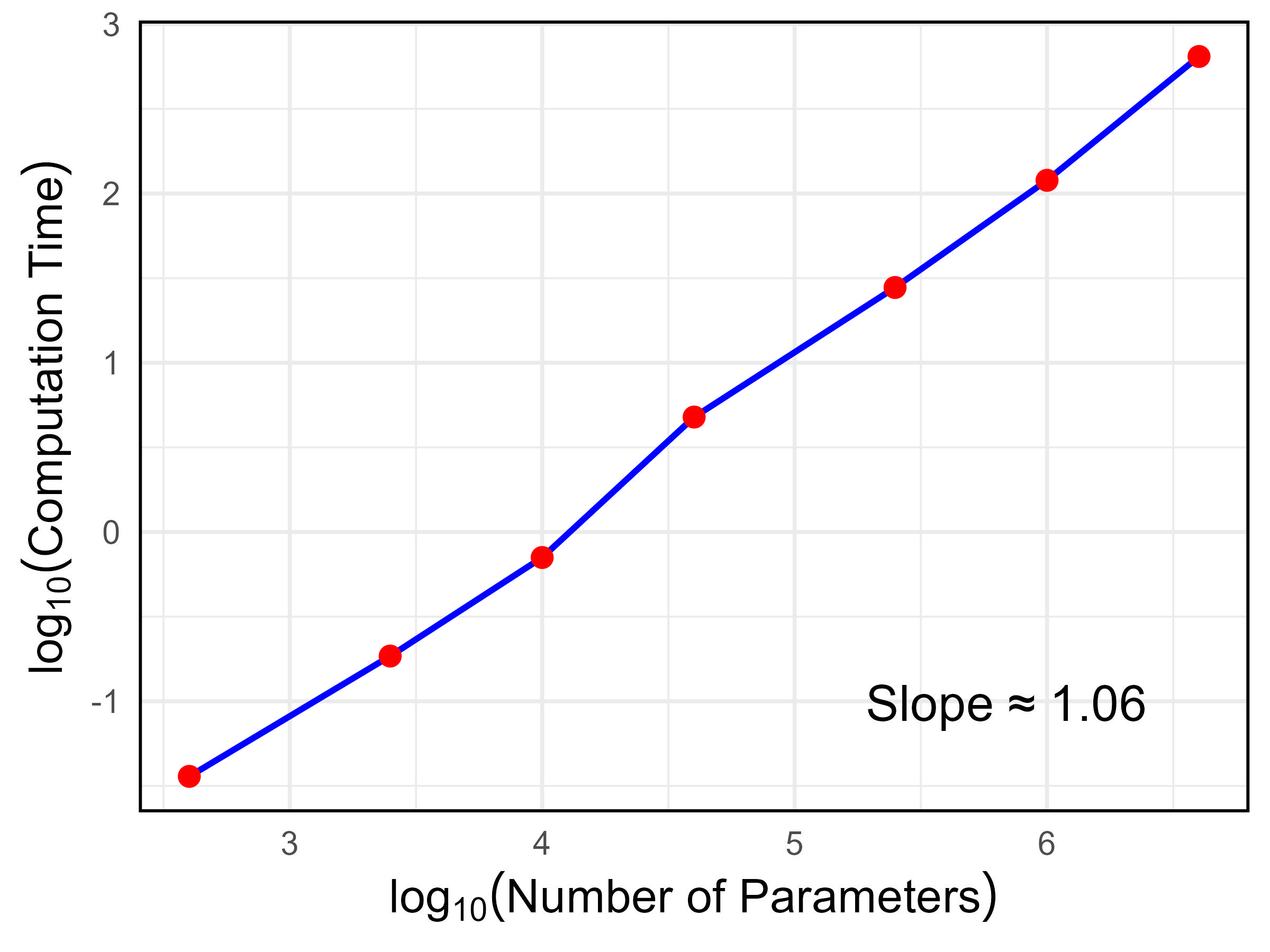} 
\caption{The plot shows $log_{10}(\text{Number of Parameters})$ versus $log_{10}(\text{Computation Time})$ for B-MASTER applied across scenarios with $P =Q=N= (20,50,100,200,500,1000,2000)$ and $\rho = 0.5$. Computation time is measured in minutes. }
\label{BMASTER_comptime}
\end{figure}

To evaluate the performance and required computation times in higher-dimensional settings for B-MASTER, we conduct another simulation study. We generate $N$ realizations of $X$ from $MVN(\bm{0},\Sigma)$, where $\Sigma = \big(\rho_{i,j} \big)_{P \times P}$ with $\rho_{i,j} = \rho^{|i-j|}$. We then generate $\bm{B}_0$ (a matrix of dimension $P \times Q$), where each element of $\bm{B}_0$ is sampled from $U\big([-5,-1] \cup [1,5])$. Hereafter, we generate an indicator matrix to impose sparse structure over the coefficient matrix adhering to the following presumptions:
\vspace{-0.3cm}
\begin{itemize}
    \item 20\% of the $X$ components influence all $Y$ components.
    \vspace{-0.3cm}
    \item Another 20\% of the $X$ components influence $100s\%$ of the $Y$ components selected at random, where $s \sim U(.4,.6)$.
    \vspace{-0.3cm}
    \item The remaining 60\% $X$ components are assumed to have no effect on $Y$.
\end{itemize}
 We construct an indicator matrix $I_0$ of dimension $P \times Q$, where 1s and 0s are assigned based on the aforementioned assumptions.  Finally, we compute $\bm{B} = \bm{B}_0 \circ I_0$, where $\circ$ denotes element-wise product. Then, we generate 
 \vspace{-0.5cm}
\begin{align*}
    Y^{i}_q = X^{i}B_q + \epsilon_{iq}
\end{align*}
for $i=1, \ldots, N$, where $B_q = (\beta_{pq})_{p = 1}^P$ denotes the $q$-th column of $\bm{B}$ and $\epsilon_{iq} \sim N(0,\sigma^2)$; we take $\sigma = 0.1$. For all considered scenarios, we set $P = Q = N$ inspired by the fact that in the real-dataset, the values of $P,Q,N$ are similar ($P = 287, Q = 249, N = 220)$. The performance of B-MASTER is evaluated for $P = 20, 50, 100, 200, 500, 1000, 2000$. All these sub-scenarios considered taking $\rho$ to be 0 (independent components) and  0.5 (highly correlated components), separately. The simulation results are provided in Table \ref{comp_time}. For each case, along with TPR, FPR, MCC, AUC, AUC20, estimated sparsity, and true sparsity (in parentheses), we also report the computation times. Although the results are based on 100 iterations with 20 burn-in, the projected time to complete 1000 iterations is also provided. Since each iteration takes a similar amount of time for a given dimensional setting, the projected time to complete 1000 iterations is approximately 10 times the time spent performing 100 iterations. Note that, for the experiments considered, we did not observe any further noticeable improvement when using 1000 iterations instead of 100. It was observed that for both cases, across all considered dimensions, B-MASTER performed well, yielding near-perfect TPR, FPR, AUC, MCC, and AUC20. The estimated sparsity was also very close to the true sparsity, showing a slight trend toward being conservative, which resulted in the estimated $\bm{B}$ being marginally sparser than the true model. The B-MASTER results for the independent 
$X$ scenario are slightly better than those for the correlated case. Furthermore, as the dimension of the parameter space increases, the computation time increases linearly, making it highly scalable for higher-dimensional models. This scalability enables us to evaluate B-MASTER performance in a model with up to 4 million coefficients within a reasonable amount of time. In Figure \ref{BMASTER_comptime}, we demonstrate the log-log plot (with base 10) of the number of parameters vs. computation time (in minutes) for B-MASTER in the $\rho=0.5$ scenario, revealing a linear dependence relationship. The approximate value of the slope is noted to be 1.06. The plot for the $\rho=0$ scenario is nearly identical, and therefore, we refrain from plotting it.

\begin{sidewaystable}[]
\resizebox{0.99\columnwidth}{!}{%
\begin{tabular}{ccccccccccccc}
\hline
$\rho$ & $P,Q,N$ & TPR & FPR & MCC & AUC & AUC20 & \begin{tabular}[c]{@{}c@{}}Sparsity\\ (true sparsity)\end{tabular} & \begin{tabular}[c]{@{}c@{}}Number of\\ parameters\end{tabular} & \begin{tabular}[c]{@{}c@{}}Number of\\ parameter\\ folds\end{tabular} & \begin{tabular}[c]{@{}c@{}}Comp. time\\ for 100 iters.\\ (in mins)\end{tabular} & \begin{tabular}[c]{@{}c@{}}Comp.time\\ folds\end{tabular} & \begin{tabular}[c]{@{}c@{}}Comp. time\\ for 1000 iters.\\ (projected,\\ in mins)\end{tabular} \\ \hline
\multirow{7}{*}{$\rho = 0$} & 20 & 0.974 & 0.000 & 0.982 & 0.999 & 0.999 & 0.715 (0.708) & 400 & 1x & 0.04 & 1x & 0.37 \\
 & 50 & 0.991 & 0.000 & 0.993 & 1.000 & 1.000 & 0.708 (0.705) & 2500 & 6.25x & 0.19 & 5.24x & 1.94 \\
 & 100 & 0.978 & 0.012 & 0.964 & 0.997 & 0.986 & 0.699 (0.700) & 10000 & 25x & 0.74 & 19.97x & 7.39 \\
 & 200 & 0.972 & 0.026 & 0.937 & 0.995 & 0.980 & 0.692 (0.702) & 40000 & 100x & 4.68 & 126.38x & 46.76 \\
 & 500 & 0.996 & 0.029 & 0.951 & 0.998 & 0.990 & 0.681 (0.699) & 250000 & 625x & 26.68 & 721.03x & 266.78 \\
 & 1000 & 0.998 & 0.006 & 0.989 & 0.999 & 0.999 & 0.697 (0.700) & 1000000 & 2500x & 108.34 & 2928.19x & 1083.43 \\
 & 2000 & 0.998 & 0.000 & 0.998 & 1.000 & 1.000 & 0.702 (0.701) & 4000000 & 10000x & 603.65 & 16314.87x & 6036.50 \\ \hline
\multirow{7}{*}{$\rho = 0.5$} & 20 & 0.966 & 0.000 & 0.976 & 1.000 & 1.000 & 0.718 (0.708) & 400 & 1x & 0.04 & 1x & 0.36 \\
 & 50 & 0.999 & 0.002 & 0.996 & 1.000 & 1.000 & 0.704 (0.705) & 2500 & 6.25x & 0.19 & 5.14x & 1.85 \\
 & 100 & 0.982 & 0.025 & 0.947 & 0.994 & 0.970 & 0.689 (0.700) & 10000 & 25x & 0.71 & 19.64x & 7.07 \\
 & 200 & 0.963 & 0.033 & 0.921 & 0.992 & 0.969 & 0.690 (0.702) & 40000 & 100x & 4.78 & 132.72x & 47.78 \\
 & 500 & 0.990 & 0.021 & 0.959 & 0.998 & 0.991 & 0.688 (0.699) & 250000 & 625x & 27.84 & 773.39x & 278.42 \\
 & 1000 & 0.984 & 0.001 & 0.988 & 1.000 & 1.000 & 0.705 (0.700) & 1000000 & 2500x & 119.65 & 3323.53x & 1196.47 \\
 & 2000 & 0.947 & 0.000 & 0.963 & 1.000 & 0.999 & 0.717 (0.701) & 4000000 & 10000x & 644.97 & 17915.75x & 6449.67 \\ \hline
\end{tabular}}
\caption{Performance evaluation of B-MASTER: The values of $P$ ($P = Q = N$), TPR, FPR, MCC, AUC, AUC20, sparsity (with true sparsity in parentheses), the number of parameters, relative folds of the number of parameters, computation time for 100 iterations (in minutes), relative folds of computation times, and the projected computation time for 1000 iterations (in minutes) are provided taking $P=Q =N= 20, 50, 100, 200, 500, 1000, 2000$ under scenarios $\rho = 0, 0.5$.  } 
\label{comp_time} 
\end{sidewaystable}


\section{Convergence Diagnostics and Hyperparameter Sensitivity}
\subsection{B-MASTER Convergence Diagnostics}
\begin{figure}[htbp]
    \centering
    \includegraphics[width=0.9\textwidth]{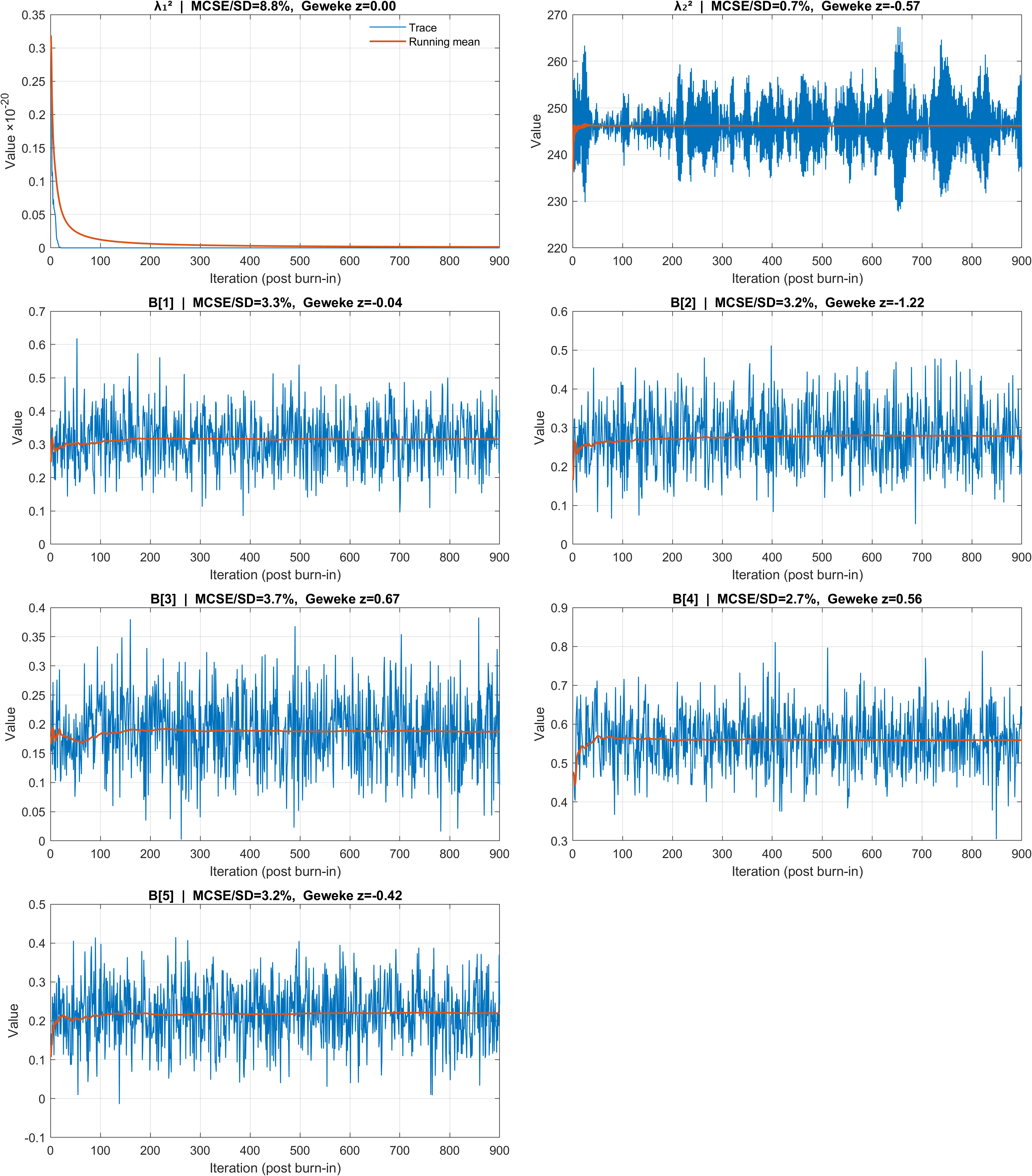}
    \caption{Trace plots and running means for $\lambda_1^2$, $\lambda_2^2$, and five representative $B$ entries ($B[1]$--$B[5]$), chosen as the five largest absolute values of $B$. 
    For each parameter, we report the Monte Carlo standard error relative to posterior variability (MCSE/SD\%) and Geweke’s stationarity $z$-score. 
    Low MCSE/SD values ($<10\%$) and Geweke statistics within $|z|\leq 2$ provide evidence of satisfactory convergence.}
    \label{fig:mcmc_traces}
\end{figure}

To evaluate MCMC convergence, we monitored Monte Carlo error and stationarity diagnostics for the two global shrinkage parameters ($\lambda_1^2$, $\lambda_2^2$) and five representative $B$ coefficients corresponding to the largest absolute entries of $B$. Figure~\ref{fig:mcmc_traces} displays trace plots with running mean overlays for these seven quantities. For each series, we report the ratio of Monte Carlo standard error to posterior standard deviation (MCSE/SD\%) as a measure of numerical accuracy and Geweke’s diagnostic $z$-statistic, which compares the first 10\% of the chain to the last 50\%. Table~\ref{tab:mcmc_mcse_geweke} summarizes these diagnostics. Across $\lambda_2^2$ and the $B$ entries, MCSE/SD ratios were consistently below 5\% and all Geweke statistics lay within $|z|\leq 2$, supporting satisfactory mixing and stationarity. The behavior of $\lambda_1^2$ was distinct: this parameter rapidly decayed to values near machine precision, yielding an MCSE/SD ratio slightly above 5\% but with a Geweke statistic of essentially zero. This reflects strong shrinkage toward zero rather than poor mixing, and is consistent with the posterior concentrating on negligible contribution from $\lambda_1^2$. This convergence analysis was conducted for the real data--based simulation study considered in Section~4.1 (main draft), using 1000 total iterations and discarding the first 100 iterations as burn-in.

\begin{table}[h]
\centering
\resizebox{0.5\columnwidth}{!}{%
\begin{tabular}{ccc}
\hline
Quantity & MCSE/SD (\%) & Geweke $z$ \\ \hline
$\lambda_1^2$   & 8.77  & $<0.01$ \\
$\lambda_2^2$   & 0.65  & -0.57 \\
$B[1]$          & 3.30  & -0.04 \\
$B[2]$          & 3.17  & -1.22 \\
$B[3]$          & 3.66  & 0.67 \\
$B[4]$          & 2.68  & 0.56 \\
$B[5]$          & 3.24  & -0.42 \\ \hline
\end{tabular}}
\caption{MCMC diagnostics (MCSE/SD\% and Geweke $z$) for $\lambda_1^2$, $\lambda_2^2$, and five representative $B$ entries (corresponding to the top five absolute values of $B$).}
\label{tab:mcmc_mcse_geweke}
\end{table}

\subsection{Sensitivity Analysis for Hyper--Priors}
\label{sec:sensitivity}
The hyper--prior specification for the global shrinkage parameters
$\lambda_1^2$ and $\lambda_2^2$ involves four constants
$r_1,\ r_2,\ \delta_1,\ \delta_2$. In our implementation, we set
$r_1=r_2=1$ and examine sensitivity primarily with respect to
$(\delta_1,\delta_2)$, which directly control the effective shrinkage
strength through the Gamma rate parameters. Specifically, the conditional
posterior distributions are given by
\[
\lambda_1^2 \mid \bm{T}^2
\sim
\mathrm{Gamma}
\left(
PQ+r_1,\;
\frac{1}{2}\sum_{p=1}^P\sum_{q=1}^Q \tau_{pq}^2+\delta_1
\right),
\]
and
\[
\lambda_2^2 \mid \bm{G}^2
\sim
\mathrm{Gamma}
\left(
\frac{PQ}{2}+r_2,\;
\frac{1}{2}\sum_{p=1}^P \gamma_p^2+\delta_2
\right).
\]

Because the shape parameters scale with the total coefficient dimension
$PQ$, the additive constants $r_1=r_2=1$ contribute negligibly in the
high--dimensional settings considered in this paper. Consequently, the
posterior behavior is driven primarily by the latent shrinkage quantities
$\{\tau_{pq}^2\}$ and $\{\gamma_p^2\}$ together with the rate parameters
$(\delta_1,\delta_2)$. Empirically, varying $r_1$ and $r_2$ over small
positive values produced nearly identical posterior summaries and
selection results. Therefore, throughout the manuscript we fix
$r_1=r_2=1$ and focus sensitivity analyses on $(\delta_1,\delta_2)$.

In contrast, the Gamma rates,
\[
b_1=\tfrac{1}{2}\sum_{p,q}\tau_{pq}^2+\delta_1, \qquad
b_2=\tfrac{1}{2}\sum_{p=1}^P \gamma_p^2+\delta_2 ,
\]
directly control the global shrinkage through $\lambda_1^2$ and
$\lambda_2^2$.  We therefore assess sensitivity to $(\delta_1,\delta_2)$
using the real data-based simulation of Section~4.1 (main draft): for each setting we
run $1{,}000$ iterations, discard the first $100$ as burn--in, and
summarize performance over $10$ replicates using
TPR, FPR, MCC, AUC, AUC20, and sparsity. We consider the default
$(\delta_1,\delta_2)=(0.1,0.1)$ together with four alternatives
$(0.01,0.01)$, $(0.01,1)$, $(1,0.01)$, and $(1,1)$.  The results are
reported in Table~\ref{tab:sensitivity}, in the same format as Table~1
of the main text (means with standard errors in parentheses).  Across all $(\delta_1,\delta_2)$ pairs, TPR, FPR, MCC, AUC, AUC20, and
sparsity remain essentially unchanged within Monte Carlo uncertainty,
with the default $(0.1,0.1)$ performing on par with the alternatives.
This indicates that posterior inference and variable--selection behavior
of \textsc{B-MASTER} are robust to moderate perturbations of the global--shrinkage
hyper--rates $\delta_1$ and $\delta_2$, while the shape hyperparameters
$r_1=r_2=1$ are effectively non--informative in the regimes studied.

\begin{table}[]
\centering
\resizebox{0.9\columnwidth}{!}{%
\begin{tabular}{ccccccc}
\hline
$(\delta_1,\delta_2)$ & TPR & FPR & MCC & AUC & AUC20 & \begin{tabular}[c]{@{}c@{}}Sparsity\\ (True = 0.76)\end{tabular} \\ \hline
Default (0.1, 0.1)  & 0.84 (0.0012) & 0.01 (0.0003) & 0.87 (0.0001) & 0.98 (0.0004) & 0.94 (0.0014) & 0.79 (0.0003) \\
(0.01, 0.01) & 0.84 (0.0003) & 0.01 (0.0001) & 0.87 (0.0002) & 0.98 (0.0001) & 0.94 (0.0002) & 0.79 (0.0001) \\
(0.01, 1)    & 0.84 (0.0003) & 0.01 (0.0001) & 0.87 (0.0003) & 0.98 (0.0002) & 0.94 (0.0004) & 0.79 (0.0001) \\
(1, 0.01)    & 0.84 (0.0004) & 0.01 (0.0000) & 0.87 (0.0003) & 0.98 (0.0001) & 0.94 (0.0002) & 0.79 (0.0001) \\
(1, 1)       & 0.84 (0.0005) & 0.01 (0.0001) & 0.87 (0.0004) & 0.98 (0.0001) & 0.94 (0.0003) & 0.79 (0.0001) \\ \hline
\end{tabular}}
\caption{Sensitivity analysis of B-MASTER under different hyperparameter choices for $(\delta_1,\delta_2)$. Mean measure values are provided, with standard errors in parentheses, based on 10 simulation experiments.}
\label{tab:sensitivity}
\end{table}

\subsection{Sensitivity to MCMC Iteration Length}
To further assess whether the reported performance is sensitive to the total number of MCMC iterations, we repeated the real data--based simulation study under three sampling lengths: 500, 1000, and 2000 total iterations, each with the first 100 iterations discarded as burn-in. This analysis directly evaluates whether the default sampling regime used in the main simulation study is sufficient for stable variable-selection performance.

\begin{table}[!h]
\centering
\resizebox{0.6\columnwidth}{!}{%
\begin{tabular}{ccccccc}
\hline
\begin{tabular}[c]{@{}c@{}}Num.\\ MCMC\\ iters\end{tabular} & TPR & FPR & MCC & AUC & AUC20 & \begin{tabular}[c]{@{}c@{}}Sparsity\\ (True = 0.76)\end{tabular} \\ \hline
500 & 0.835 & 0.009 & 0.870 & 0.981 & 0.936 & 0.791 \\
1000 & 0.836 & 0.009 & 0.871 & 0.982 & 0.937 & 0.791 \\
2000 & 0.836 & 0.008 & 0.870 & 0.981 & 0.936 & 0.791 \\ \hline
\end{tabular}}
\caption{Sensitivity analysis for the total number of MCMC iterations in the B-MASTER simulation study. Results are reported for MCMC iteration lengths of 500, 1000, and 2000, each using a fixed burn-in of 100 iterations. The metrics include true positive rate (TPR), false positive rate (FPR), Matthews correlation coefficient (MCC), area under the ROC curve (AUC), partial AUC restricted to low false positive regions (AUC20), and estimated sparsity of the coefficient matrix. The true sparsity level in the simulated setting was 0.76.}
\label{tab:ITER_sensitivity_simulation}
\end{table}

\begin{figure}[!t]
    \centering
    \includegraphics[width=0.9\textwidth]{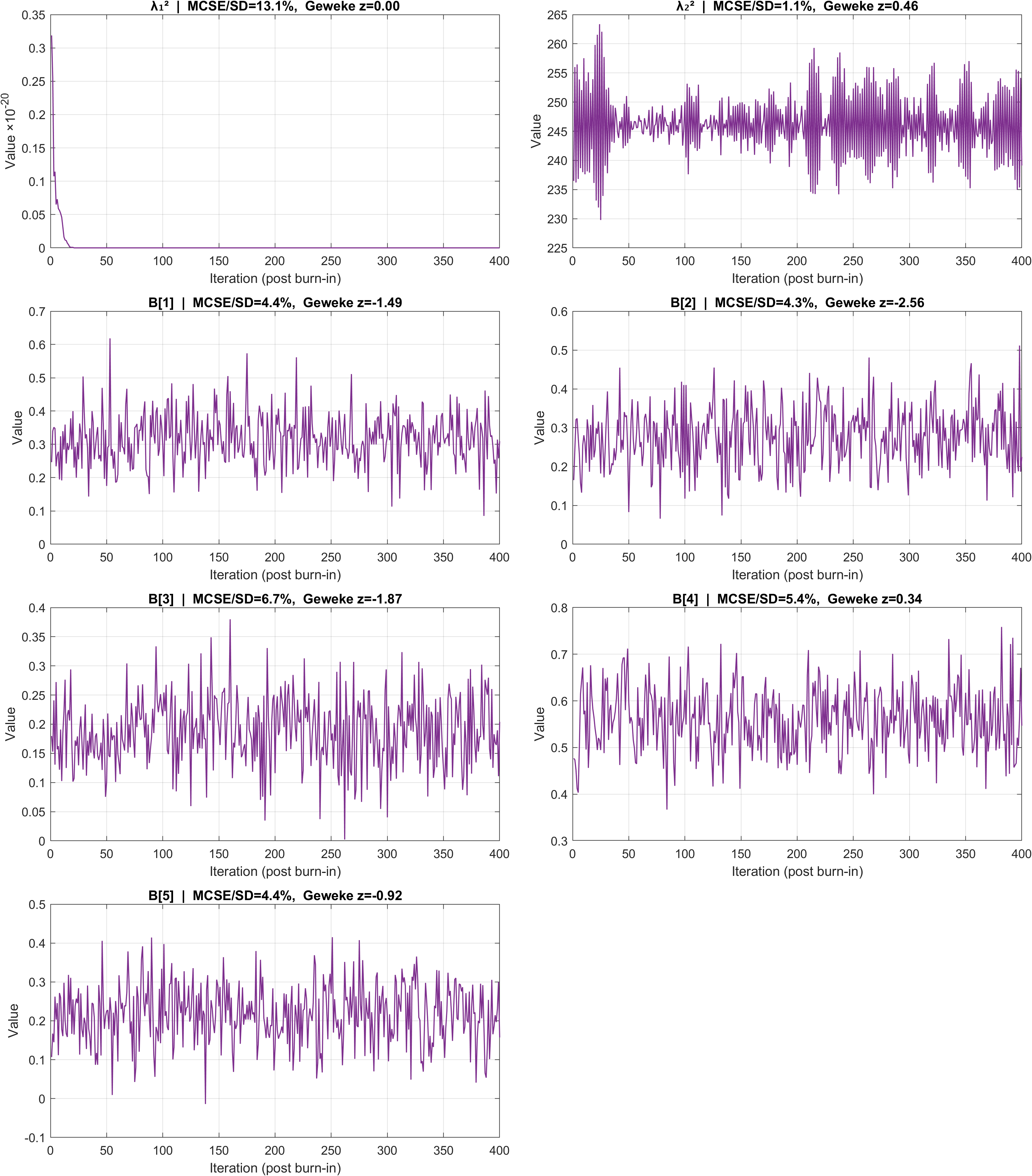}
    \caption{Trace plots from the B-MASTER real-data analysis using 500 total MCMC iterations with the first 100 iterations discarded as burn-in. Shown are the chains for $\lambda_1^2$, $\lambda_2^2$, and five representative regression coefficients ($B[1]$--$B[5]$), selected as the five largest absolute entries of the estimated coefficient matrix $B$. For each parameter, the Monte Carlo standard error relative to posterior variability (MCSE/SD\%) and Geweke stationarity diagnostic $z$-score are reported in the panel titles. Across most displayed parameters, the chains exhibit stable post--burn-in behavior, relatively low MCSE/SD values ($<10\%$), and Geweke statistics close to the commonly used range $|z|\leq 2$, indicating generally satisfactory mixing and convergence behavior for this shorter MCMC run.}
    \label{fig:ITER_sensitivity_mcmc_traces_iter_500}
\end{figure}

\begin{figure}[!t]
    \centering
    \includegraphics[width=0.9\textwidth]{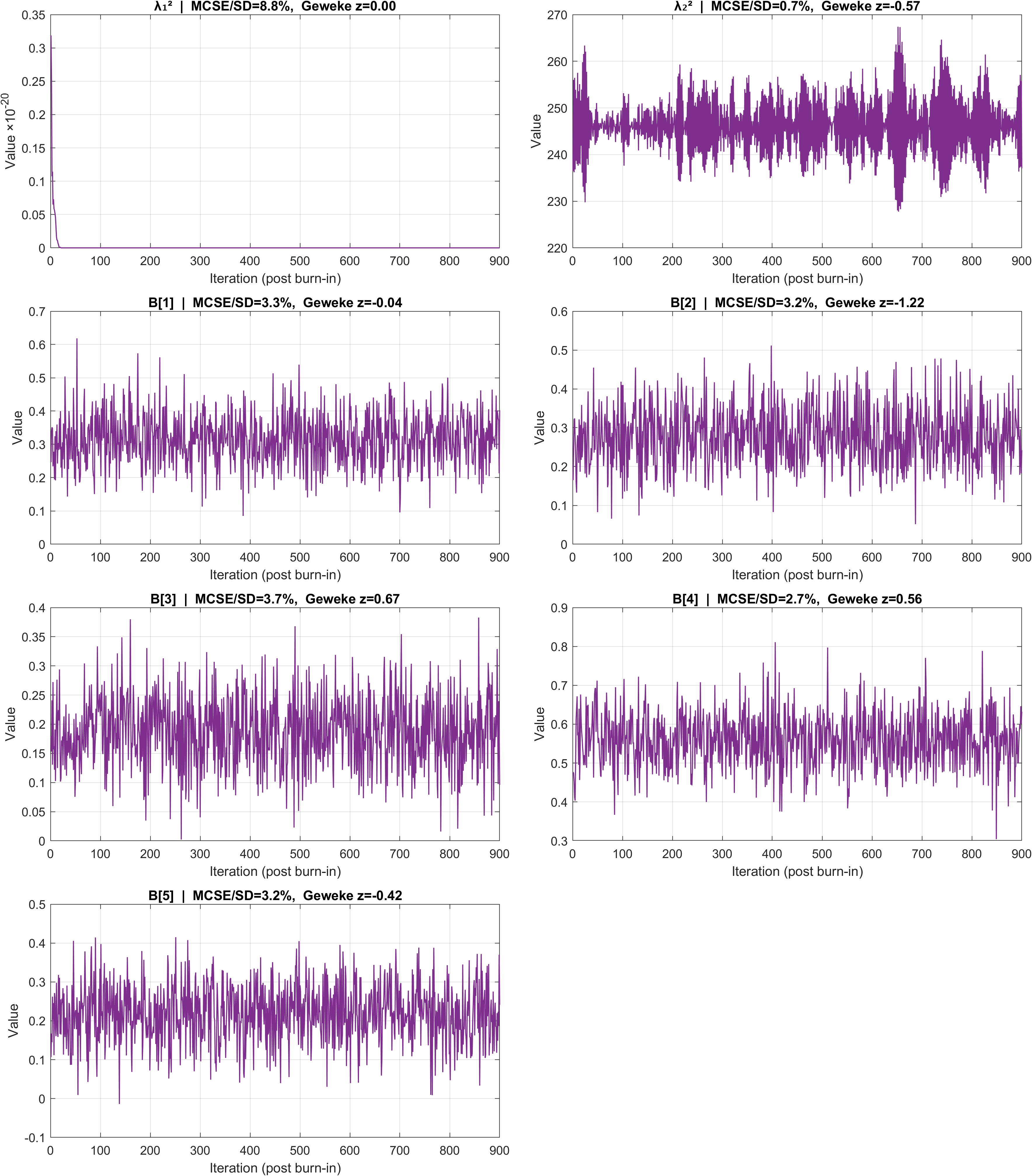}
    \caption{Trace plots from the B-MASTER real-data analysis using 1000 total MCMC iterations with the first 100 iterations discarded as burn-in. Shown are the chains for $\lambda_1^2$, $\lambda_2^2$, and five representative regression coefficients ($B[1]$--$B[5]$), selected as the five largest absolute entries of the estimated coefficient matrix $B$. For each parameter, the Monte Carlo standard error relative to posterior variability (MCSE/SD\%) and Geweke stationarity diagnostic $z$-score are reported in the panel titles. Across all displayed parameters, the chains exhibit stable post--burn-in behavior, low relative MCSE values, and Geweke statistics within the commonly used range $|z|\leq 2$, supporting satisfactory mixing and convergence of the Gibbs sampler.}
    \label{fig:ITER_sensitivity_mcmc_traces_iter_1000}
\end{figure}

\begin{figure}[!t]
    \centering
    \includegraphics[width=0.9\textwidth]{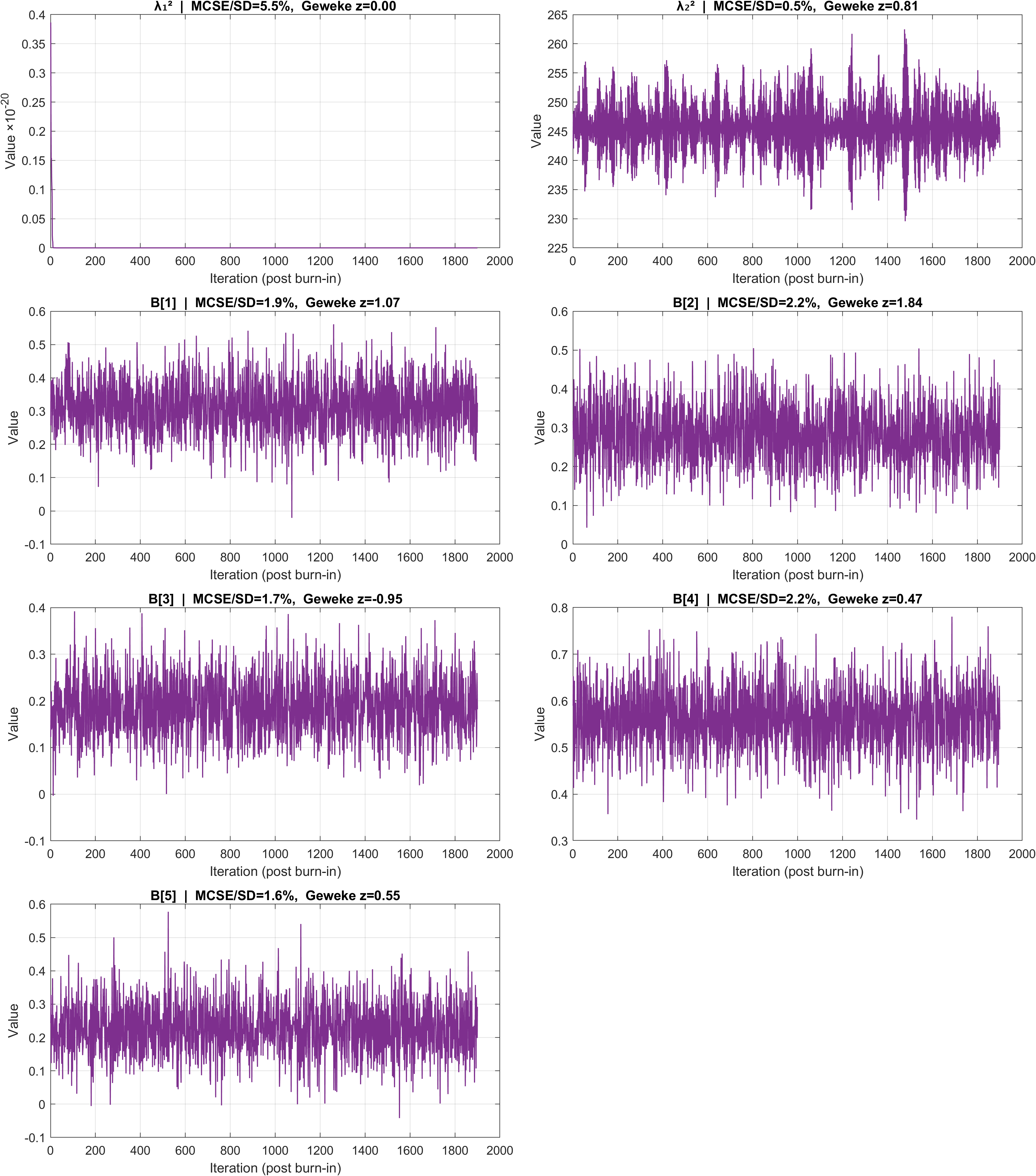}
    \caption{Trace plots from the B-MASTER real-data analysis using 2000 total MCMC iterations with the first 100 iterations discarded as burn-in. Shown are the chains for $\lambda_1^2$, $\lambda_2^2$, and five representative regression coefficients ($B[1]$--$B[5]$), selected as the five largest absolute entries of the estimated coefficient matrix $B$. For each parameter, the Monte Carlo standard error relative to posterior variability (MCSE/SD\%) and Geweke stationarity diagnostic $z$-score are reported in the panel titles. Across all displayed parameters, the chains exhibit stable post--burn-in behavior, low relative MCSE values ($<10\%$), and Geweke statistics within the commonly used range $|z|\leq 2$, supporting satisfactory mixing and convergence of the Gibbs sampler.}
    \label{fig:ITER_sensitivity_mcmc_traces_iter_2000}
\end{figure}

Figures~\ref{fig:ITER_sensitivity_mcmc_traces_iter_500}--\ref{fig:ITER_sensitivity_mcmc_traces_iter_2000} show trace plots for the two global shrinkage parameters, $\lambda_1^2$ and $\lambda_2^2$, and five representative regression coefficients corresponding to the largest absolute entries of the estimated coefficient matrix. Across the three iteration lengths, the displayed chains show stable post--burn-in behavior. The global shrinkage parameter $\lambda_1^2$ rapidly concentrates near zero, reflecting strong shrinkage rather than lack of convergence, while $\lambda_2^2$ and the representative regression coefficients fluctuate around stable posterior regions. The MCSE/SD ratios and Geweke diagnostics reported in the figure panels provide additional evidence that the posterior summaries are not driven by poor mixing.

Table~\ref{tab:ITER_sensitivity_simulation} summarizes the corresponding variable-selection performance. The performance metrics are nearly unchanged across 500, 1000, and 2000 iterations: TPR remains approximately 0.835--0.836, FPR remains approximately 0.008--0.009, MCC remains approximately 0.870--0.871, AUC remains approximately 0.981--0.982, AUC20 remains approximately 0.936--0.937, and the estimated sparsity remains 0.791 across all three settings. These results indicate that the posterior edge-selection results are stable with respect to the total number of MCMC iterations within the range considered. Thus, the computational advantages reported in the main text do not appear to be an artifact of an insufficient sampling length.

\FloatBarrier
\section{B-MASTER Case Study}
\subsection{Descriptive statistics and diagnostics}
\begin{figure}[]
\centering
\begin{subfigure}{0.45\textwidth}
  \includegraphics[width=\linewidth]{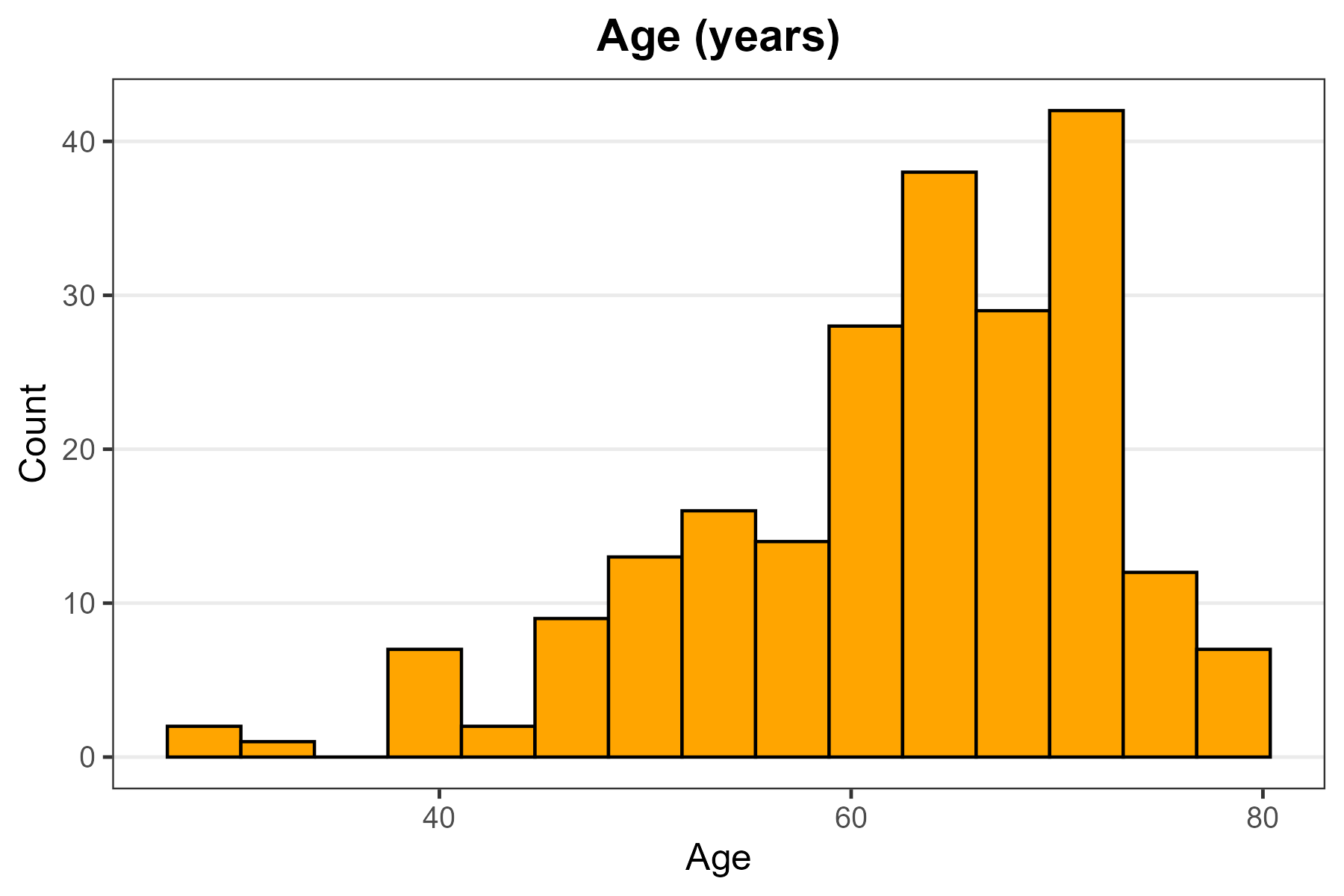}
  \caption{Age distribution}
\end{subfigure}
\begin{subfigure}{0.45\textwidth}
  \includegraphics[width=\linewidth]{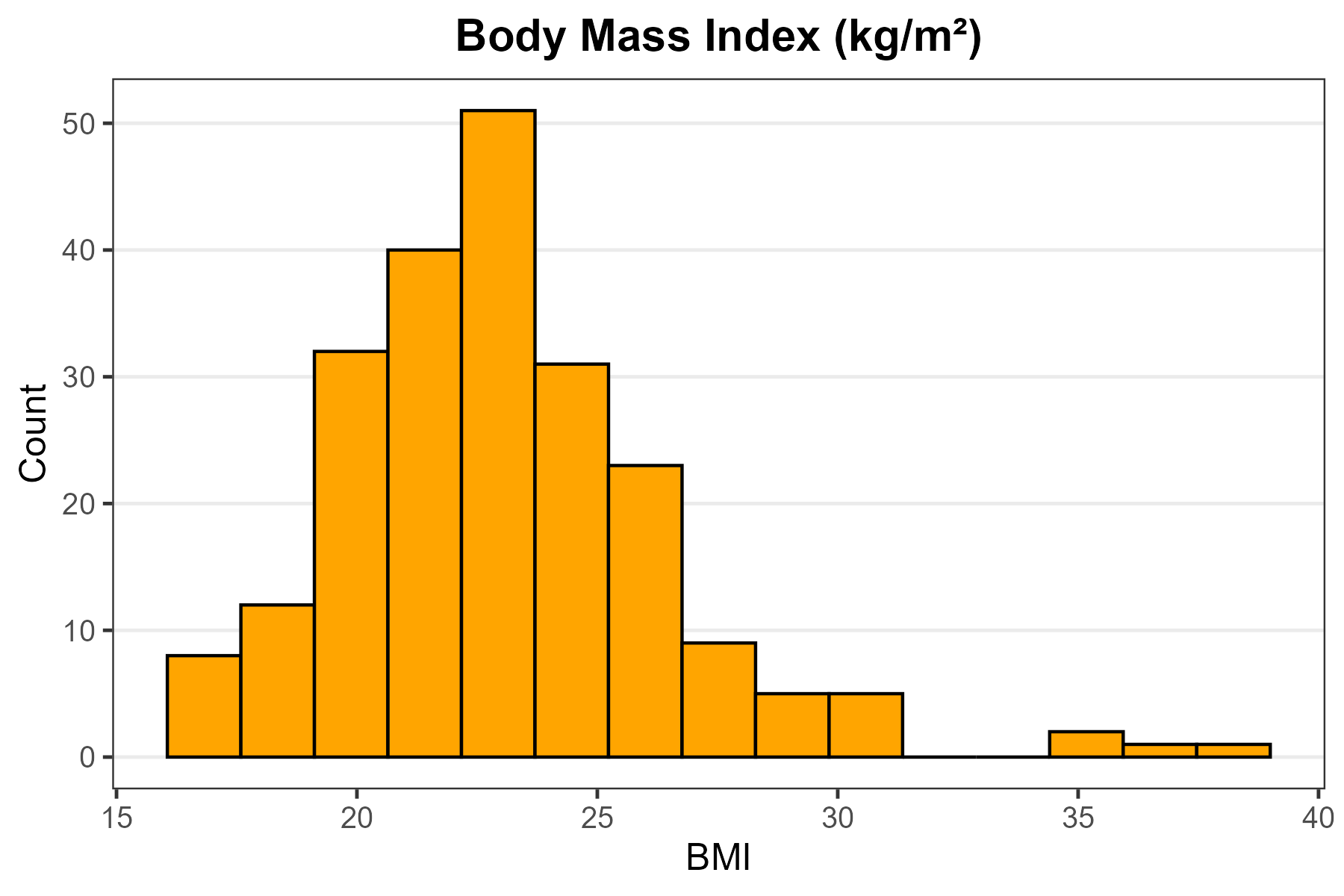}
  \caption{Body Mass Index (BMI)}
\end{subfigure}

\begin{subfigure}{0.45\textwidth}
  \includegraphics[width=\linewidth]{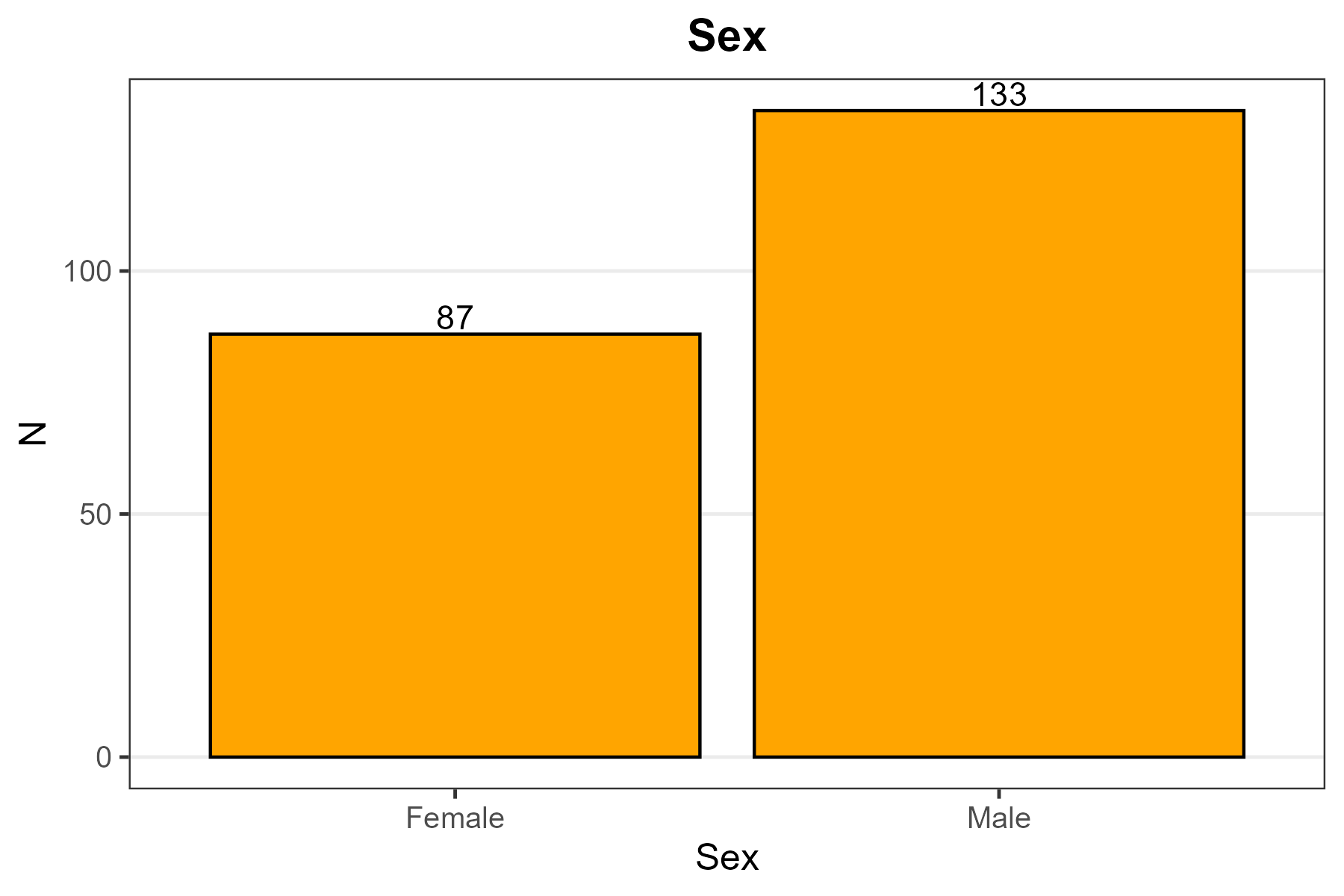}
  \caption{Sex distribution}
\end{subfigure}
\begin{subfigure}{0.45\textwidth}
  \includegraphics[width=\linewidth]{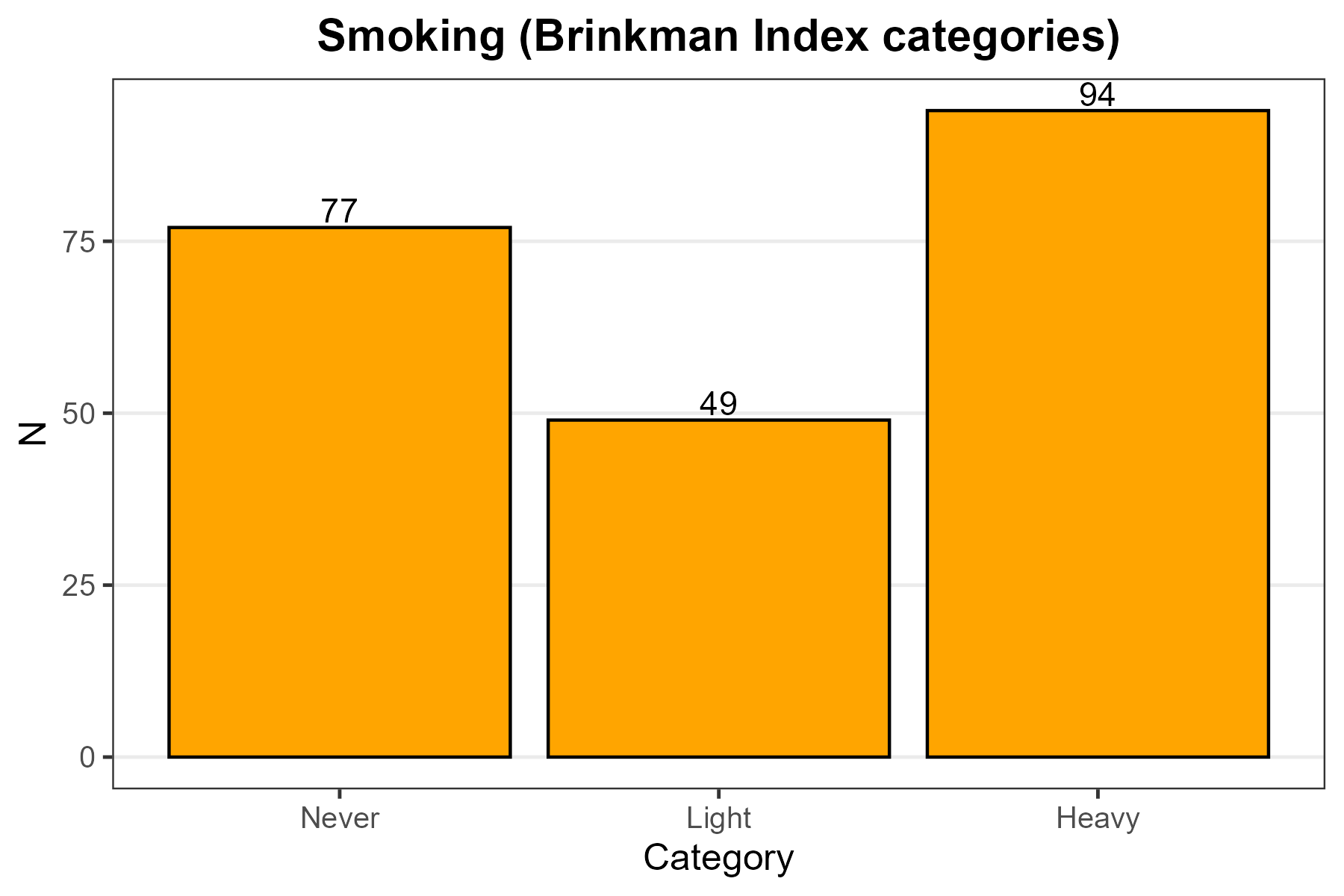}
  \caption{Smoking categories}
\end{subfigure}

\begin{subfigure}{0.45\textwidth}
  \includegraphics[width=\linewidth]{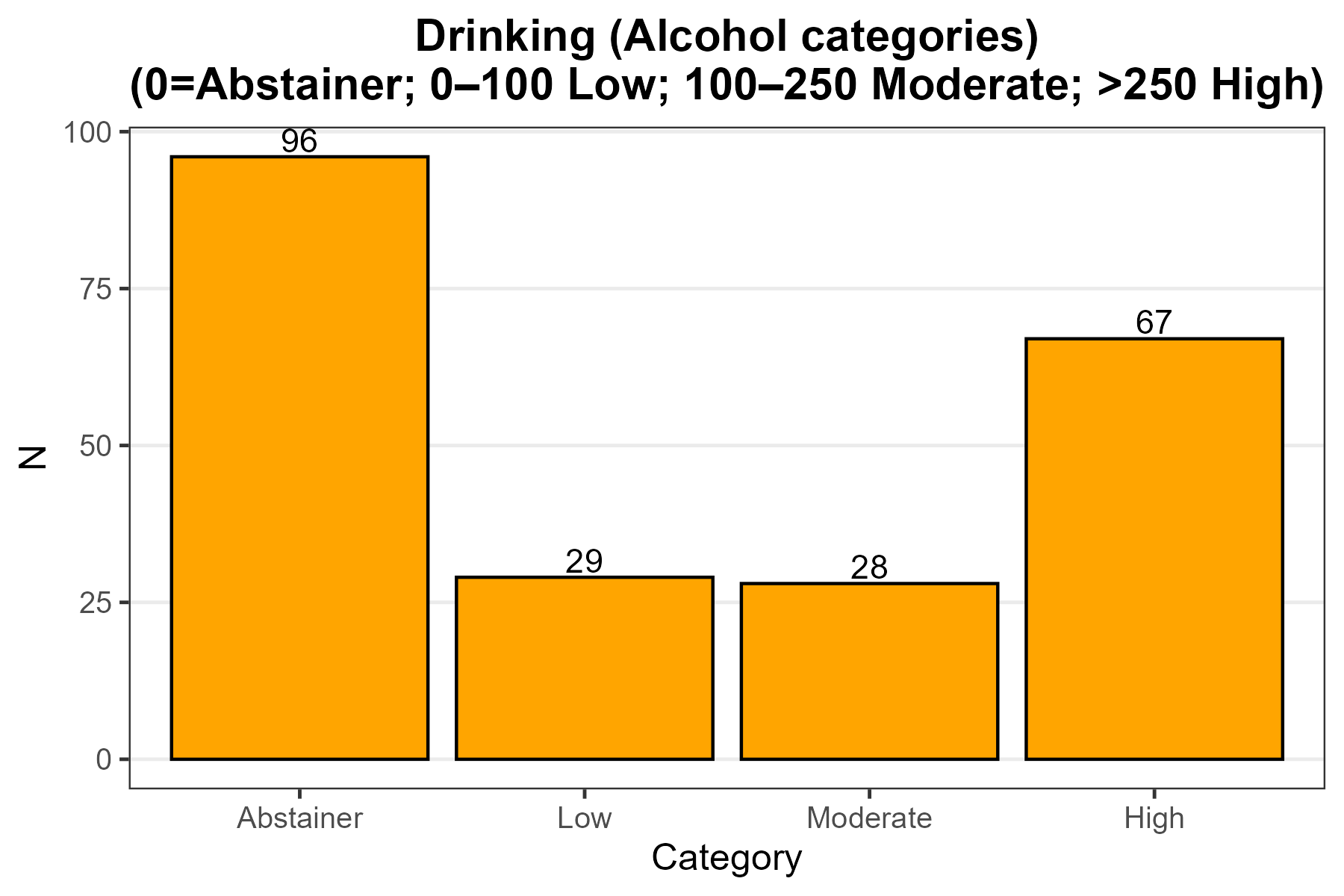}
  \caption{Alcohol consumption categories}
\end{subfigure}
\begin{subfigure}{0.45\textwidth}
\end{subfigure}

\caption{Supplementary descriptive plots of patient characteristics in the case study cohort. Distributions are shown for (a) age, (b) BMI, (c) sex, (d) smoking, and (e) drinking.}
\label{fig:supp_patient_characteristics}
\end{figure}

\begin{figure}[]
    \centering
    \begin{subfigure}[t]{0.48\textwidth}
        \centering
        \includegraphics[width=\linewidth]{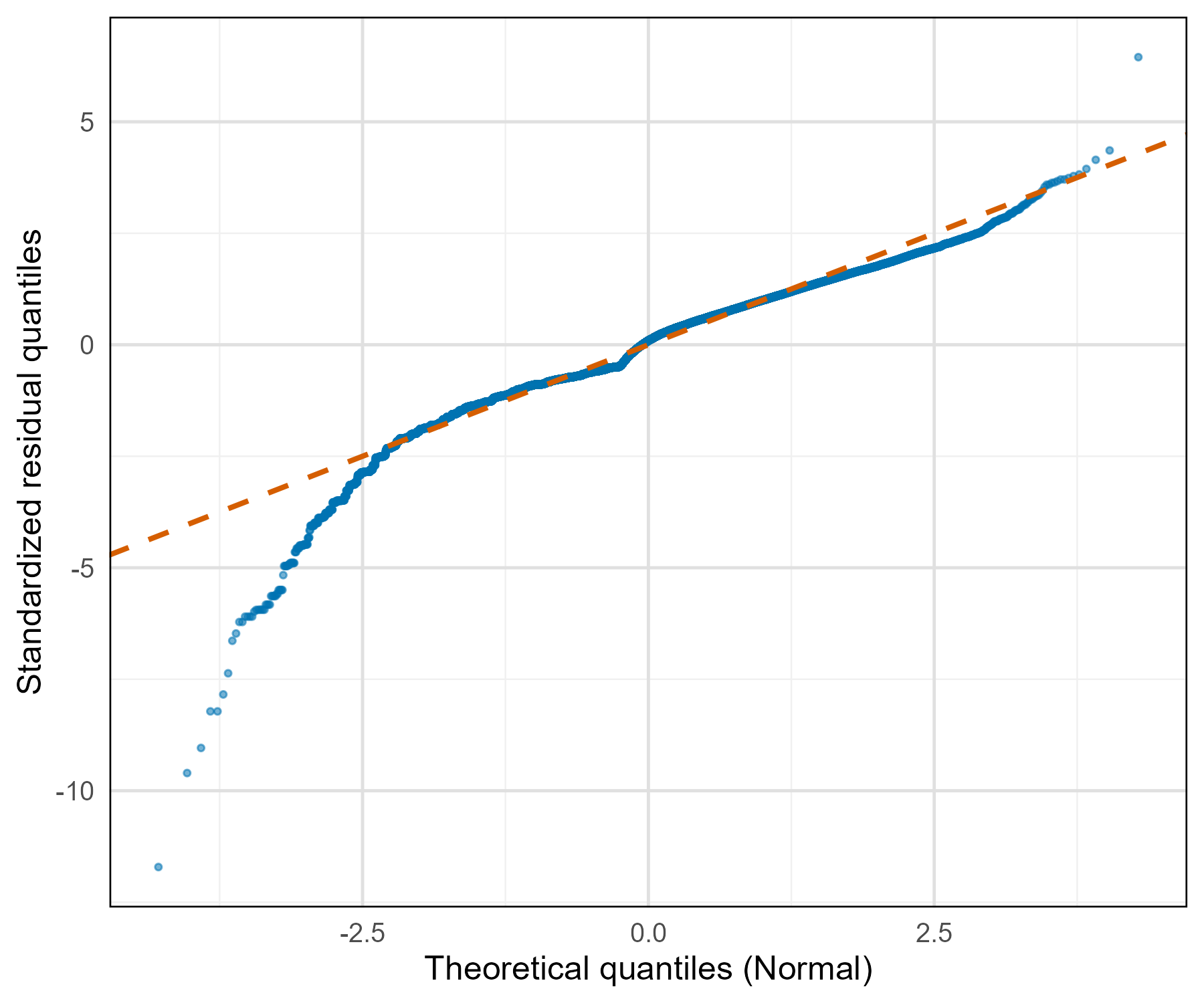}
        \caption{QQ-plot of standardized residuals showing heavy tails relative to Gaussian.}
        \label{fig:qqplot}
    \end{subfigure}
    \hfill
    \begin{subfigure}[t]{0.48\textwidth}
        \centering
        \includegraphics[width=\linewidth]{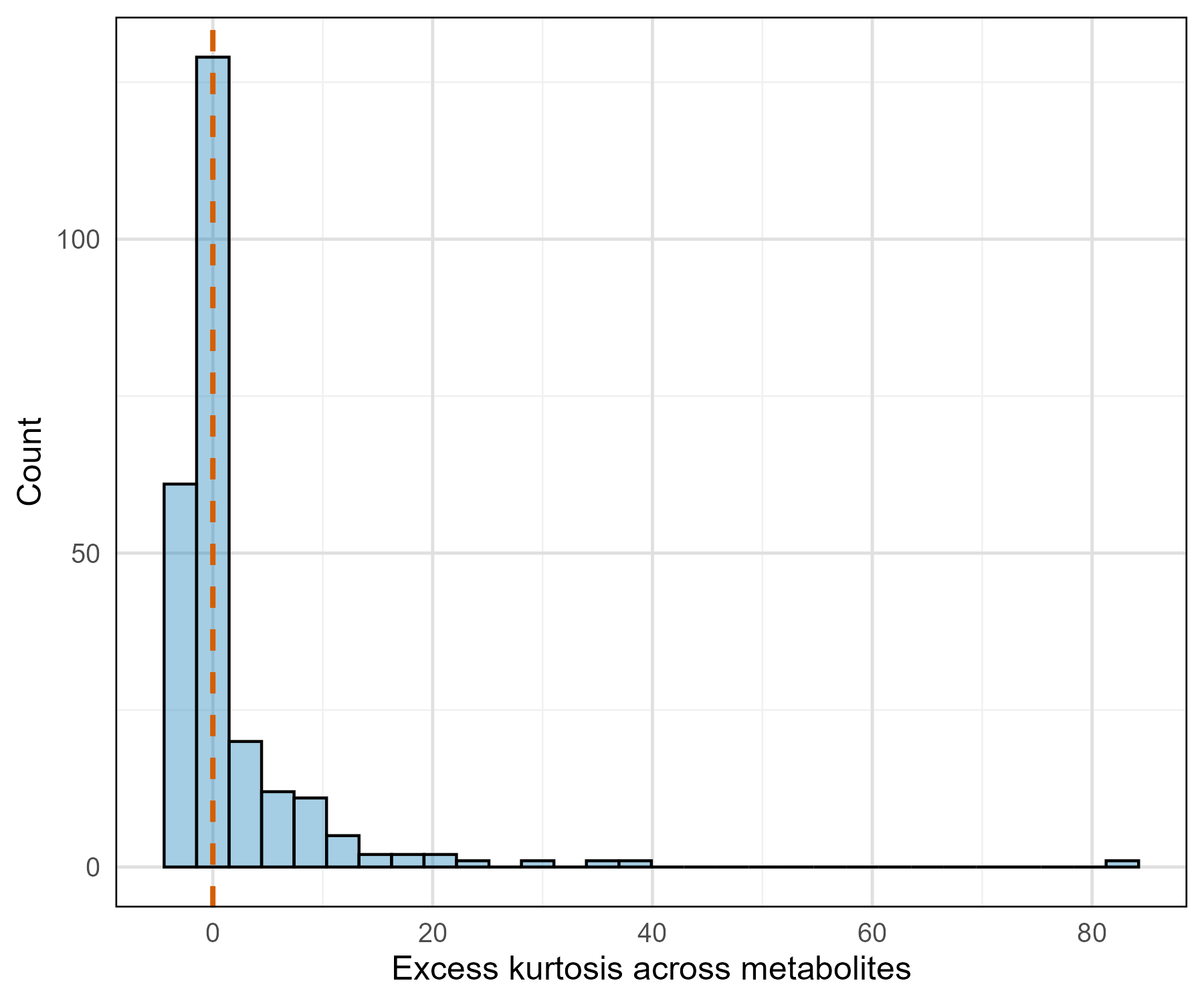}
        \caption{Histogram of residual kurtosis across metabolites.}
        \label{fig:kurtosis}
    \end{subfigure}
    \caption{Normality diagnostics of residuals. Panel (a) highlights deviations from normality in a pooled QQ-plot, while panel (b) shows excess kurtosis across metabolites.}
    \label{fig:normality_diagnostics}
\end{figure}

The CRC case study cohort consisted of $N=220$ subjects with paired microbiome and metabolomic measurements, including $P=287$ microbial genera and $Q=249$ metabolites. Supplementary Figure~\ref{fig:supp_patient_characteristics} summarizes the distributions of key demographic and lifestyle variables. The cohort exhibited broad age and BMI distributions, while the sex distribution was moderately balanced. Smoking and alcohol-consumption variables also showed substantial heterogeneity across subjects. Supplementary Figure~\ref{fig:normality_diagnostics} presents residual normality diagnostics obtained from preliminary marginal regression analyses. The pooled QQ-plot in Figure~\ref{fig:qqplot} demonstrates noticeable departures from Gaussian behavior in the tails, while the metabolite-wise kurtosis distribution in Figure~\ref{fig:kurtosis} indicates frequent excess kurtosis across outcomes. These observations suggest that the metabolomic responses exhibit moderate heavy-tailed behavior and dependence heterogeneity.

\begin{figure}[]
    \centering
    \begin{subfigure}[t]{0.48\textwidth}
        \centering
        \includegraphics[width=\textwidth]{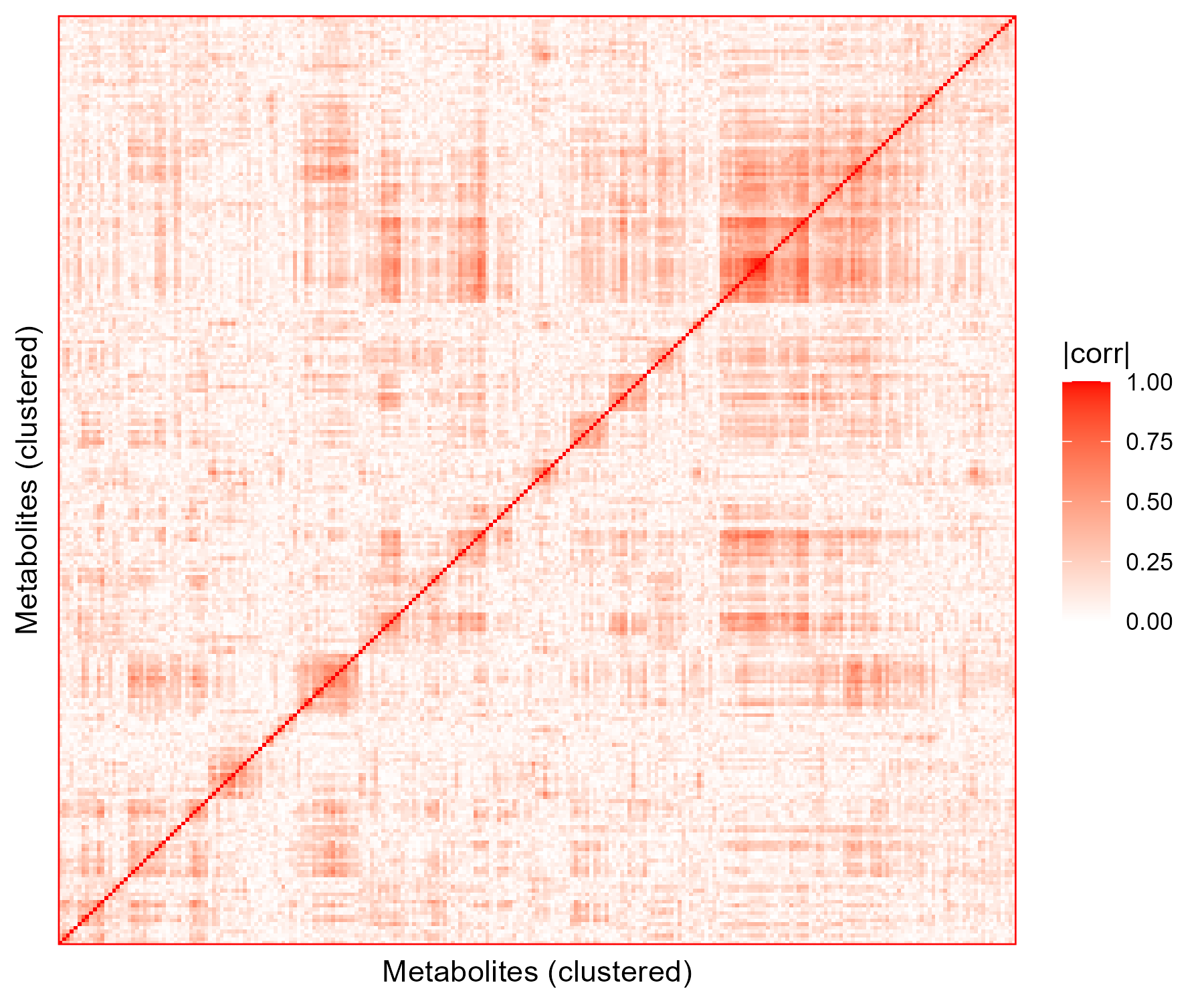}
        \caption{Absolute correlation heatmap among all metabolites ($Y$).}
        \label{fig:heatmap_Y}
    \end{subfigure}
    \hfill
    \begin{subfigure}[t]{0.48\textwidth}
        \centering
        \includegraphics[width=\textwidth]{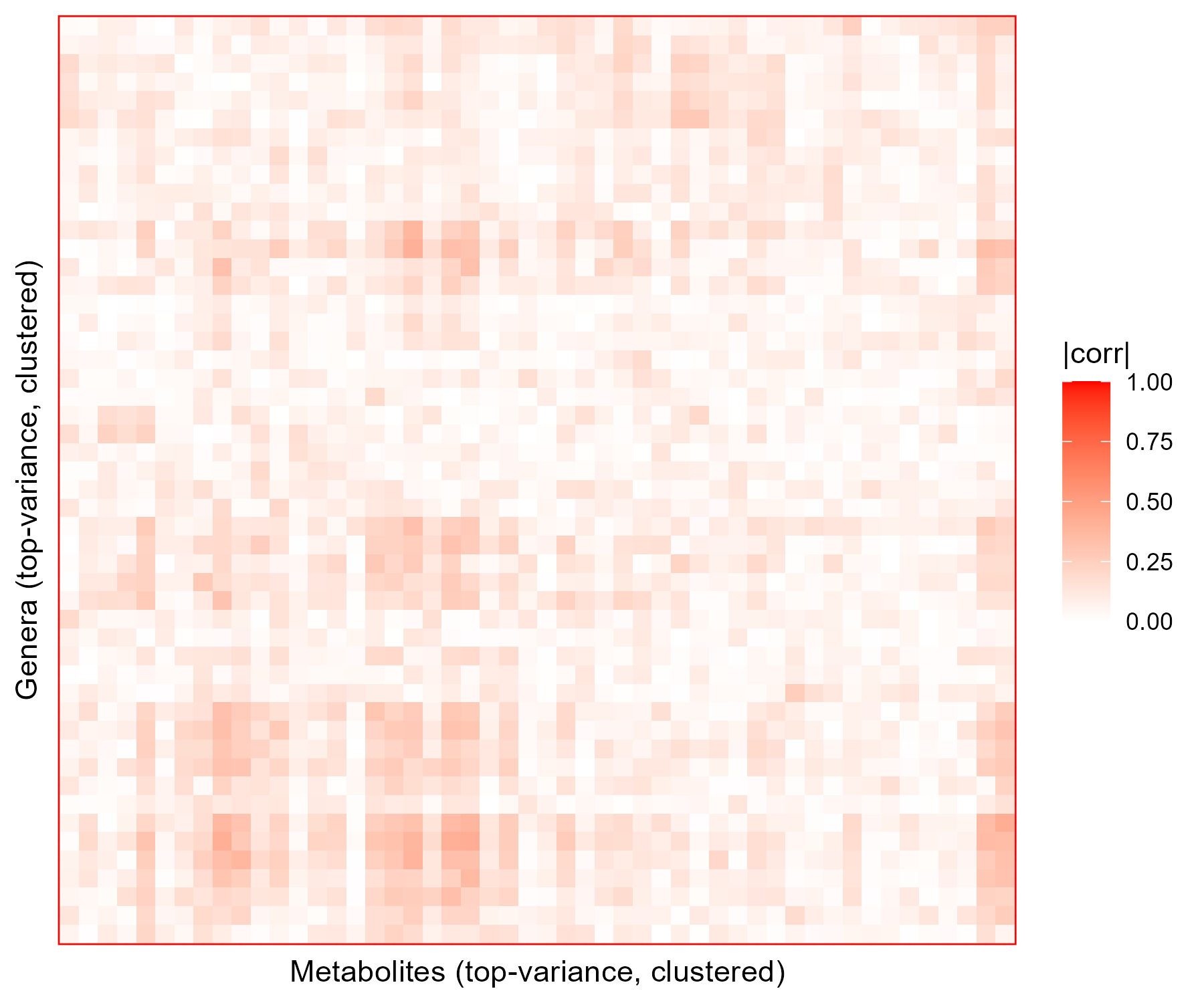}
        \caption{Absolute correlation between the top 50 variance-selected genera ($X$) and top 50 variance-selected metabolites ($Y$).}
        \label{fig:heatmap_XY}
    \end{subfigure}
    \caption{
    Correlation diagnostics for the CRC dataset.  
    (a) Heatmap of absolute correlations across all metabolites, showing block structures indicative of dependence.  
    (b) Cross-correlation heatmap between the top 50 genera and top 50 metabolites, selected by variance and clustered within each domain.  
    These plots reveal strong dependence both within and across microbiome and metabolome features, motivating a framework capable of handling correlated predictors and responses.
    }
    \label{fig:correlation_diagnostics}
\end{figure}

\begin{figure}[]
\centering
\includegraphics[width=.90\linewidth]{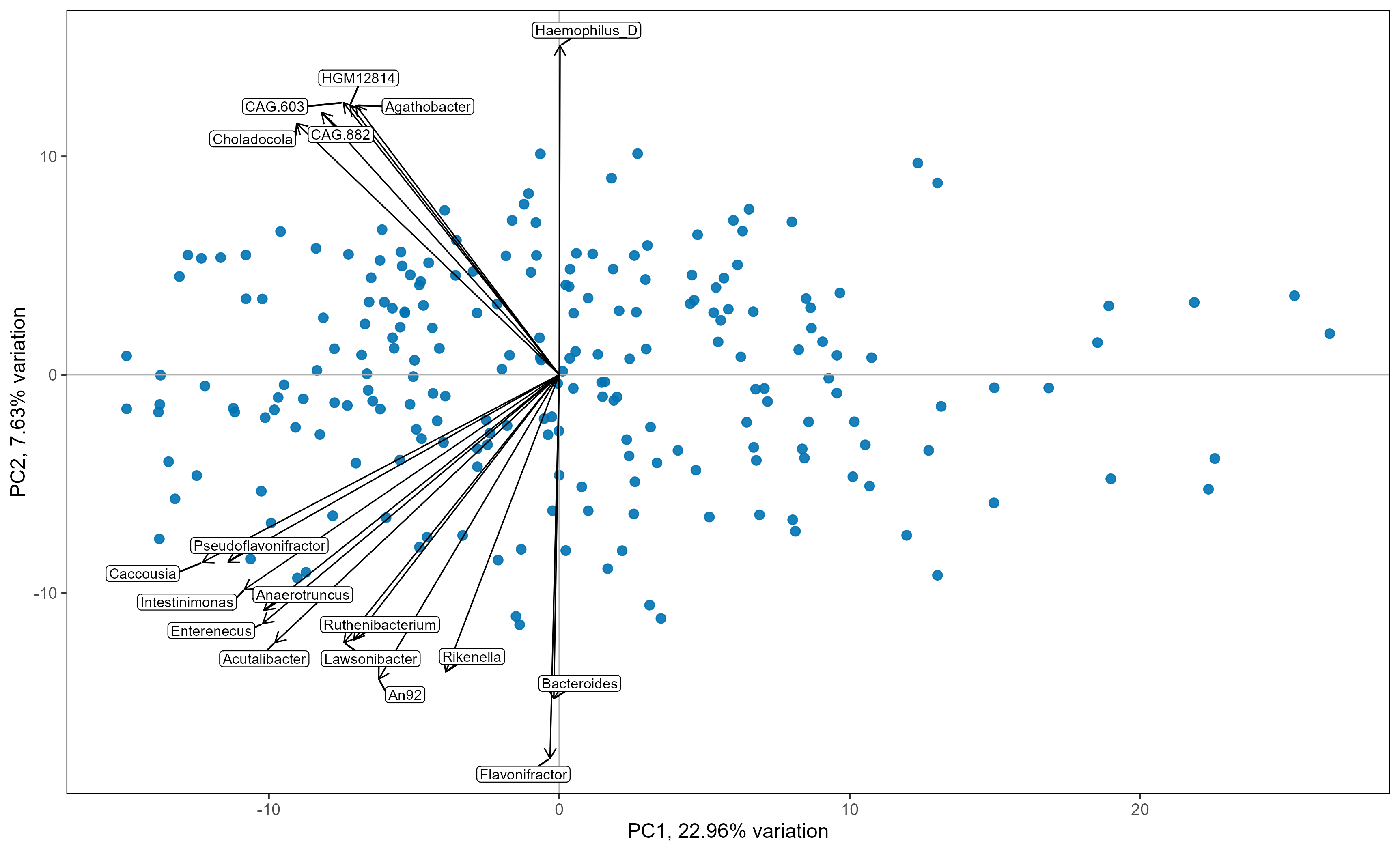} 
\includegraphics[width=.90\linewidth]{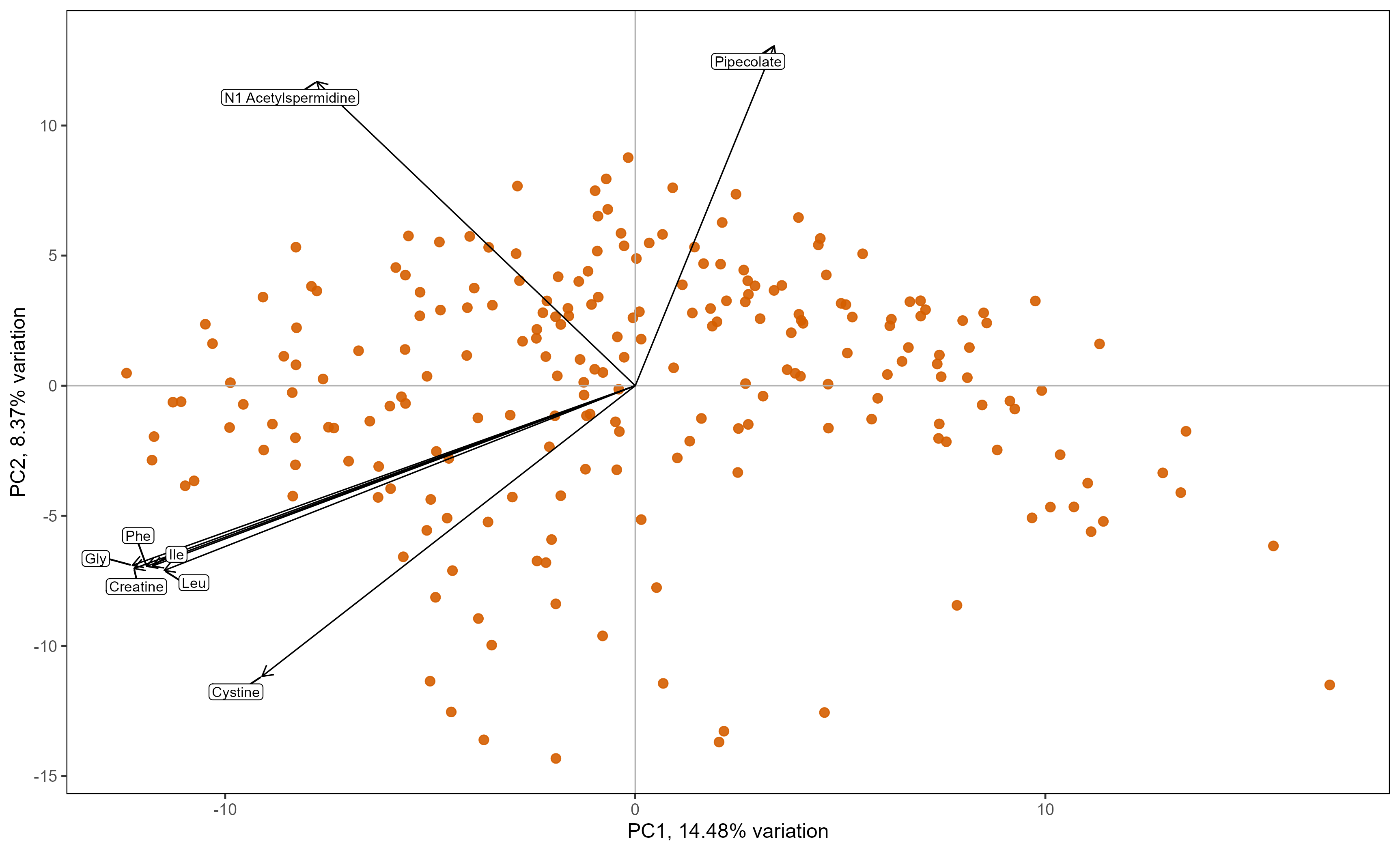} 
\caption{Score plots from Principal Component Analysis (PCA) illustrating the distribution of $P = 287$ microbial genera (top) and $Q = 249$ metabolites (bottom) based on fecal samples collected from $N = 220$ patients.}
\label{PCA}
\end{figure}

To further investigate dependence structures within the dataset, Supplementary Figure~\ref{fig:correlation_diagnostics} displays correlation heatmaps among metabolites and between microbiome and metabolomic features. Figure~\ref{fig:heatmap_Y} reveals strong block-wise correlation structures among metabolites, suggesting the presence of coordinated metabolic pathways and substantial cross-response dependence. Figure~\ref{fig:heatmap_XY} similarly demonstrates nontrivial cross-domain association patterns between microbial genera and metabolites, with localized clusters of elevated correlations. These observations motivate a joint multivariate modeling strategy rather than independent univariate analyses across metabolites. Supplementary Figure~\ref{PCA} presents principal component analysis (PCA) score plots for the microbiome and metabolomic feature spaces. The PCA projections suggest substantial variability and heterogeneity across subjects without strong low-dimensional separation patterns dominating the dataset. This diffuse structure supports the need for sparse high-dimensional modeling approaches capable of identifying localized association structures embedded within complex multivariate dependence patterns, rather than relying solely on low-rank global latent representations.

\subsection{Fractional Influence Score (FIS)}

To provide an interpretable measure of how influence is distributed across metabolites, we introduce a \emph{Fractional Influence Score (FIS)} for each microbial genus identified by B-MASTER. For any metabolite outcome $Y_q$, suppose it is influenced by $h_q$ genera (as determined by the B-MASTER selection step). Instead of attributing a full unit of influence to every selected genus, we allocate equal shares: each contributing genus receives a score of $1/h_q$. Summing these fractional contributions across all metabolites yields the overall score for a genus,
\[
\mathrm{FIS}(g) \;=\; \sum_{q=1}^{Q} \frac{\mathbf{1}\{g \in \mathcal{I}_q\}}{h_q},
\]
where $\mathcal{I}_q$ is the set of influencing genera for metabolite $q$. Because $h_q$ appears in the denominator, this score naturally distinguishes genera that uniquely regulate a metabolite from those that appear only as part of larger sets of co-predictors. 
The full distribution of genus-specific Fractional Influence Scores (FIS) across all 287 genera is summarized in Tables~\ref{tab:FIS_summary_table_1_to_50}--\ref{tab:FIS_summary_table_251_to_287}. These tables were included to provide a comprehensive characterization of the inferred influence hierarchy beyond the subset displayed in the main manuscript figures. In particular, they allow assessment of how influence is distributed across the complete microbiome set, the extent to which high-ranking genera are separated from lower-ranking genera, and the robustness of the inferred rankings under increasingly stringent posterior selection criteria.

Several observations emerge from these summaries. First, although the FIS values decrease gradually rather than exhibiting a single abrupt cutoff, the cumulative FIS percentages indicate meaningful concentration of influence among the top-ranked genera. Specifically, the top 50 genera accounted for approximately 24.85\% of the total cumulative FIS despite representing only about 17\% of all genera. This enrichment suggests that influence is not uniformly distributed across the microbiome network. Rather, a relatively small subset of genera repeatedly appears across many metabolite-specific posterior selection sets and therefore accumulates disproportionately large influence contributions.

Second, the ranking structure exhibits substantial stability under stricter posterior selection thresholds. Across Tables~\ref{tab:FIS_summary_table_1_to_50}--\ref{tab:FIS_summary_table_251_to_287}, many of the highest-ranked genera under the default 90\% posterior criterion remain highly ranked under the 95\% and 99\% posterior criteria. For example, several top-ranked genera including \textit{Alistipes\_A}, \textit{Ligilactobacillus}, \textit{Veillonella}, \textit{Mogibacterium}, and \textit{Massiliomicrobiota} continued to exhibit strong FIS values and relatively stable rankings under stricter posterior thresholds. This stability provides an additional layer of uncertainty quantification for the inferred master predictors and suggests that the dominant signals are not driven by isolated weak posterior selections.

Third, the median Bayesian posterior p-values reported in Tables~\ref{tab:FIS_summary_table_1_to_50}--\ref{tab:FIS_summary_table_251_to_287} provide a complementary measure of posterior edge-level support. Genera with high FIS values generally exhibited small median posterior p-values across their selected metabolite associations, indicating that the observed influence structure is supported not only by repeated selection frequency but also by strong posterior evidence at the metabolite-association level. Importantly, the Bayesian p-values summarize uncertainty for the selected microbiome-metabolite relationships, while the multi-threshold FIS summaries (90\%, 95\%, and 99\%) provide an indirect uncertainty assessment for the genus-level influence rankings themselves.

The choice to focus on the top 50 genera in the main manuscript was therefore intended primarily as a visualization and interpretability device rather than as a strict biological cutoff. The full tables demonstrate that the FIS hierarchy extends continuously across all genera, while still revealing a practically meaningful concentration of influence among the highest-ranked genera. Together, these results support the utility of the FIS framework as a quantitative measure for identifying influential microbial master predictors within the microbiome-metabolite network inferred by B-MASTER.

\begin{table}[!t]
\centering
\renewcommand{\arraystretch}{1.1}
\resizebox{0.9\textwidth}{!}{%
\begin{tabular}{lccccccccc}
\hline
Genus & \begin{tabular}[c]{@{}c@{}}FIS rank \\ (default; \\ 90\% CI \\ based)\end{tabular} & \begin{tabular}[c]{@{}c@{}}FIS \\ (default; \\ 90\% CI\\ based)\end{tabular} & \begin{tabular}[c]{@{}c@{}}Cumulative \\ FIS (\%)\end{tabular} & \begin{tabular}[c]{@{}c@{}}Num. \\ Selected \\ metabolites\end{tabular} & \begin{tabular}[c]{@{}c@{}}Median \\ Bayesian \\ p-value \\ (default; \\ 90\%  CI-based)\end{tabular} & \begin{tabular}[c]{@{}c@{}}FIS \\ (95\% \\ CI \\ based)\end{tabular} & \begin{tabular}[c]{@{}c@{}}Rank \\ (95\% \\ CI\\ based)\end{tabular} & \begin{tabular}[c]{@{}c@{}}FIS \\ (99\% \\ CI \\ based)\end{tabular} & \begin{tabular}[c]{@{}c@{}}Rank\\  (99\% \\ CI \\ based)\end{tabular} \\ \hline
CAG.196 & 1 & 1.45 & 0.58 & 136 & 0.009 & 1.58 & 3 & 1.94 & 4 \\
Choladocola & 2 & 1.42 & 1.15 & 133 & 0.009 & 1.44 & 12 & 1.87 & 8 \\
Alistipes\_A & 3 & 1.4 & 1.72 & 131 & 0.002 & 1.63 & 1 & 2.32 & 1 \\
Veillonella & 4 & 1.4 & 2.28 & 130 & 0.008 & 1.5 & 8 & 1.93 & 5 \\
Ligilactobacillus & 5 & 1.38 & 2.83 & 129 & 0.004 & 1.6 & 2 & 2.1 & 2 \\
Agathobacter & 6 & 1.36 & 3.38 & 126 & 0.009 & 1.55 & 4 & 1.69 & 17 \\
Mogibacterium & 7 & 1.36 & 3.92 & 127 & 0.009 & 1.54 & 5 & 1.88 & 6 \\
Massiliomicrobiota & 8 & 1.35 & 4.46 & 126 & 0.007 & 1.46 & 10 & 1.96 & 3 \\
Caccousia & 9 & 1.34 & 5 & 126 & 0.007 & 1.39 & 18 & 1.73 & 15 \\
Succinivibrio & 10 & 1.32 & 5.53 & 123 & 0.011 & 1.45 & 11 & 1.65 & 19 \\
Phil1 & 11 & 1.31 & 6.06 & 123 & 0.011 & 1.41 & 17 & 1.68 & 18 \\
Gemmiger & 12 & 1.31 & 6.58 & 123 & 0.007 & 1.51 & 7 & 1.85 & 9 \\
Anaerotignum & 13 & 1.31 & 7.11 & 123 & 0.004 & 1.54 & 6 & 1.87 & 7 \\
Egerieisoma & 14 & 1.3 & 7.63 & 121 & 0.009 & 1.39 & 19 & 1.76 & 11 \\
UBA3402 & 15 & 1.29 & 8.15 & 121 & 0.013 & 1.35 & 21 & 1.53 & 27 \\
Fimenecus & 16 & 1.28 & 8.67 & 120 & 0.013 & 1.44 & 14 & 1.56 & 23 \\
Avimicrobium & 17 & 1.28 & 9.18 & 120 & 0.009 & 1.44 & 13 & 1.63 & 20 \\
Merdisoma & 18 & 1.28 & 9.7 & 119 & 0.009 & 1.43 & 15 & 1.75 & 13 \\
Enterocloster & 19 & 1.27 & 10.21 & 119 & 0.016 & 1.32 & 26 & 1.43 & 39 \\
CAG.83 & 20 & 1.26 & 10.72 & 118 & 0.009 & 1.49 & 9 & 1.74 & 14 \\
UBA737 & 21 & 1.25 & 11.22 & 117 & 0.011 & 1.38 & 20 & 1.53 & 26 \\
Gemella & 22 & 1.24 & 11.72 & 116 & 0.018 & 1.19 & 51 & 1.35 & 51 \\
Pauljensenia & 23 & 1.22 & 12.21 & 114 & 0.007 & 1.33 & 24 & 1.75 & 12 \\
HGM12814 & 24 & 1.22 & 12.7 & 114 & 0.01 & 1.41 & 16 & 1.53 & 25 \\
Butyricicoccus\_A & 25 & 1.21 & 13.18 & 113 & 0.007 & 1.34 & 22 & 1.78 & 10 \\
CAG.317 & 26 & 1.2 & 13.67 & 112 & 0.013 & 1.25 & 36 & 1.37 & 49 \\
Streptococcus & 27 & 1.2 & 14.15 & 112 & 0.013 & 1.29 & 28 & 1.36 & 50 \\
Bariatricus & 28 & 1.2 & 14.63 & 112 & 0.009 & 1.22 & 45 & 1.57 & 22 \\
Anaerotruncus & 29 & 1.19 & 15.11 & 111 & 0.011 & 1.25 & 39 & 1.46 & 33 \\
Fusobacterium\_B & 30 & 1.19 & 15.59 & 111 & 0.013 & 1.29 & 30 & 1.4 & 43 \\
Faecalimonas & 31 & 1.19 & 16.06 & 112 & 0.007 & 1.34 & 23 & 1.72 & 16 \\
Onthousia & 32 & 1.19 & 16.54 & 110 & 0.016 & 1.23 & 42 & 1.31 & 54 \\
Limosilactobacillus & 33 & 1.19 & 17.02 & 111 & 0.013 & 1.16 & 57 & 1.4 & 41 \\
Amedibacterium & 34 & 1.18 & 17.49 & 111 & 0.009 & 1.26 & 35 & 1.52 & 29 \\
Mitsuokella & 35 & 1.18 & 17.97 & 110 & 0.014 & 1.25 & 37 & 1.5 & 30 \\
CAG.95 & 36 & 1.18 & 18.44 & 110 & 0.009 & 1.21 & 46 & 1.52 & 28 \\
Turicibacter & 37 & 1.17 & 18.91 & 109 & 0.016 & 1.27 & 32 & 1.34 & 52 \\
CAG.177 & 38 & 1.16 & 19.38 & 108 & 0.018 & 1.16 & 58 & 1.12 & 85 \\
Phascolarctobacterium\_A & 39 & 1.15 & 19.84 & 108 & 0.009 & 1.27 & 34 & 1.48 & 31 \\
Eubacterium\_F & 40 & 1.15 & 20.3 & 107 & 0.007 & 1.27 & 33 & 1.6 & 21 \\
Romboutsia & 41 & 1.14 & 20.76 & 105 & 0.018 & 1.13 & 63 & 1.2 & 72 \\
Butyricicoccus & 42 & 1.14 & 21.21 & 107 & 0.018 & 1.12 & 66 & 1.26 & 65 \\
Merdicola & 43 & 1.14 & 21.67 & 105 & 0.011 & 1.22 & 44 & 1.4 & 42 \\
Clostridium\_AQ & 44 & 1.14 & 22.13 & 106 & 0.014 & 1.2 & 48 & 1.23 & 69 \\
Sellimonas & 45 & 1.14 & 22.58 & 107 & 0.011 & 1.3 & 27 & 1.47 & 32 \\
Bacteroides & 46 & 1.13 & 23.04 & 106 & 0.013 & 1.18 & 54 & 1.3 & 55 \\
Lawsonibacter & 47 & 1.13 & 23.49 & 106 & 0.011 & 1.29 & 29 & 1.38 & 47 \\
Alistipes & 48 & 1.13 & 23.95 & 105 & 0.018 & 1.1 & 69 & 1.29 & 57 \\
Granulicatella & 49 & 1.13 & 24.4 & 105 & 0.016 & 1.22 & 43 & 1.39 & 46 \\
CAG.267 & 50 & 1.13 & 24.85 & 105 & 0.016 & 1.18 & 53 & 1.28 & 59 \\ \hline
\end{tabular}}
\caption{Top 50 genera ranked by Fractional Influence Score (FIS) under the default 90\% posterior selection criterion in the CRC cohort. The table reports genus-specific FIS rankings, cumulative FIS contribution, number of selected metabolites connected to each genus, median Bayesian posterior edge significance, and corresponding FIS/rank summaries obtained under stricter 95\% and 99\% posterior selection criteria. These results illustrate the dominant CRC-associated master predictors identified by B-MASTER and demonstrate substantial ranking stability across posterior selection thresholds.}
\label{tab:FIS_summary_table_1_to_50}
\end{table}

\begin{table}[!t]
\centering
\renewcommand{\arraystretch}{1.1}
\resizebox{0.9\textwidth}{!}{%
\begin{tabular}{lccccccccc}
\hline
Genus & \begin{tabular}[c]{@{}c@{}}FIS rank \\ (default; \\ 90\% CI \\ based)\end{tabular} & \begin{tabular}[c]{@{}c@{}}FIS \\ (default; \\ 90\% CI\\ based)\end{tabular} & \begin{tabular}[c]{@{}c@{}}Cumulative \\ FIS (\%)\end{tabular} & \begin{tabular}[c]{@{}c@{}}Num. \\ Selected \\ metabolites\end{tabular} & \begin{tabular}[c]{@{}c@{}}Median \\ Bayesian \\ p-value \\ (default; \\ 90\%  CI-based)\end{tabular} & \begin{tabular}[c]{@{}c@{}}FIS \\ (95\% \\ CI \\ based)\end{tabular} & \begin{tabular}[c]{@{}c@{}}Rank \\ (95\% \\ CI\\ based)\end{tabular} & \begin{tabular}[c]{@{}c@{}}FIS \\ (99\% \\ CI \\ based)\end{tabular} & \begin{tabular}[c]{@{}c@{}}Rank\\  (99\% \\ CI \\ based)\end{tabular} \\ \hline
Lancefieldella & 51 & 1.13 & 25.3 & 105 & 0.011 & 1.14 & 62 & 1.39 & 45 \\
Copromonas & 52 & 1.12 & 25.75 & 105 & 0.018 & 1.17 & 55 & 1.28 & 60 \\
1XD42.69 & 53 & 1.12 & 26.2 & 105 & 0.009 & 1.32 & 25 & 1.44 & 35 \\
Enterococcus & 54 & 1.12 & 26.65 & 103 & 0.009 & 1.25 & 40 & 1.45 & 34 \\
UMGS872 & 55 & 1.12 & 27.1 & 105 & 0.011 & 1.21 & 47 & 1.41 & 40 \\
Gordonibacter & 56 & 1.11 & 27.55 & 104 & 0.011 & 1.25 & 41 & 1.38 & 48 \\
Intestinibacter & 57 & 1.11 & 28 & 103 & 0.016 & 1.14 & 60 & 1.29 & 56 \\
CAG.239.g\_\_51.20 & 58 & 1.11 & 28.44 & 103 & 0.009 & 1.25 & 38 & 1.56 & 24 \\
CAG.603 & 59 & 1.11 & 28.89 & 104 & 0.013 & 1.18 & 52 & 1.27 & 62 \\
Butyribacter & 60 & 1.11 & 29.33 & 104 & 0.013 & 1.16 & 56 & 1.34 & 53 \\
Slackia\_A & 61 & 1.1 & 29.78 & 103 & 0.016 & 1.15 & 59 & 1.19 & 73 \\
UBA4372 & 62 & 1.1 & 30.22 & 103 & 0.013 & 1.13 & 64 & 1.22 & 71 \\
Oliverpabstia & 63 & 1.09 & 30.65 & 101 & 0.007 & 1.2 & 49 & 1.44 & 37 \\
Mailhella & 64 & 1.09 & 31.09 & 101 & 0.009 & 1.28 & 31 & 1.44 & 36 \\
Eubacterium\_R & 65 & 1.07 & 31.52 & 100 & 0.02 & 1.02 & 90 & 1.15 & 78 \\
Porphyromonas & 66 & 1.07 & 31.95 & 100 & 0.018 & 1.03 & 88 & 0.98 & 105 \\
Ruminococcus\_D & 67 & 1.06 & 32.38 & 99 & 0.013 & 1.14 & 61 & 1.26 & 64 \\
Angelakisella & 68 & 1.06 & 32.8 & 99 & 0.022 & 1.09 & 72 & 1.01 & 99 \\
An92 & 69 & 1.06 & 33.23 & 99 & 0.013 & 1.11 & 67 & 1.25 & 66 \\
Lachnospiraceae.g\_\_14.2 & 70 & 1.06 & 33.65 & 99 & 0.016 & 1.03 & 86 & 1.13 & 81 \\
CAG.1031 & 71 & 1.05 & 34.07 & 98 & 0.014 & 1.06 & 83 & 1.12 & 83 \\
Acetatifactor & 72 & 1.05 & 34.49 & 98 & 0.018 & 1.08 & 73 & 1.01 & 98 \\
MWCK01 & 73 & 1.04 & 34.91 & 97 & 0.013 & 1.11 & 68 & 1.01 & 100 \\
Bilophila & 74 & 1.04 & 35.33 & 96 & 0.009 & 1.19 & 50 & 1.4 & 44 \\
Copromorpha & 75 & 1.03 & 35.74 & 97 & 0.011 & 1.06 & 80 & 1.27 & 61 \\
Streptomyces & 76 & 1.02 & 36.15 & 96 & 0.018 & 1.09 & 70 & 1.08 & 90 \\
Oxalobacter & 77 & 1.02 & 36.56 & 95 & 0.02 & 1.06 & 82 & 1.16 & 76 \\
F23.B02 & 78 & 1.01 & 36.97 & 95 & 0.013 & 1.08 & 75 & 1.15 & 77 \\
Flavonifractor & 79 & 1.01 & 37.37 & 94 & 0.016 & 1 & 93 & 1.05 & 93 \\
Faecalitalea & 80 & 1 & 37.78 & 94 & 0.028 & 0.9 & 123 & 0.92 & 117 \\
UMGS856 & 81 & 1 & 38.18 & 93 & 0.022 & 1.03 & 84 & 0.98 & 106 \\
Fimivicinus & 82 & 1 & 38.58 & 93 & 0.02 & 0.99 & 100 & 1 & 102 \\
Agathobaculum & 83 & 1 & 38.98 & 94 & 0.018 & 1.07 & 79 & 1.13 & 82 \\
Odoribacter & 84 & 1 & 39.38 & 94 & 0.011 & 1.09 & 71 & 1.23 & 68 \\
Clostridium & 85 & 1 & 39.78 & 93 & 0.013 & 1.06 & 81 & 1.13 & 80 \\
CAG.488 & 86 & 1 & 40.19 & 93 & 0.022 & 1.08 & 77 & 0.77 & 148 \\
Eggerthella & 87 & 1 & 40.59 & 93 & 0.011 & 1.13 & 65 & 1.24 & 67 \\
Lactococcus & 88 & 1 & 40.99 & 93 & 0.018 & 0.97 & 103 & 1.08 & 89 \\
Intestinimonas & 89 & 0.99 & 41.38 & 93 & 0.013 & 1.08 & 74 & 1.14 & 79 \\
Blautia\_A & 90 & 0.99 & 41.78 & 93 & 0.027 & 0.93 & 117 & 0.8 & 138 \\
Schaedlerella & 91 & 0.99 & 42.18 & 92 & 0.013 & 1.03 & 87 & 1.03 & 95 \\
Phil12 & 92 & 0.98 & 42.57 & 91 & 0.022 & 0.91 & 120 & 0.78 & 145 \\
NSJ.51 & 93 & 0.98 & 42.96 & 91 & 0.009 & 1.01 & 91 & 1.28 & 58 \\
NSJ.61 & 94 & 0.98 & 43.36 & 91 & 0.016 & 0.99 & 97 & 1.11 & 86 \\
Evtepia & 95 & 0.97 & 43.75 & 91 & 0.011 & 0.98 & 101 & 1.18 & 74 \\
Lacrimispora & 96 & 0.97 & 44.14 & 91 & 0.016 & 0.95 & 110 & 0.98 & 107 \\
CAG.266 & 97 & 0.97 & 44.53 & 90 & 0.016 & 1 & 94 & 1.08 & 91 \\
Holdemania & 98 & 0.96 & 44.91 & 90 & 0.013 & 1 & 96 & 1.05 & 94 \\
Limisoma & 99 & 0.96 & 45.3 & 90 & 0.016 & 1.03 & 85 & 1.03 & 96 \\
Dorea & 100 & 0.96 & 45.69 & 90 & 0.007 & 1.07 & 78 & 1.43 & 38 \\ \hline
\end{tabular}}
\caption{Genera ranked 51--100 by Fractional Influence Score (FIS) under the default 90\% posterior selection criterion in the CRC cohort. The table reports genus-specific FIS rankings, cumulative FIS contribution, number of selected metabolites connected to each genus, median Bayesian posterior edge significance, and corresponding FIS/rank summaries obtained under stricter 95\% and 99\% posterior selection criteria.}
\label{tab:FIS_summary_table_51_to_100}
\end{table}

\begin{table}[!t]
\centering
\renewcommand{\arraystretch}{1.1}
\resizebox{0.9\textwidth}{!}{%
\begin{tabular}{lccccccccc}
\hline
Genus & \begin{tabular}[c]{@{}c@{}}FIS rank \\ (default; \\ 90\% CI \\ based)\end{tabular} & \begin{tabular}[c]{@{}c@{}}FIS \\ (default; \\ 90\% CI\\ based)\end{tabular} & \begin{tabular}[c]{@{}c@{}}Cumulative \\ FIS (\%)\end{tabular} & \begin{tabular}[c]{@{}c@{}}Num. \\ Selected \\ metabolites\end{tabular} & \begin{tabular}[c]{@{}c@{}}Median \\ Bayesian \\ p-value \\ (default; \\ 90\%  CI-based)\end{tabular} & \begin{tabular}[c]{@{}c@{}}FIS \\ (95\% \\ CI \\ based)\end{tabular} & \begin{tabular}[c]{@{}c@{}}Rank \\ (95\% \\ CI\\ based)\end{tabular} & \begin{tabular}[c]{@{}c@{}}FIS \\ (99\% \\ CI \\ based)\end{tabular} & \begin{tabular}[c]{@{}c@{}}Rank\\  (99\% \\ CI \\ based)\end{tabular} \\ \hline
Parvimonas & 101 & 0.96 & 46.07 & 89 & 0.029 & 0.89 & 125 & 0.94 & 116 \\
Tidjanibacter & 102 & 0.96 & 46.46 & 89 & 0.016 & 1 & 95 & 0.96 & 112 \\
Coprobacter & 103 & 0.95 & 46.84 & 89 & 0.013 & 1.08 & 76 & 1.11 & 87 \\
CAG.41 & 104 & 0.95 & 47.22 & 89 & 0.018 & 0.95 & 111 & 0.88 & 126 \\
Negativibacillus & 105 & 0.95 & 47.6 & 89 & 0.011 & 0.95 & 107 & 1.22 & 70 \\
Scatomorpha & 106 & 0.95 & 47.98 & 88 & 0.009 & 0.99 & 98 & 1.27 & 63 \\
Lachnospira & 107 & 0.94 & 48.36 & 88 & 0.017 & 0.99 & 99 & 0.96 & 111 \\
Fusobacterium & 108 & 0.94 & 48.74 & 88 & 0.018 & 0.93 & 116 & 0.95 & 114 \\
UBA1417 & 109 & 0.93 & 49.11 & 87 & 0.011 & 1.02 & 89 & 1.12 & 84 \\
CAG.269 & 110 & 0.93 & 49.49 & 87 & 0.024 & 0.86 & 137 & 0.7 & 165 \\
Rothia & 111 & 0.93 & 49.86 & 87 & 0.016 & 0.9 & 121 & 1.07 & 92 \\
PeH17 & 112 & 0.93 & 50.23 & 86 & 0.014 & 1.01 & 92 & 0.97 & 110 \\
Megasphaera & 113 & 0.92 & 50.6 & 86 & 0.02 & 0.95 & 109 & 0.84 & 133 \\
Bulleidia & 114 & 0.92 & 50.97 & 86 & 0.016 & 0.98 & 102 & 0.97 & 108 \\
Faecousia & 115 & 0.92 & 51.34 & 86 & 0.019 & 0.96 & 106 & 0.86 & 129 \\
Cryptobacteroides & 116 & 0.92 & 51.71 & 86 & 0.016 & 0.95 & 112 & 1 & 103 \\
Thauera & 117 & 0.92 & 52.08 & 85 & 0.018 & 0.96 & 104 & 0.9 & 121 \\
Zag111 & 118 & 0.91 & 52.45 & 85 & 0.011 & 0.89 & 126 & 1.02 & 97 \\
Faecalibacterium & 119 & 0.9 & 52.81 & 84 & 0.02 & 0.94 & 113 & 0.9 & 119 \\
Faecimorpha & 120 & 0.9 & 53.17 & 84 & 0.022 & 0.85 & 141 & 0.9 & 118 \\
Pelethomonas & 121 & 0.9 & 53.53 & 84 & 0.021 & 0.95 & 108 & 0.89 & 124 \\
Acutalibacter & 122 & 0.9 & 53.9 & 85 & 0.024 & 0.87 & 135 & 0.78 & 146 \\
Clostridium\_Q & 123 & 0.9 & 54.26 & 85 & 0.029 & 0.87 & 130 & 0.53 & 212 \\
UBA3818 & 124 & 0.9 & 54.62 & 84 & 0.009 & 0.96 & 105 & 1.17 & 75 \\
Barnesiella & 125 & 0.9 & 54.98 & 84 & 0.028 & 0.81 & 154 & 0.76 & 155 \\
Gallimonas & 126 & 0.9 & 55.34 & 84 & 0.029 & 0.76 & 171 & 0.74 & 161 \\
Aphodousia & 127 & 0.9 & 55.7 & 84 & 0.024 & 0.85 & 140 & 0.66 & 173 \\
Cloacibacillus & 128 & 0.9 & 56.06 & 84 & 0.013 & 0.87 & 136 & 0.98 & 104 \\
Eubacterium\_I & 129 & 0.89 & 56.42 & 82 & 0.02 & 0.88 & 129 & 0.77 & 149 \\
Stercorousia & 130 & 0.88 & 56.77 & 83 & 0.033 & 0.77 & 166 & 0.67 & 169 \\
Faecalicoccus & 131 & 0.88 & 57.13 & 83 & 0.018 & 0.82 & 152 & 0.95 & 115 \\
Limivivens & 132 & 0.88 & 57.48 & 81 & 0.02 & 0.83 & 147 & 0.77 & 151 \\
UBA738 & 133 & 0.88 & 57.84 & 82 & 0.02 & 0.86 & 139 & 0.89 & 123 \\
Pseudoruminococcus\_A & 134 & 0.88 & 58.19 & 81 & 0.022 & 0.89 & 127 & 0.85 & 131 \\
Monoglobus & 135 & 0.88 & 58.54 & 81 & 0.016 & 0.93 & 114 & 0.9 & 120 \\
Lachnoclostridium\_B & 136 & 0.88 & 58.89 & 82 & 0.017 & 0.87 & 133 & 0.95 & 113 \\
CAG.492 & 137 & 0.87 & 59.24 & 81 & 0.016 & 0.92 & 118 & 1 & 101 \\
Anaerobutyricum & 138 & 0.87 & 59.59 & 81 & 0.024 & 0.9 & 122 & 0.71 & 163 \\
Mediterraneibacter\_A & 139 & 0.86 & 59.94 & 80 & 0.018 & 0.85 & 142 & 0.79 & 141 \\
UMGS1071 & 140 & 0.86 & 60.28 & 81 & 0.016 & 0.9 & 124 & 0.89 & 122 \\
CAG.1427 & 141 & 0.86 & 60.63 & 80 & 0.023 & 0.83 & 148 & 0.75 & 159 \\
Phocaeicola\_A & 142 & 0.86 & 60.97 & 80 & 0.036 & 0.87 & 134 & 0.7 & 164 \\
Emergencia & 143 & 0.86 & 61.32 & 80 & 0.027 & 0.77 & 164 & 0.66 & 171 \\
Lachnoclostridium\_A & 144 & 0.85 & 61.66 & 80 & 0.013 & 0.93 & 115 & 1.08 & 88 \\
Anaeromassilibacillus & 145 & 0.85 & 62 & 80 & 0.019 & 0.84 & 144 & 0.97 & 109 \\
UBA11524 & 146 & 0.85 & 62.34 & 80 & 0.02 & 0.87 & 131 & 0.82 & 136 \\
Limadaptatus & 147 & 0.85 & 62.69 & 79 & 0.033 & 0.74 & 185 & 0.62 & 189 \\
Ventrimonas & 148 & 0.85 & 63.03 & 79 & 0.02 & 0.86 & 138 & 0.76 & 156 \\
Megamonas & 149 & 0.85 & 63.37 & 79 & 0.036 & 0.76 & 176 & 0.48 & 222 \\
Dialister & 150 & 0.85 & 63.71 & 79 & 0.022 & 0.87 & 132 & 0.47 & 226 \\ \hline
\end{tabular}}
\caption{Genera ranked 101--150 by Fractional Influence Score (FIS) under the default 90\% posterior selection criterion in the CRC cohort. The table reports genus-specific FIS rankings, cumulative FIS contribution, number of selected metabolites connected to each genus, median Bayesian posterior edge significance, and corresponding FIS/rank summaries obtained under stricter 95\% and 99\% posterior selection criteria.}
\label{tab:FIS_summary_table_101_to_150}
\end{table}

\begin{table}[!t]
\centering
\renewcommand{\arraystretch}{1.1}
\resizebox{0.9\textwidth}{!}{%
\begin{tabular}{lccccccccc}
\hline
Genus & \begin{tabular}[c]{@{}c@{}}FIS rank \\ (default; \\ 90\% CI \\ based)\end{tabular} & \begin{tabular}[c]{@{}c@{}}FIS \\ (default; \\ 90\% CI\\ based)\end{tabular} & \begin{tabular}[c]{@{}c@{}}Cumulative \\ FIS (\%)\end{tabular} & \begin{tabular}[c]{@{}c@{}}Num. \\ Selected \\ metabolites\end{tabular} & \begin{tabular}[c]{@{}c@{}}Median \\ Bayesian \\ p-value \\ (default; \\ 90\%  CI-based)\end{tabular} & \begin{tabular}[c]{@{}c@{}}FIS \\ (95\% \\ CI \\ based)\end{tabular} & \begin{tabular}[c]{@{}c@{}}Rank \\ (95\% \\ CI\\ based)\end{tabular} & \begin{tabular}[c]{@{}c@{}}FIS \\ (99\% \\ CI \\ based)\end{tabular} & \begin{tabular}[c]{@{}c@{}}Rank\\  (99\% \\ CI \\ based)\end{tabular} \\ \hline
Prevotella & 151 & 0.84 & 64.04 & 78 & 0.016 & 0.92 & 119 & 0.87 & 127 \\
Muricomes & 152 & 0.84 & 64.38 & 77 & 0.022 & 0.84 & 145 & 0.71 & 162 \\
Choladousia & 153 & 0.84 & 64.72 & 79 & 0.024 & 0.77 & 165 & 0.66 & 170 \\
Coproplasma & 154 & 0.83 & 65.05 & 78 & 0.029 & 0.79 & 160 & 0.77 & 147 \\
NSJ.62 & 155 & 0.83 & 65.39 & 78 & 0.021 & 0.81 & 155 & 0.8 & 139 \\
Limiplasma & 156 & 0.83 & 65.72 & 77 & 0.018 & 0.84 & 143 & 0.74 & 160 \\
CAG.882 & 157 & 0.82 & 66.05 & 77 & 0.02 & 0.79 & 158 & 0.87 & 128 \\
Victivallis & 158 & 0.82 & 66.38 & 76 & 0.031 & 0.72 & 189 & 0.57 & 205 \\
UMGS775 & 159 & 0.82 & 66.71 & 77 & 0.022 & 0.78 & 162 & 0.82 & 137 \\
Parabacteroides & 160 & 0.82 & 67.03 & 76 & 0.029 & 0.76 & 172 & 0.61 & 195 \\
Pelethocola & 161 & 0.82 & 67.36 & 76 & 0.019 & 0.74 & 184 & 0.69 & 167 \\
CAG.170 & 162 & 0.82 & 67.69 & 77 & 0.018 & 0.79 & 161 & 0.8 & 140 \\
ER4 & 163 & 0.82 & 68.02 & 77 & 0.018 & 0.76 & 175 & 0.88 & 125 \\
Mesosutterella & 164 & 0.8 & 68.34 & 76 & 0.022 & 0.83 & 149 & 0.61 & 191 \\
Clostridium\_P & 165 & 0.8 & 68.66 & 75 & 0.016 & 0.89 & 128 & 0.84 & 134 \\
Catenibacterium & 166 & 0.8 & 68.98 & 75 & 0.027 & 0.75 & 179 & 0.48 & 220 \\
Sutterella & 167 & 0.8 & 69.31 & 75 & 0.022 & 0.77 & 167 & 0.63 & 185 \\
Pseudobutyricicoccus & 168 & 0.78 & 69.62 & 73 & 0.029 & 0.75 & 180 & 0.59 & 200 \\
UBA7173 & 169 & 0.78 & 69.93 & 73 & 0.029 & 0.83 & 150 & 0.56 & 209 \\
Hungatella & 170 & 0.78 & 70.25 & 73 & 0.018 & 0.76 & 174 & 0.77 & 152 \\
UMGS1375 & 171 & 0.78 & 70.56 & 73 & 0.016 & 0.83 & 146 & 0.85 & 132 \\
Marseille.P3106 & 172 & 0.78 & 70.88 & 73 & 0.02 & 0.8 & 156 & 0.83 & 135 \\
CAG.115 & 173 & 0.78 & 71.19 & 72 & 0.027 & 0.62 & 224 & 0.61 & 194 \\
Pelethousia & 174 & 0.78 & 71.5 & 72 & 0.023 & 0.81 & 153 & 0.79 & 144 \\
CAG.353 & 175 & 0.78 & 71.81 & 72 & 0.023 & 0.69 & 201 & 0.64 & 180 \\
RUG115 & 176 & 0.77 & 72.12 & 73 & 0.02 & 0.75 & 178 & 0.76 & 154 \\
UBA6398 & 177 & 0.77 & 72.43 & 72 & 0.033 & 0.64 & 217 & 0.56 & 210 \\
CAG.245 & 178 & 0.77 & 72.74 & 73 & 0.02 & 0.82 & 151 & 0.79 & 143 \\
HGM10836 & 179 & 0.77 & 73.05 & 72 & 0.027 & 0.68 & 205 & 0.61 & 190 \\
Pseudoscilispira & 180 & 0.77 & 73.36 & 72 & 0.03 & 0.71 & 195 & 0.57 & 204 \\
Clostridium\_A & 181 & 0.77 & 73.67 & 71 & 0.029 & 0.68 & 204 & 0.63 & 186 \\
Enterenecus & 182 & 0.76 & 73.98 & 71 & 0.02 & 0.78 & 163 & 0.63 & 182 \\
Desulfovibrio & 183 & 0.76 & 74.28 & 72 & 0.023 & 0.73 & 187 & 0.45 & 232 \\
Enterobacter & 184 & 0.76 & 74.59 & 71 & 0.018 & 0.75 & 181 & 0.69 & 166 \\
Clostridium\_AP & 185 & 0.76 & 74.89 & 71 & 0.027 & 0.76 & 170 & 0.66 & 172 \\
Phascolarctobacterium & 186 & 0.76 & 75.2 & 70 & 0.023 & 0.69 & 202 & 0.65 & 176 \\
Faecalibacillus & 187 & 0.76 & 75.5 & 71 & 0.027 & 0.76 & 173 & 0.64 & 181 \\
Collinsella & 188 & 0.76 & 75.8 & 71 & 0.02 & 0.69 & 200 & 0.59 & 198 \\
Ruthenibacterium & 189 & 0.76 & 76.11 & 70 & 0.021 & 0.72 & 188 & 0.59 & 199 \\
Haemophilus\_D & 190 & 0.75 & 76.41 & 70 & 0.021 & 0.75 & 183 & 0.63 & 184 \\
Merdimorpha & 191 & 0.75 & 76.71 & 70 & 0.033 & 0.7 & 199 & 0.49 & 218 \\
Coprococcus & 192 & 0.75 & 77.01 & 70 & 0.019 & 0.79 & 159 & 0.63 & 183 \\
Brachyspira & 193 & 0.75 & 77.31 & 71 & 0.024 & 0.76 & 177 & 0.61 & 193 \\
UMGS1370 & 194 & 0.75 & 77.61 & 70 & 0.031 & 0.66 & 211 & 0.65 & 177 \\
Fimisoma & 195 & 0.75 & 77.91 & 69 & 0.018 & 0.71 & 191 & 0.58 & 202 \\
Anaerostipes & 196 & 0.74 & 78.21 & 69 & 0.018 & 0.62 & 222 & 0.79 & 142 \\
Eisenbergiella & 197 & 0.74 & 78.51 & 69 & 0.027 & 0.65 & 216 & 0.58 & 203 \\
Metalachnospira & 198 & 0.74 & 78.81 & 69 & 0.02 & 0.77 & 168 & 0.65 & 178 \\
Onthomonas & 199 & 0.73 & 79.1 & 69 & 0.027 & 0.66 & 210 & 0.59 & 201 \\
Scatocola & 200 & 0.73 & 79.4 & 69 & 0.029 & 0.64 & 218 & 0.56 & 208 \\ \hline
\end{tabular}}
\caption{Genera ranked 151--200 by Fractional Influence Score (FIS) under the default 90\% posterior selection criterion in the CRC cohort. The table reports genus-specific FIS rankings, cumulative FIS contribution, number of selected metabolites connected to each genus, median Bayesian posterior edge significance, and corresponding FIS/rank summaries obtained under stricter 95\% and 99\% posterior selection criteria.}
\label{tab:FIS_summary_table_151_to_200}
\end{table}

\begin{table}[!t]
\centering
\renewcommand{\arraystretch}{1.1}
\resizebox{0.9\textwidth}{!}{%
\begin{tabular}{lccccccccc}
\hline
Genus & \begin{tabular}[c]{@{}c@{}}FIS rank \\ (default; \\ 90\% CI \\ based)\end{tabular} & \begin{tabular}[c]{@{}c@{}}FIS \\ (default; \\ 90\% CI\\ based)\end{tabular} & \begin{tabular}[c]{@{}c@{}}Cumulative \\ FIS (\%)\end{tabular} & \begin{tabular}[c]{@{}c@{}}Num. \\ Selected \\ metabolites\end{tabular} & \begin{tabular}[c]{@{}c@{}}Median \\ Bayesian \\ p-value \\ (default; \\ 90\%  CI-based)\end{tabular} & \begin{tabular}[c]{@{}c@{}}FIS \\ (95\% \\ CI \\ based)\end{tabular} & \begin{tabular}[c]{@{}c@{}}Rank \\ (95\% \\ CI\\ based)\end{tabular} & \begin{tabular}[c]{@{}c@{}}FIS \\ (99\% \\ CI \\ based)\end{tabular} & \begin{tabular}[c]{@{}c@{}}Rank\\  (99\% \\ CI \\ based)\end{tabular} \\ \hline
Sodaliphilus & 201 & 0.73 & 79.69 & 69 & 0.031 & 0.62 & 220 & 0.61 & 192 \\
Peptostreptococcus & 202 & 0.73 & 79.99 & 68 & 0.026 & 0.71 & 194 & 0.6 & 197 \\
Veillonella\_A & 203 & 0.73 & 80.28 & 68 & 0.016 & 0.67 & 206 & 0.75 & 158 \\
Rikenella & 204 & 0.73 & 80.57 & 68 & 0.017 & 0.79 & 157 & 0.77 & 150 \\
Fusobacterium\_A & 205 & 0.73 & 80.86 & 68 & 0.018 & 0.7 & 197 & 0.61 & 196 \\
CAZU01 & 206 & 0.73 & 81.15 & 67 & 0.024 & 0.73 & 186 & 0.5 & 217 \\
Fusicatenibacter & 207 & 0.72 & 81.44 & 68 & 0.028 & 0.68 & 203 & 0.48 & 219 \\
CAG.485 & 208 & 0.72 & 81.73 & 67 & 0.02 & 0.66 & 209 & 0.47 & 225 \\
Niameybacter & 209 & 0.72 & 82.02 & 67 & 0.016 & 0.71 & 192 & 0.65 & 174 \\
Adlercreutzia & 210 & 0.72 & 82.31 & 67 & 0.02 & 0.76 & 169 & 0.76 & 157 \\
UBA5026 & 211 & 0.72 & 82.6 & 67 & 0.024 & 0.7 & 198 & 0.51 & 215 \\
Senegalimassilia & 212 & 0.71 & 82.89 & 67 & 0.02 & 0.71 & 190 & 0.65 & 175 \\
Frisingicoccus & 213 & 0.71 & 83.17 & 66 & 0.031 & 0.67 & 207 & 0.46 & 228 \\
Akkermansia & 214 & 0.7 & 83.45 & 66 & 0.011 & 0.75 & 182 & 0.86 & 130 \\
Duodenibacillus & 215 & 0.7 & 83.74 & 66 & 0.027 & 0.65 & 213 & 0.52 & 214 \\
AM51.8 & 216 & 0.7 & 84.02 & 65 & 0.022 & 0.66 & 212 & 0.67 & 168 \\
Eubacterium\_G & 217 & 0.69 & 84.3 & 65 & 0.022 & 0.7 & 196 & 0.64 & 179 \\
CAG.127 & 218 & 0.69 & 84.57 & 65 & 0.038 & 0.58 & 232 & 0.41 & 242 \\
Roseburia & 219 & 0.69 & 84.85 & 65 & 0.031 & 0.65 & 214 & 0.42 & 238 \\
Enterococcus\_B & 220 & 0.69 & 85.12 & 64 & 0.024 & 0.71 & 193 & 0.44 & 234 \\
CAG.217 & 221 & 0.69 & 85.4 & 64 & 0.034 & 0.6 & 227 & 0.57 & 207 \\
Aeromonas & 222 & 0.69 & 85.68 & 64 & 0.032 & 0.65 & 215 & 0.48 & 223 \\
Merdimonas & 223 & 0.68 & 85.95 & 64 & 0.024 & 0.61 & 226 & 0.62 & 188 \\
Citrobacter\_B & 224 & 0.68 & 86.22 & 64 & 0.031 & 0.63 & 219 & 0.46 & 229 \\
Paraprevotella & 225 & 0.67 & 86.49 & 63 & 0.031 & 0.53 & 246 & 0.48 & 221 \\
Ruminococcus\_B & 226 & 0.67 & 86.76 & 63 & 0.033 & 0.55 & 240 & 0.51 & 216 \\
Ventrisoma & 227 & 0.67 & 87.03 & 62 & 0.033 & 0.54 & 244 & 0.47 & 224 \\
Mediterraneibacter & 228 & 0.66 & 87.3 & 62 & 0.024 & 0.67 & 208 & 0.57 & 206 \\
CAG.873 & 229 & 0.66 & 87.56 & 62 & 0.021 & 0.58 & 231 & 0.62 & 187 \\
Scatavimonas & 230 & 0.66 & 87.83 & 62 & 0.031 & 0.59 & 229 & 0.41 & 241 \\
Ruminococcus & 231 & 0.66 & 88.1 & 62 & 0.038 & 0.6 & 228 & 0.41 & 240 \\
CAG.273 & 232 & 0.65 & 88.36 & 62 & 0.031 & 0.56 & 238 & 0.44 & 236 \\
Spyradocola & 233 & 0.65 & 88.62 & 61 & 0.042 & 0.54 & 245 & 0.3 & 261 \\
Fimivivens & 234 & 0.65 & 88.88 & 61 & 0.033 & 0.57 & 236 & 0.38 & 245 \\
CAG.349 & 235 & 0.65 & 89.14 & 61 & 0.022 & 0.62 & 223 & 0.77 & 153 \\
UMGS1975 & 236 & 0.64 & 89.4 & 60 & 0.032 & 0.58 & 230 & 0.43 & 237 \\
Butyricimonas & 237 & 0.64 & 89.66 & 60 & 0.033 & 0.55 & 241 & 0.44 & 233 \\
Dysosmobacter & 238 & 0.64 & 89.92 & 60 & 0.033 & 0.56 & 237 & 0.32 & 256 \\
Pullichristensenella & 239 & 0.63 & 90.17 & 59 & 0.04 & 0.49 & 256 & 0.46 & 231 \\
Alcanivorax & 240 & 0.63 & 90.42 & 59 & 0.038 & 0.49 & 254 & 0.3 & 259 \\
Marvinbryantia & 241 & 0.63 & 90.67 & 58 & 0.027 & 0.61 & 225 & 0.46 & 230 \\
Faecimonas & 242 & 0.62 & 90.93 & 58 & 0.031 & 0.62 & 221 & 0.34 & 255 \\
HGM13006 & 243 & 0.62 & 91.18 & 58 & 0.037 & 0.54 & 242 & 0.38 & 246 \\
UMGS946 & 244 & 0.61 & 91.42 & 58 & 0.024 & 0.57 & 233 & 0.55 & 211 \\
COE1 & 245 & 0.61 & 91.67 & 57 & 0.027 & 0.55 & 239 & 0.47 & 227 \\
CAG.590 & 246 & 0.6 & 91.91 & 56 & 0.036 & 0.54 & 243 & 0.25 & 270 \\
Dorea\_A & 247 & 0.6 & 92.15 & 56 & 0.034 & 0.51 & 251 & 0.28 & 264 \\
Erysipelatoclostridium & 248 & 0.6 & 92.39 & 56 & 0.036 & 0.45 & 264 & 0.38 & 247 \\
Caecibacter & 249 & 0.58 & 92.62 & 55 & 0.04 & 0.44 & 265 & 0.38 & 244 \\
CAG.103 & 250 & 0.58 & 92.86 & 54 & 0.037 & 0.51 & 249 & 0.53 & 213 \\ \hline
\end{tabular}}
\caption{Genera ranked 201--250 by Fractional Influence Score (FIS) under the default 90\% posterior selection criterion in the CRC cohort. The table reports genus-specific FIS rankings, cumulative FIS contribution, number of selected metabolites connected to each genus, median Bayesian posterior edge significance, and corresponding FIS/rank summaries obtained under stricter 95\% and 99\% posterior selection criteria.}
\label{tab:FIS_summary_table_201_to_250}
\end{table}

\begin{table}[!t]
\centering
\renewcommand{\arraystretch}{1.1}
\resizebox{0.9\textwidth}{!}{%
\begin{tabular}{lccccccccc}
\hline
Genus & \begin{tabular}[c]{@{}c@{}}FIS rank \\ (default; \\ 90\% CI \\ based)\end{tabular} & \begin{tabular}[c]{@{}c@{}}FIS \\ (default; \\ 90\% CI\\ based)\end{tabular} & \begin{tabular}[c]{@{}c@{}}Cumulative \\ FIS (\%)\end{tabular} & \begin{tabular}[c]{@{}c@{}}Num. \\ Selected \\ metabolites\end{tabular} & \begin{tabular}[c]{@{}c@{}}Median \\ Bayesian \\ p-value \\ (default; \\ 90\%  CI-based)\end{tabular} & \begin{tabular}[c]{@{}c@{}}FIS \\ (95\% \\ CI \\ based)\end{tabular} & \begin{tabular}[c]{@{}c@{}}Rank \\ (95\% \\ CI\\ based)\end{tabular} & \begin{tabular}[c]{@{}c@{}}FIS \\ (99\% \\ CI \\ based)\end{tabular} & \begin{tabular}[c]{@{}c@{}}Rank\\  (99\% \\ CI \\ based)\end{tabular} \\ \hline
Ruminococcus\_E & 251 & 0.58 & 93.09 & 54 & 0.029 & 0.57 & 235 & 0.35 & 253 \\
CAG.238 & 252 & 0.58 & 93.32 & 53 & 0.036 & 0.48 & 258 & 0.36 & 250 \\
Holdemanella & 253 & 0.57 & 93.55 & 54 & 0.024 & 0.51 & 252 & 0.37 & 248 \\
Allisonella & 254 & 0.57 & 93.78 & 54 & 0.039 & 0.44 & 266 & 0.24 & 273 \\
Citrobacter & 255 & 0.57 & 94.01 & 53 & 0.038 & 0.48 & 260 & 0.35 & 251 \\
Bifidobacterium & 256 & 0.57 & 94.24 & 53 & 0.033 & 0.53 & 247 & 0.42 & 239 \\
Klebsiella & 257 & 0.56 & 94.46 & 53 & 0.042 & 0.43 & 269 & 0.2 & 276 \\
Acidaminococcus & 258 & 0.56 & 94.69 & 53 & 0.018 & 0.57 & 234 & 0.44 & 235 \\
SZUA.378 & 259 & 0.56 & 94.91 & 52 & 0.032 & 0.51 & 250 & 0.39 & 243 \\
SFEL01 & 260 & 0.55 & 95.13 & 51 & 0.036 & 0.45 & 263 & 0.2 & 275 \\
UBA11774 & 261 & 0.54 & 95.35 & 51 & 0.049 & 0.36 & 276 & 0.31 & 258 \\
CAG.302 & 262 & 0.54 & 95.57 & 50 & 0.038 & 0.42 & 270 & 0.26 & 266 \\
Blautia & 263 & 0.54 & 95.78 & 50 & 0.036 & 0.45 & 262 & 0.21 & 274 \\
Azonexus & 264 & 0.53 & 95.99 & 49 & 0.033 & 0.52 & 248 & 0.29 & 262 \\
Oribacterium & 265 & 0.53 & 96.21 & 50 & 0.034 & 0.48 & 257 & 0.26 & 268 \\
UBA1394 & 266 & 0.52 & 96.42 & 49 & 0.024 & 0.5 & 253 & 0.29 & 263 \\
UBA1685 & 267 & 0.51 & 96.62 & 48 & 0.03 & 0.4 & 274 & 0.35 & 252 \\
Scatacola\_A & 268 & 0.51 & 96.83 & 48 & 0.027 & 0.49 & 255 & 0.25 & 272 \\
Parasutterella & 269 & 0.51 & 97.03 & 48 & 0.033 & 0.43 & 267 & 0.34 & 254 \\
Massilistercora & 270 & 0.5 & 97.23 & 47 & 0.036 & 0.47 & 261 & 0.17 & 278 \\
Ruminococcus\_C & 271 & 0.5 & 97.44 & 47 & 0.04 & 0.38 & 275 & 0.16 & 279 \\
Phocaeicola & 272 & 0.5 & 97.64 & 46 & 0.034 & 0.48 & 259 & 0.25 & 271 \\
RUG705 & 273 & 0.49 & 97.83 & 46 & 0.059 & 0.29 & 281 & 0.26 & 267 \\
KLE1615 & 274 & 0.47 & 98.02 & 44 & 0.027 & 0.42 & 271 & 0.32 & 257 \\
Chitinophaga & 275 & 0.46 & 98.21 & 43 & 0.031 & 0.43 & 268 & 0.3 & 260 \\
Ruminiclostridium\_E & 276 & 0.45 & 98.39 & 42 & 0.034 & 0.4 & 272 & 0.28 & 265 \\
Limivicinus & 277 & 0.43 & 98.56 & 40 & 0.039 & 0.35 & 277 & 0.11 & 286 \\
Pseudoflavonifractor & 278 & 0.42 & 98.73 & 39 & 0.051 & 0.27 & 283 & 0.26 & 269 \\
GCA.900066755 & 279 & 0.41 & 98.89 & 38 & 0.032 & 0.33 & 279 & 0.16 & 281 \\
Escherichia & 280 & 0.41 & 99.06 & 38 & 0.032 & 0.4 & 273 & 0.37 & 249 \\
Anaeroglobus & 281 & 0.4 & 99.22 & 37 & 0.04 & 0.34 & 278 & 0.11 & 285 \\
Ventricola & 282 & 0.4 & 99.38 & 37 & 0.038 & 0.33 & 280 & 0.2 & 277 \\
Fournierella & 283 & 0.4 & 99.54 & 37 & 0.053 & 0.23 & 285 & 0.16 & 282 \\
Butyrivibrio\_A & 284 & 0.35 & 99.68 & 32 & 0.034 & 0.28 & 282 & 0.16 & 280 \\
Coprococcus\_A & 285 & 0.34 & 99.81 & 31 & 0.04 & 0.27 & 284 & 0.13 & 283 \\
Kluyvera & 286 & 0.27 & 99.92 & 25 & 0.055 & 0.17 & 286 & 0.1 & 287 \\
Alloprevotella & 287 & 0.2 & 100 & 19 & 0.033 & 0.17 & 287 & 0.12 & 284 \\ \hline
\end{tabular}}
\caption{Genera ranked 251--287 by Fractional Influence Score (FIS) under the default 90\% posterior selection criterion in the CRC cohort. The table reports genus-specific FIS rankings, cumulative FIS contribution, number of selected metabolites connected to each genus, median Bayesian posterior edge significance, and corresponding FIS/rank summaries obtained under stricter 95\% and 99\% posterior selection criteria.}
\label{tab:FIS_summary_table_251_to_287}
\end{table}

\clearpage
\subsection{Case Study Inference}

The B-MASTER analysis provides detailed insights into the associations between microbial
genera and metabolites in two biologically relevant subsets. For clarity, we describe each
finding with explicit direction and Bayesian $p$-values, as summarized in
Tables~\ref{tab:pvalues_subset1} and \ref{tab:pvalues_subset2}. Positive associations are
shown in red, negative associations in blue, and non-significant effects ($p>0.05$) are
denoted by ``$-$''.

For the abundant metabolite subset (Subset~1), several genera show broad regulatory
influence. \textit{Veillonella} demonstrates consistent and statistically significant
associations across multiple metabolites. Specifically, it is negatively associated with
propionate ($p=0.033$, negative), butyrate ($p=0.016$, negative), glutamate ($p<0.001$,
negative), 5-aminopentanoate ($p=0.004$, negative), lysine ($p<0.001$, negative), and
alanine ($p=0.036$, negative). In contrast, it exhibits positive associations with
dihydrouracil ($p=0.002$, positive) and urea ($p=0.013$, positive). These results indicate
a mixed regulatory profile in which \textit{Veillonella} simultaneously promotes nitrogen
metabolism while suppressing several amino acid pathways.

\begin{table}[ht]
\centering
\resizebox{\textwidth}{!}{%
\begin{tabular}{lcccccccccc}
\hline
 & Propionate & Butyrate & Dihydrouracil & Glutamate & Urea & Succinate & 5-Aminopentanoate & Valerate & Lysine & Alanine \\ \hline
Veillonella      & \textcolor{blue}{$0.033$} & \textcolor{blue}{$0.016$} & \textcolor{red}{$0.002$} & \textcolor{blue}{$<0.001$} & \textcolor{red}{$0.013$} & $-$ & \textcolor{blue}{$0.004$} & $-$ & \textcolor{blue}{$<0.001$} & \textcolor{blue}{$0.036$} \\
Oliverpabstia    & $-$ & \textcolor{blue}{$0.031$} & \textcolor{red}{$0.011$} & \textcolor{blue}{$<0.001$} & \textcolor{red}{$<0.001$} & $-$ & \textcolor{red}{$<0.001$} & $-$ & \textcolor{blue}{$0.004$} & \textcolor{red}{$0.002$} \\
Merdicola        & $-$ & \textcolor{blue}{$<0.001$} & \textcolor{blue}{$0.042$} & \textcolor{blue}{$<0.001$} & \textcolor{red}{$0.038$} & \textcolor{red}{$0.022$} & $-$ & \textcolor{blue}{$<0.001$} & \textcolor{red}{$0.011$} & $-$ \\
Sellimonas       & \textcolor{red}{$<0.001$} & \textcolor{red}{$0.013$} & \textcolor{red}{$0.002$} & $-$ & $-$ & $-$ & \textcolor{red}{$0.013$} & \textcolor{red}{$<0.001$} & $-$ & \textcolor{blue}{$0.002$} \\
Avimicrobium     & $-$ & $-$ & \textcolor{red}{$0.002$} & \textcolor{red}{$<0.001$} & $-$ & \textcolor{blue}{$0.024$} & \textcolor{red}{$0.002$} & $-$ & \textcolor{red}{$<0.001$} & \textcolor{red}{$0.007$} \\
UBA4372          & \textcolor{blue}{$0.020$} & \textcolor{blue}{$<0.001$} & $-$ & \textcolor{blue}{$0.002$} & \textcolor{red}{$0.009$} & \textcolor{red}{$0.011$} & $-$ & \textcolor{blue}{$<0.001$} & $-$ & $-$ \\
Enterocloster    & \textcolor{red}{$0.002$} & $-$ & \textcolor{blue}{$<0.001$} & $-$ & \textcolor{blue}{$0.022$} & \textcolor{blue}{$0.011$} & $-$ & $-$ & \textcolor{blue}{$0.018$} & \textcolor{red}{$0.007$} \\
Mogibacterium    & \textcolor{red}{$0.007$} & $-$ & $-$ & \textcolor{red}{$<0.001$} & \textcolor{blue}{$<0.001$} & \textcolor{blue}{$0.047$} & $-$ & \textcolor{red}{$<0.001$} & $-$ & \textcolor{blue}{$0.011$} \\
Lancefieldella   & $-$ & $-$ & \textcolor{blue}{$0.018$} & \textcolor{red}{$<0.001$} & $-$ & $-$ & \textcolor{red}{$<0.001$} & \textcolor{red}{$0.049$} & \textcolor{red}{$<0.001$} & \textcolor{red}{$0.002$} \\
Turicibacter     & \textcolor{blue}{$0.024$} & $-$ & \textcolor{blue}{$<0.001$} & $-$ & $-$ & \textcolor{blue}{$0.038$} & $-$ & \textcolor{blue}{$0.018$} & \textcolor{red}{$<0.001$} & \textcolor{blue}{$<0.001$} \\
Intestinibacter  & \textcolor{blue}{$0.031$} & \textcolor{blue}{$0.004$} & \textcolor{red}{$0.002$} & \textcolor{blue}{$0.002$} & \textcolor{red}{$0.033$} & $-$ & $-$ & $-$ & \textcolor{blue}{$0.011$} & $-$ \\
Limiplasma       & \textcolor{red}{$0.002$} & \textcolor{red}{$0.002$} & \textcolor{red}{$0.040$} & $-$ & \textcolor{blue}{$0.024$} & \textcolor{blue}{$0.007$} & $-$ & $-$ & $-$ & \textcolor{red}{$0.036$} \\
Anaerotignum     & $-$ & $-$ & $-$ & \textcolor{blue}{$0.047$} & \textcolor{blue}{$0.016$} & \textcolor{blue}{$0.049$} & \textcolor{blue}{$0.004$} & $-$ & \textcolor{blue}{$0.002$} & \textcolor{blue}{$<0.001$} \\
Pauljensenia     & $-$ & \textcolor{blue}{$0.049$} & \textcolor{red}{$0.024$} & \textcolor{blue}{$0.036$} & $-$ & $-$ & \textcolor{blue}{$0.002$} & \textcolor{blue}{$<0.001$} & $-$ & \textcolor{blue}{$0.009$} \\
An92             & \textcolor{red}{$0.002$} & \textcolor{red}{$0.042$} & $-$ & $-$ & \textcolor{red}{$0.002$} & \textcolor{red}{$0.042$} & $-$ & $-$ & \textcolor{red}{$0.027$} & \textcolor{blue}{$0.009$} \\ \hline
\end{tabular}}
\caption{Bayesian $p$-values for associations between genera (rows) and metabolites (columns) in subset 1 (most abundant metabolites). Values in \textcolor{red}{red} indicate positive associations, values in \textcolor{blue}{blue} indicate negative associations, and “$-$” denotes $p>0.05$ or missing. Values equal to $0$ are displayed as “$<0.001$”.}
\label{tab:pvalues_subset1}
\end{table}

\begin{figure}[]
\centering
\includegraphics[width=.42\linewidth]{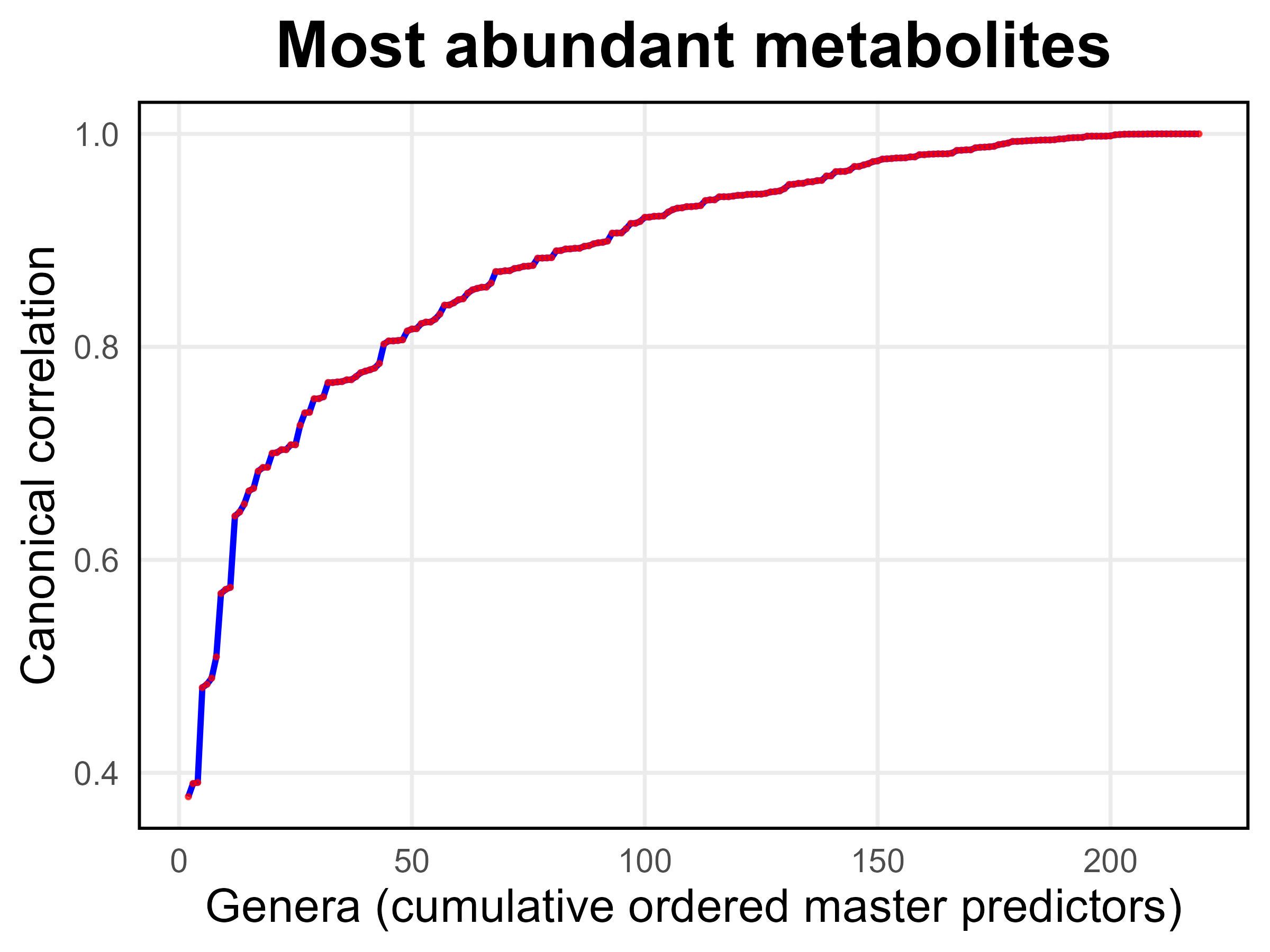}
\hspace{0.04\linewidth}
\includegraphics[width=.42\linewidth]{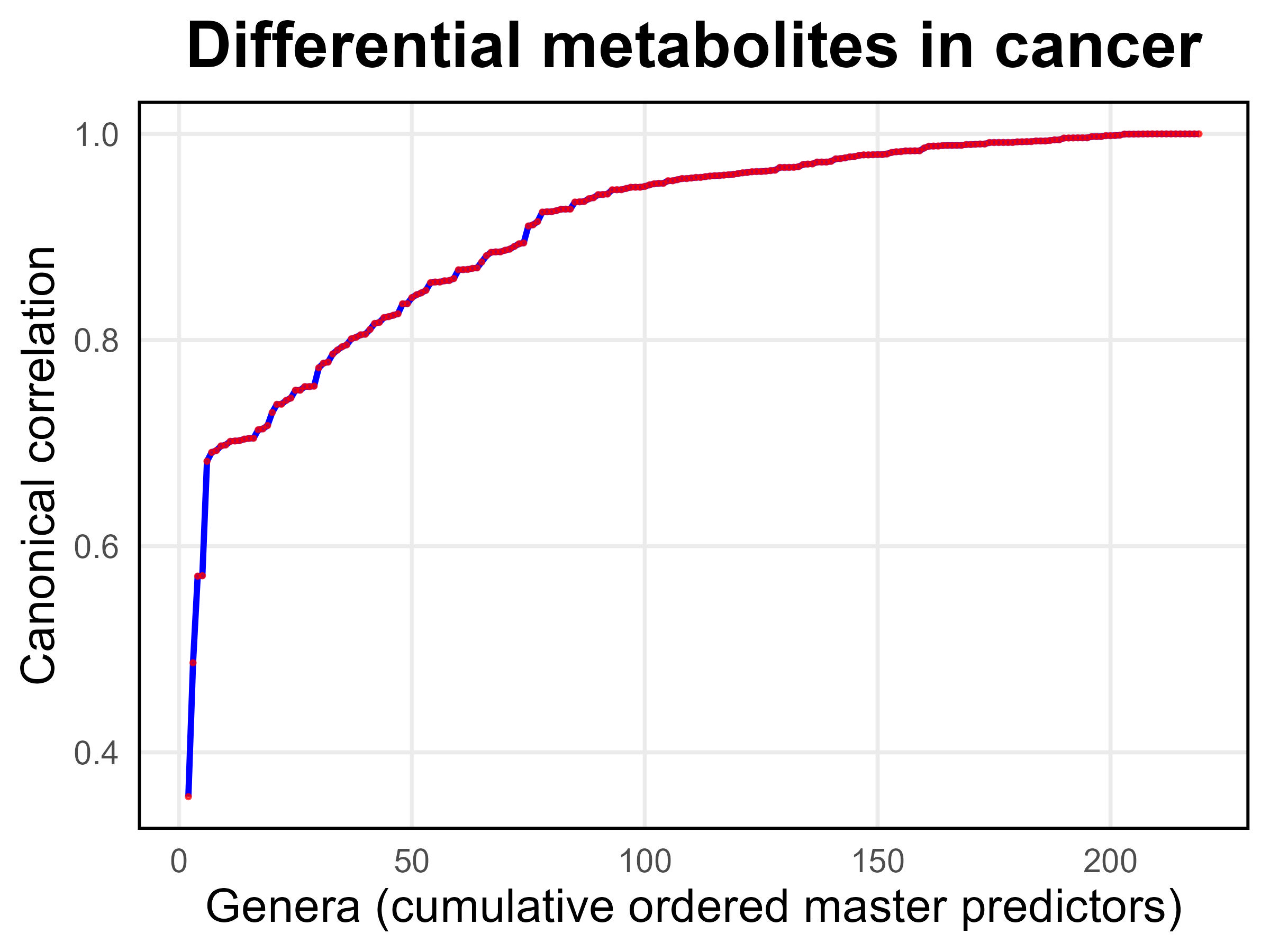} 
\caption{The canonical correlation between metabolites in Subset 1 (left) and Subset 2 (right) and the cumulatively included master predictors, arranged by their rank order, is visually depicted. The ranking of the predictors based on their prevalence is derived separately for each corresponding subset of metabolites.}
\label{CumCCA}
\end{figure}

\textit{Oliverpabstia} is negatively associated with butyrate ($p=0.031$, negative),
glutamate ($p<0.001$, negative), and lysine ($p=0.004$, negative), while positively
associated with dihydrouracil ($p=0.011$, positive), urea ($p<0.001$, positive), and
5-aminopentanoate ($p<0.001$, positive). These results highlight both SCFA and amino
acid regulation. \textit{Merdicola} shows negative associations with butyrate ($p<0.001$,
negative), dihydrouracil ($p=0.042$, negative), glutamate ($p<0.001$, negative), and
valerate ($p<0.001$, negative), while exhibiting positive links with urea ($p=0.038$,
positive), succinate ($p=0.022$, positive), and lysine ($p=0.011$, positive). Taken
together, \textit{Merdicola} appears to shift metabolism toward nitrogenous compounds at
the expense of SCFAs.

\textit{Sellimonas} is exclusively positively associated: propionate ($p<0.001$, positive),
butyrate ($p=0.013$, positive), dihydrouracil ($p=0.002$, positive), 5-aminopentanoate
($p=0.013$, positive), and valerate ($p<0.001$, positive), although it is negatively
associated with alanine ($p=0.002$, negative). \textit{Avimicrobium} demonstrates positive
effects on dihydrouracil ($p=0.002$, positive), glutamate ($p<0.001$, positive),
5-aminopentanoate ($p=0.002$, positive), lysine ($p<0.001$, positive), and alanine
($p=0.007$, positive), while exerting a negative effect on succinate ($p=0.024$, negative).

\textit{UBA4372} exhibits negative associations with propionate ($p=0.020$, negative),
butyrate ($p<0.001$, negative), glutamate ($p=0.002$, negative), and valerate
($p<0.001$, negative), while positively associated with urea ($p=0.009$, positive) and
succinate ($p=0.011$, positive). Similarly, \textit{Enterocloster} shows positive
associations with propionate ($p=0.002$, positive) and alanine ($p=0.007$, positive),
but negative associations with dihydrouracil ($p<0.001$, negative), urea ($p=0.022$,
negative), succinate ($p=0.011$, negative), and lysine ($p=0.018$, negative). 
\textit{Mogibacterium} is positively associated with propionate ($p=0.007$, positive),
glutamate ($p<0.001$, positive), and valerate ($p<0.001$, positive), while negatively
associated with urea ($p<0.001$, negative), succinate ($p=0.047$, negative), and alanine
($p=0.011$, negative).

\textit{Lancefieldella} contributes strong positive associations with glutamate
($p<0.001$, positive), 5-aminopentanoate ($p<0.001$, positive), valerate ($p=0.049$,
positive), lysine ($p<0.001$, positive), and alanine ($p=0.002$, positive), along with a
negative association with dihydrouracil ($p=0.018$, negative). \textit{Turicibacter}
exhibits predominantly negative associations: propionate ($p=0.024$, negative),
dihydrouracil ($p<0.001$, negative), succinate ($p=0.038$, negative), valerate
($p=0.018$, negative), and alanine ($p<0.001$, negative), but a positive association with
lysine ($p<0.001$, positive). \textit{Intestinibacter} shows negative associations with
propionate ($p=0.031$, negative), butyrate ($p=0.004$, negative), glutamate
($p=0.002$, negative), and lysine ($p=0.011$, negative), while positively associated with
dihydrouracil ($p=0.002$, positive) and urea ($p=0.033$, positive). \textit{Limiplasma}
is positively associated with propionate ($p=0.002$, positive), butyrate ($p=0.002$,
positive), dihydrouracil ($p=0.040$, positive), and alanine ($p=0.036$, positive), while
negatively associated with urea ($p=0.024$, negative) and succinate ($p=0.007$,
negative).

\textit{Anaerotignum} shows exclusively negative associations: glutamate ($p=0.047$,
negative), urea ($p=0.016$, negative), succinate ($p=0.049$, negative),
5-aminopentanoate ($p=0.004$, negative), lysine ($p=0.002$, negative), and alanine
($p<0.001$, negative). \textit{Pauljensenia} demonstrates mixed effects, including
negative associations with butyrate ($p=0.049$, negative), glutamate ($p=0.036$,
negative), 5-aminopentanoate ($p=0.002$, negative), valerate ($p<0.001$, negative), and
alanine ($p=0.009$, negative), but a positive association with dihydrouracil ($p=0.024$,
positive). Finally, \textit{An92} is positively associated with propionate ($p=0.002$,
positive), butyrate ($p=0.042$, positive), urea ($p=0.002$, positive), succinate
($p=0.042$, positive), and lysine ($p=0.027$, positive), while negatively associated with
alanine ($p=0.009$, negative).

\begin{table}[ht]
\centering
\resizebox{\textwidth}{!}{%
\begin{tabular}{lccccccccccc}
\hline
 & $X\_DCA$ & Glycocholate & Taurocholate & Isovalerate & L-Isoleucine & L-Leucine & L-Valine & L-Phenylalanine & L-Tyrosine & L-Serine & Glycine \\ \hline
Gemmiger & \textcolor{blue}{$<0.001$} & \textcolor{red}{$<0.001$} & \textcolor{red}{$<0.001$} & $-$ & \textcolor{red}{$<0.001$} & \textcolor{red}{$<0.001$} & \textcolor{red}{$<0.001$} & \textcolor{red}{$<0.001$} & \textcolor{red}{$<0.001$} & \textcolor{red}{$<0.001$} & \textcolor{red}{$<0.001$} \\
Ligilactobacillus & $-$ & \textcolor{red}{$0.002$} & \textcolor{red}{$0.022$} & \textcolor{red}{$0.011$} & \textcolor{red}{$<0.001$} & \textcolor{red}{$<0.001$} & \textcolor{red}{$<0.001$} & \textcolor{red}{$<0.001$} & \textcolor{red}{$<0.001$} & \textcolor{red}{$<0.001$} & \textcolor{red}{$<0.001$} \\
Faecalimonas & $-$ & \textcolor{red}{$0.009$} & \textcolor{red}{$0.009$} & \textcolor{red}{$<0.001$} & \textcolor{red}{$<0.001$} & \textcolor{red}{$<0.001$} & \textcolor{red}{$<0.001$} & \textcolor{red}{$0.002$} & \textcolor{red}{$0.024$} & \textcolor{red}{$0.009$} & \textcolor{red}{$<0.001$} \\
Bacteroides & $-$ & \textcolor{blue}{$0.040$} & \textcolor{blue}{$0.027$} & $-$ & \textcolor{blue}{$<0.001$} & \textcolor{blue}{$<0.001$} & \textcolor{blue}{$<0.001$} & \textcolor{blue}{$<0.001$} & \textcolor{blue}{$<0.001$} & \textcolor{blue}{$0.007$} & \textcolor{blue}{$0.002$} \\
Agathobacter & \textcolor{red}{$0.013$} & \textcolor{red}{$<0.001$} & $-$ & \textcolor{red}{$0.031$} & \textcolor{red}{$<0.001$} & \textcolor{red}{$<0.001$} & \textcolor{red}{$<0.001$} & \textcolor{red}{$<0.001$} & \textcolor{red}{$0.002$} & $-$ & \textcolor{red}{$0.036$} \\
Caccousia & $-$ & $-$ & \textcolor{red}{$0.002$} & \textcolor{red}{$0.002$} & \textcolor{red}{$<0.001$} & \textcolor{red}{$<0.001$} & \textcolor{red}{$<0.001$} & $-$ & \textcolor{red}{$0.002$} & \textcolor{red}{$<0.001$} & \textcolor{red}{$<0.001$} \\
CAG.83 & \textcolor{blue}{$0.002$} & $-$ & $-$ & \textcolor{blue}{$0.009$} & \textcolor{blue}{$0.024$} & \textcolor{blue}{$<0.001$} & \textcolor{blue}{$0.002$} & \textcolor{blue}{$<0.001$} & \textcolor{blue}{$<0.001$} & $-$ & \textcolor{blue}{$0.002$} \\
Anaerotignum & $-$ & $-$ & \textcolor{blue}{$0.002$} & \textcolor{blue}{$<0.001$} & \textcolor{blue}{$<0.001$} & \textcolor{blue}{$0.004$} & \textcolor{blue}{$<0.001$} & \textcolor{blue}{$0.020$} & $-$ & \textcolor{blue}{$<0.001$} & \textcolor{blue}{$0.049$} \\
Flavonifractor & $-$ & \textcolor{red}{$<0.001$} & \textcolor{red}{$0.004$} & $-$ & \textcolor{blue}{$0.036$} & \textcolor{blue}{$0.031$} & \textcolor{blue}{$0.009$} & \textcolor{blue}{$<0.001$} & \textcolor{blue}{$<0.001$} & \textcolor{red}{$0.020$} & $-$ \\
Amedibacterium & $-$ & \textcolor{blue}{$0.020$} & $-$ & \textcolor{blue}{$0.020$} & \textcolor{blue}{$<0.001$} & \textcolor{blue}{$<0.001$} & \textcolor{blue}{$<0.001$} & \textcolor{blue}{$0.004$} & $-$ & \textcolor{blue}{$0.044$} & \textcolor{blue}{$0.016$} \\
1XD42.69 & $-$ & $-$ & \textcolor{blue}{$0.013$} & $-$ & \textcolor{blue}{$<0.001$} & \textcolor{blue}{$0.002$} & \textcolor{blue}{$0.002$} & \textcolor{blue}{$<0.001$} & \textcolor{blue}{$0.002$} & \textcolor{blue}{$<0.001$} & $-$ \\
Evtepia & $-$ & $-$ & $-$ & \textcolor{red}{$<0.001$} & \textcolor{red}{$0.009$} & \textcolor{red}{$0.011$} & \textcolor{red}{$0.007$} & \textcolor{red}{$<0.001$} & \textcolor{red}{$<0.001$} & \textcolor{red}{$0.004$} & $-$ \\
Choladocola & $-$ & \textcolor{blue}{$0.016$} & $-$ & \textcolor{red}{$<0.001$} & \textcolor{blue}{$0.004$} & \textcolor{blue}{$0.002$} & \textcolor{blue}{$<0.001$} & \textcolor{blue}{$<0.001$} & \textcolor{blue}{$0.011$} & $-$ & $-$ \\
Lachnospira & $-$ & \textcolor{red}{$0.016$} & $-$ & $-$ & \textcolor{blue}{$<0.001$} & \textcolor{blue}{$<0.001$} & \textcolor{blue}{$<0.001$} & \textcolor{blue}{$0.002$} & \textcolor{blue}{$<0.001$} & \textcolor{blue}{$0.042$} & $-$ \\
Merdicola & $-$ & \textcolor{red}{$<0.001$} & \textcolor{red}{$0.007$} & $-$ & \textcolor{red}{$0.002$} & \textcolor{red}{$<0.001$} & \textcolor{red}{$0.007$} & \textcolor{red}{$<0.001$} & \textcolor{red}{$0.047$} & $-$ & $-$ \\ \hline
\end{tabular}}
\caption{Bayesian $p$-values for associations between top 15 genera (rows) and metabolites (columns) in subset~2 (differentially abundant metabolites in cancer vs control). Values in \textcolor{red}{red} indicate positive associations, values in \textcolor{blue}{blue} indicate negative associations, and ``$-$'' denotes $p>0.05$. Values $<0.001$ are displayed as ``$<0.001$''.}
\label{tab:pvalues_subset2}
\end{table}

For the differentially abundant metabolite subset (Subset~2), the pattern is more
systemic, with many genera influencing nearly all metabolites. \textit{Gemmiger} exhibits
negative association with $X_{\mathrm{DCA}}$ ($p<0.001$, negative) but positive associations
with glycocholate ($p<0.001$, positive), taurocholate ($p<0.001$, positive),
L-isoleucine ($p<0.001$, positive), L-leucine ($p<0.001$, positive), L-valine ($p<0.001$,
positive), L-phenylalanine ($p<0.001$, positive), L-tyrosine ($p<0.001$, positive),
L-serine ($p<0.001$, positive), and glycine ($p<0.001$, positive). \textit{Ligilactobacillus}
shows uniformly positive associations, including glycocholate ($p=0.002$, positive),
taurocholate ($p=0.022$, positive), isovalerate ($p=0.011$, positive), L-isoleucine
($p<0.001$, positive), L-leucine ($p<0.001$, positive), and all other amino acids
($p<0.001$, positive).

\textit{Faecalimonas} is positively associated with glycocholate ($p=0.009$, positive),
taurocholate ($p=0.009$, positive), isovalerate ($p<0.001$, positive), and all seven amino
acids ($p\leq0.024$, positive). In contrast, \textit{Bacteroides} exhibits exclusively
negative associations, including glycocholate ($p=0.040$, negative), taurocholate
($p=0.027$, negative), and all seven amino acids ($p\leq0.007$, negative). \textit{Agathobacter}
is uniformly positive, including $X_{\mathrm{DCA}}$ ($p=0.013$, positive), glycocholate
($p<0.001$, positive), isovalerate ($p=0.031$, positive), and multiple amino acids
($p\leq0.036$, positive). Similarly, \textit{Caccousia} exhibits positive associations with
taurocholate ($p=0.002$, positive), isovalerate ($p=0.002$, positive), L-isoleucine
($p<0.001$, positive), and all subsequent amino acids ($p<0.001$, positive).

\textit{CAG.83} demonstrates consistent negative associations, including $X\_DCA$
($p=0.002$, negative), isovalerate ($p=0.009$, negative), L-isoleucine ($p=0.024$,
negative), L-leucine ($p<0.001$, negative), L-valine ($p=0.002$, negative),
L-phenylalanine ($p<0.001$, negative), L-tyrosine ($p<0.001$, negative), and glycine
($p=0.002$, negative). \textit{Anaerotignum} similarly shows broad negative associations,
including taurocholate ($p=0.002$, negative), isovalerate ($p<0.001$, negative),
L-isoleucine ($p<0.001$, negative), L-leucine ($p=0.004$, negative), L-valine ($p<0.001$,
negative), L-phenylalanine ($p=0.020$, negative), L-serine ($p<0.001$, negative), and
glycine ($p=0.049$, negative).

\textit{Flavonifractor} shows mixed directionality, with positive associations for
glycocholate ($p<0.001$, positive), taurocholate ($p=0.004$, positive), and L-serine
($p=0.020$, positive), but negative associations for L-isoleucine ($p=0.036$, negative),
L-leucine ($p=0.031$, negative), L-valine ($p=0.009$, negative), L-phenylalanine
($p<0.001$, negative), and L-tyrosine ($p<0.001$, negative). Other genera with uniform
negative associations include \textit{Amedibacterium} (glycocholate $p=0.020$, negative;
isovalerate $p=0.020$, negative; and several amino acids $p\leq0.044$, negative) and
\textit{1XD42.69} (taurocholate $p=0.013$, negative; multiple amino acids $p\leq0.002$,
negative). \textit{Evtepia} is strongly positive across metabolites, including isovalerate
($p<0.001$, positive), L-isoleucine ($p=0.009$, positive), L-leucine ($p=0.011$,
positive), L-valine ($p=0.007$, positive), L-phenylalanine ($p<0.001$, positive),
L-tyrosine ($p<0.001$, positive), and L-serine ($p=0.004$, positive).

\textit{Choladocola} exhibits a predominantly negative profile, including glycocholate ($p=0.016$, negative), L-isoleucine ($p=0.004$, negative), L-leucine ($p=0.002$, negative), L-valine ($p<0.001$, negative), L-phenylalanine ($p<0.001$, negative), and L-tyrosine ($p=0.011$, negative), with only one positive association observed for isovalerate ($p<0.001$, positive). \textit{Lachnospira} also shows a mixed pattern, with a single positive association to glycocholate ($p=0.016$, positive) but negative associations with several amino acids, including L-isoleucine ($p<0.001$, negative), L-leucine ($p<0.001$, negative), L-valine ($p<0.001$, negative), L-phenylalanine ($p=0.002$, negative), L-tyrosine ($p<0.001$, negative), and L-serine ($p=0.042$, negative). Finally, \textit{Merdicola} demonstrates exclusively positive associations, including glycocholate ($p<0.001$, positive), taurocholate ($p=0.007$, positive), L-isoleucine ($p=0.002$, positive), L-leucine ($p<0.001$, positive), L-valine ($p=0.007$, positive), L-phenylalanine ($p<0.001$, positive), and L-tyrosine ($p=0.047$, positive).

In summary, Subset~1 results reveal genera with targeted effects on SCFA and amino acid
pathways, while Subset~2 uncovers systemic positive or negative regulators of CRC-linked
metabolites. Explicit reporting of Bayesian $p$-values and directionality highlights the
statistical robustness of these associations and provides a foundation for downstream
biological validation.

\subsection{Disease-stratified comparison of master predictors}
To examine whether analogous master predictors could also be identified among healthy subjects, we performed an additional B-MASTER analysis using the healthy control samples ($n=\,$127) while retaining the original CRC-based analysis presented in the main manuscript unchanged. The same microbiome-metabolite modeling framework, posterior selection rule, and Fractional Influence Score (FIS) definition were applied to the healthy cohort. This enabled a direct comparison of genus-level master predictor rankings and influence structures between CRC and healthy subjects, allowing us to assess whether the inferred microbiome-metabolite influence architecture was shared or disease-specific.

\begin{figure}[htbp]
    \centering
    \includegraphics[width=0.99\textwidth]{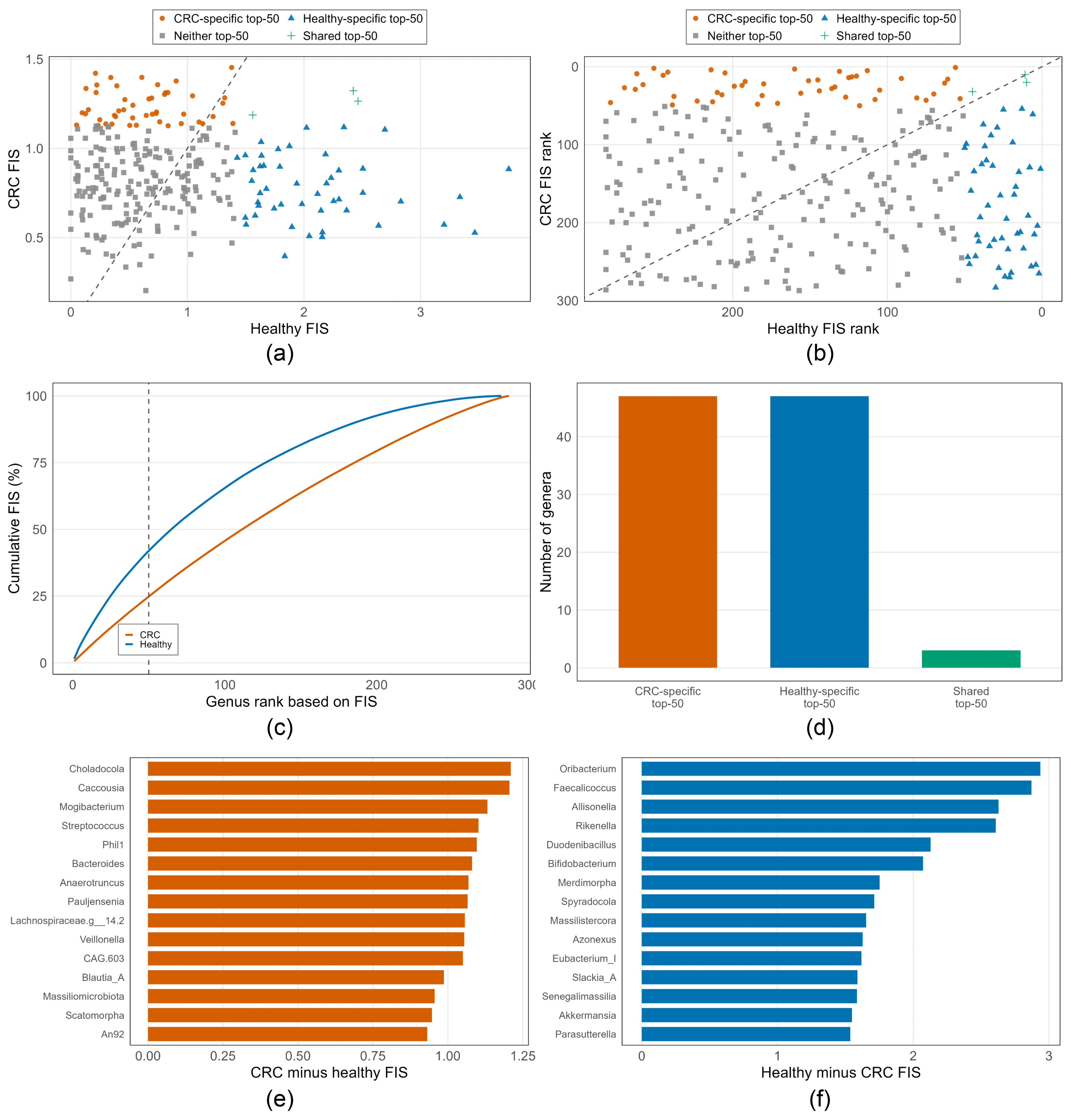}
    \caption{
Comparison of master predictor structures identified by B-MASTER in colorectal cancer (CRC) and healthy cohorts. 
(a) Scatterplot comparing genus-level Fractional Influence Scores (FIS) between healthy and CRC cohorts, highlighting CRC-specific top-50 genera, healthy-specific top-50 genera, shared top-50 genera, and remaining genera. 
(b) Comparison of genus rankings based on FIS between healthy and CRC cohorts. The diagonal dashed line denotes identical rankings across cohorts. 
(c) Cumulative FIS concentration curves for CRC and healthy cohorts. The dashed vertical line marks the top-50 genera threshold. 
(d) Number of genera classified as CRC-specific top-50, healthy-specific top-50, or shared top-50 across cohorts. 
(e) Top genera exhibiting the largest positive differential influence scores (CRC minus healthy FIS), representing CRC-enriched master predictors. 
(f) Top genera exhibiting the largest negative differential influence scores (healthy minus CRC FIS), representing healthy-enriched master predictors.
}
\label{fig:CRC_HEALTHY_master_predictor_comparison_edited}
\end{figure}

Figure~\ref{fig:CRC_HEALTHY_master_predictor_comparison_edited} summarizes this comparison. The scatterplot of CRC versus healthy FIS values in Figure~\ref{fig:CRC_HEALTHY_master_predictor_comparison_edited}(a) shows substantial separation between CRC-specific and healthy-specific top-ranked genera. Only three genera were shared between the CRC and healthy top-50 FIS rankings, whereas 47 genera were CRC-specific top-50 genera and 47 were healthy-specific top-50 genera. This corresponds to a Jaccard overlap of only 0.0309, indicating that the highest-ranking master predictors were largely distinct across the two cohorts.

The rank comparison in Figure~\ref{fig:CRC_HEALTHY_master_predictor_comparison_edited}(b) provides a complementary view of this disease-specific reordering. Many CRC-specific top-50 genera had substantially lower ranks in the healthy cohort, and conversely many healthy-specific top-50 genera were ranked much lower in CRC. This pattern supports the interpretation that B-MASTER is not merely identifying genera with uniformly high influence across all subjects, but rather detecting cohort-specific microbiome-metabolite regulatory patterns.

The cumulative FIS curves in Figure~\ref{fig:CRC_HEALTHY_master_predictor_comparison_edited}(c) further suggest that the distribution of influence differs between groups. The top 50 genera accounted for 24.85\% of total FIS in the CRC cohort, compared with 41.98\% in the healthy cohort. Thus, the healthy cohort exhibited a more concentrated influence structure, whereas the CRC cohort showed a more diffuse distribution of microbiome-metabolite influence across genera. The corresponding top-50 classification counts are shown in Figure~\ref{fig:CRC_HEALTHY_master_predictor_comparison_edited}(d).

\begin{table}[!t]
\centering
\renewcommand{\arraystretch}{1.2}
\resizebox{0.85\textwidth}{!}{%
\begin{tabular}{|cccccccc|}
\hline
Category & Genus & \begin{tabular}[c]{@{}c@{}}CRC\\ rank\end{tabular} & \begin{tabular}[c]{@{}c@{}}CRC\\ FIS\end{tabular} & \begin{tabular}[c]{@{}c@{}}Healthy\\ rank\end{tabular} & \begin{tabular}[c]{@{}c@{}}Healthy\\ FIS\end{tabular} & \begin{tabular}[c]{@{}c@{}}$\Delta$FIS\\ (CRC - Healthy)\end{tabular} & \begin{tabular}[c]{@{}c@{}}Percentile\\ shift\end{tabular} \\ \hline
\multicolumn{1}{|c|}{\multirow{10}{*}{\begin{tabular}[c]{@{}c@{}}CRC-enriched\\ (Top 10)\end{tabular}}} & \multicolumn{1}{c|}{Choladocola} & 2 & 1.42 & 251 & 0.21 & 1.21 & 87.1 \\
\multicolumn{1}{|c|}{} & \multicolumn{1}{c|}{Caccousia} & 9 & 1.34 & 262 & 0.13 & 1.21 & 88.5 \\
\multicolumn{1}{|c|}{} & \multicolumn{1}{c|}{Mogibacterium} & 7 & 1.36 & 242 & 0.22 & 1.13 & 82.2 \\
\multicolumn{1}{|c|}{} & \multicolumn{1}{c|}{Streptococcus} & 27 & 1.20 & 270 & 0.10 & 1.10 & 85.0 \\
\multicolumn{1}{|c|}{} & \multicolumn{1}{c|}{Phil1} & 11 & 1.31 & 245 & 0.22 & 1.10 & 81.8 \\
\multicolumn{1}{|c|}{} & \multicolumn{1}{c|}{Bacteroides} & 46 & 1.13 & 279 & 0.05 & 1.08 & 81.5 \\
\multicolumn{1}{|c|}{} & \multicolumn{1}{c|}{Anaerotruncus} & 29 & 1.19 & 263 & 0.13 & 1.07 & 81.8 \\
\multicolumn{1}{|c|}{} & \multicolumn{1}{c|}{Pauljensenia} & 23 & 1.22 & 258 & 0.15 & 1.07 & 82.2 \\
\multicolumn{1}{|c|}{} & \multicolumn{1}{c|}{Lachnospiraceae.g\_14.2} & 70 & 1.06 & 282 & 0.00 & 1.06 & 74.1 \\
\multicolumn{1}{|c|}{} & \multicolumn{1}{c|}{Veillonella} & 4 & 1.40 & 214 & 0.34 & 1.05 & 73.4 \\ \hline
\multicolumn{1}{|c|}{\multirow{10}{*}{\begin{tabular}[c]{@{}c@{}}Healthy-enriched\\ (Top 10)\end{tabular}}} & \multicolumn{1}{c|}{Oribacterium} & 265 & 0.53 & 2 & 3.47 & -2.94 & -92.0 \\
\multicolumn{1}{|c|}{} & \multicolumn{1}{c|}{Faecalicoccus} & 131 & 0.88 & 1 & 3.76 & -2.87 & -45.5 \\
\multicolumn{1}{|c|}{} & \multicolumn{1}{c|}{Allisonella} & 254 & 0.57 & 4 & 3.20 & -2.63 & -87.4 \\
\multicolumn{1}{|c|}{} & \multicolumn{1}{c|}{Rikenella} & 204 & 0.73 & 3 & 3.34 & -2.61 & -70.3 \\
\multicolumn{1}{|c|}{} & \multicolumn{1}{c|}{Duodenibacillus} & 215 & 0.70 & 5 & 2.83 & -2.13 & -73.4 \\
\multicolumn{1}{|c|}{} & \multicolumn{1}{c|}{Bifidobacterium} & 256 & 0.57 & 7 & 2.64 & -2.07 & -87.1 \\
\multicolumn{1}{|c|}{} & \multicolumn{1}{c|}{Merdimorpha} & 191 & 0.75 & 9 & 2.50 & -1.75 & -63.6 \\
\multicolumn{1}{|c|}{} & \multicolumn{1}{c|}{Spyradocola} & 233 & 0.65 & 12 & 2.37 & -1.71 & -77.3 \\
\multicolumn{1}{|c|}{} & \multicolumn{1}{c|}{Massilistercora} & 270 & 0.50 & 21 & 2.16 & -1.65 & -87.1 \\
\multicolumn{1}{|c|}{} & \multicolumn{1}{c|}{Azonexus} & 264 & 0.53 & 20 & 2.16 & -1.63 & -85.3 \\ \hline
\multicolumn{1}{|c|}{\multirow{3}{*}{Shared top-50}} & \multicolumn{1}{c|}{Succinivibrio} & 10 & 1.32 & 11 & 2.42 & -1.10 & 0.3 \\
\multicolumn{1}{|c|}{} & \multicolumn{1}{c|}{CAG.83} & 20 & 1.26 & 10 & 2.46 & -1.20 & -3.5 \\
\multicolumn{1}{|c|}{} & \multicolumn{1}{c|}{Onthousia} & 32 & 1.19 & 45 & 1.56 & -0.37 & 4.5 \\ \hline
\end{tabular}}
\caption{
Differential comparison of master predictors identified by B-MASTER in colorectal cancer (CRC) and healthy cohorts. 
The table summarizes the top CRC-enriched genera (largest positive $\Delta$FIS), top healthy-enriched genera (largest negative $\Delta$FIS), and representative genera shared among the top-50 FIS rankings in both cohorts. 
For each genus, we report the cohort-specific Fractional Influence Score (FIS), corresponding FIS rank, differential FIS ($\Delta$FIS = FIS$_{\mathrm{CRC}}$ - FIS$_{\mathrm{Healthy}}$), and percentile rank shift between cohorts. 
Large positive percentile shifts indicate substantially stronger influence in CRC relative to healthy controls, whereas large negative shifts indicate stronger influence in healthy subjects.
}
\label{tab:CRC_HEALTHY_master_predictor_comparison_table}
\end{table}

Table~\ref{tab:CRC_HEALTHY_master_predictor_comparison_table} lists the leading differentially enriched master predictors. Among CRC-enriched genera, \textit{Choladocola}, \textit{Caccousia}, \textit{Mogibacterium}, \textit{Streptococcus}, and \textit{Bacteroides} showed large positive differential FIS values and pronounced upward rank shifts in CRC relative to healthy controls. Conversely, \textit{Oribacterium}, \textit{Faecalicoccus}, \textit{Allisonella}, \textit{Rikenella}, and \textit{Bifidobacterium} showed substantially larger FIS values in healthy subjects. These patterns are also visualized in Figure~\ref{fig:CRC_HEALTHY_master_predictor_comparison_edited}(e)--(f), which display the largest CRC-enriched and healthy-enriched differential FIS values, respectively.

The target-set comparison was defined consistently with the master predictor analysis: for a given genus, its target set consists of the metabolites selected as associated with that genus under the posterior selection rule used for the FIS calculation. Under this definition, differentially enriched master predictors also exhibited marked differences in target-set breadth. For example, several CRC-enriched genera had high CRC FIS values and broad CRC target sets while having little or no corresponding influence in healthy controls. This provides additional biological support for disease-associated restructuring of microbiome-metabolite relationships.

Overall, this disease-stratified analysis indicates that B-MASTER identifies both CRC-enriched and healthy-enriched master predictors, with limited overlap among the highest-ranking genera. These findings complement the main CRC-focused analysis and suggest that the inferred master predictor structure reflects disease-associated changes in system-level microbiome-metabolite regulation rather than a generic influence pattern shared uniformly across cohorts.

\clearpage
\bibliography{reference}